%% file: Thesis.tex
\documentclass[12pt, oneside]{report}

\usepackage{graphicx,color}

\usepackage{setspace}                   
\newcommand{\arctg}{\mathop{\rm atan}\nolimits}
\newcommand{\tr}{\mathop{\rm tr}\nolimits}
\newcommand{\p}{{\bf p}}
\begin{document}

\input{Titolo.tex}

\pagenumbering{roman}
\setcounter{page}{1}


\tableofcontents

\newpage
\doublespacing
\pagenumbering{arabic}
\setcounter{page}{1}

\chapter*{Introduzione}
\addcontentsline{toc}{chapter}{Introduzione}
\input{Introduzione.tex}
\clearpage

\chapter*{Introduction}
\addcontentsline{toc}{chapter}{Introduction}
\input{Introduction.tex}
\clearpage

\chapter{Dilute Bose gas}
\input{DBGintro.tex}

\section{Mean-field description: Gross-Pitaevskii equation}
\input{GPE.tex}

\section{Beyond mean-field: Bogoliubov theory}
\input{Bogoliubov.tex}

\chapter{Dilute Bose gas with disorder: perturbation expansion}
\section{Introduction}
\input{DBGDintro.tex}
\section{Bogoliubov theory in the presence of disorder
\label{Bogoliubov theory in disorder} \label{BogDis}}
\input{disorder.tex}
\input{BogoliubovDis.tex}
\section{Superfluid density}
\input{SD.tex}
\input{GPsd.tex}

\chapter{Quantum Monte Carlo Method}
\section{Diffusion Monte Carlo}
\input{DMC.tex}
\section{Homogeneous Bose Gas}
\input{HBG.tex}
\input{values.tex}
\input{errors.tex}

\chapter{Dilute Bose gas with disorder: a Diffusion Monte Carlo study}
\section{Introduction}
\input{intro.tex}
\section{Trial wavefunction}
\input{trial.tex}
\section{Average over disorder}
\input{disorder2.tex}

\section{Results}
\input{results.tex}

\addcontentsline{toc}{chapter}{Referencies}

\input{refer.tex}
\begin{appendix}
\chapter{Appendix}
\input{appendix.tex}
\end{appendix}

\chapter*{Acknowledgements}
\addcontentsline{toc}{chapter}{Acknowledgements}
\input{acknow.tex}

\end{document}

%% file: Titolo.tex
\thispagestyle{empty}
{\bf
\Large
\centerline{UNIVERSIT\`A DEGLI STUDI DI TRENTO}
\vspace{1.0cm}
\centerline{FACOLT\`A DI SCIENZE MATEMATICHE}
\centerline{FISICHE E NATURALI}
\vspace{1.5cm}
\large
\centerline{TESI DI LAUREA IN FISICA}
}
\vspace{3.0cm}

\Large
{\bf
\centerline{Propriet\`a di un Gas di Bose}
\vspace{0.5cm}
\centerline{in Presenza di Disordine}
}
\vspace{1.5cm}

\Large
{\bf
\centerline{Properties of a Bose Gas}
\vspace{0.5cm}
\centerline{in the Presence of Disorder}
}

\vspace{3.0cm}

\normalsize
$$
\begin{array}{p{6cm}p{3cm}p{6cm}}
Relatori:&&Laureando:\\
{\large Dr. Stefano Giorgini}&&{\large Grigori Astrakhartchik}\\
{\large Prof. Lev P. Pitaevskii}&&
\end{array}
$$

\vspace{2cm}

\centerline{ANNO ACCADEMICO 2000-2001}

%% file: Introduzione.tex
La condensazione di Bose-Einstein (BEC) bench\`e proposta da
Einstein \cite{Einstein 1924, Einstein 1925} per un gas ideale
quantistico molto tempo fa (1924), rimase solo un artificio
matematico fino al 1938 quando London \cite{London} la utilizz\`o
per spiegare la superfluidita dell $^4$He liquido. Recentamente
(1995), dopo numerosi tentativi, la BEC venne osservata in una
serie di esperimenti sui vapori di metalli alcalini \cite{JILA,
MIT}. Da allora si \`e sviluppato notevolemente un interesse a
livello mondiale per lo studio dei gas di Bose diluiti sia dal
punto di vista sperimentale che teorico.

Negli ultimi anni, inoltre una grande attenzione \`e stata dedicata
allo studio dei sistemi di Bose disordinati. La realizzazione
sperimentale di questi sistemi \`e ottenuta per mezzo dell'
assorbimento di $^4$He liquido da parte di vari materiali porosi
(quali {\em vycor} o {\em aerogel}). Questi sistemi mostrano varie
interessanti propriet\`a non ancora comprese a fondo a livello
teorico, quali la soppressione della superfluidit\`a \cite{Reppy},
la grande variet\`a di eccitazione elementari \cite{Kanwal} ed il
comportamento critico, diverso da quello di {\em bulk}, vicino
alla transizione di fase \cite{Wong}.

In questa tesi \`e stato studiato un gas di Bose in presenza di
impurit\`a fisse. Questo modello costituisce una buona
approssimazione dell' $^4$He liquido assorbito in materiali porosi
e pu\`o diventare importante per descrivere un condensato di Bose
in presenza di impurit\`a pesanti.

A temperatura zero il sistema \`e descritto dai seguenti parametri:

a) $na^3$ (parametro del gas), dove $n$ \`e la densit\`a di particelle
ed $a$ \`e la lunghezza di {\em scattering} in onda $s$.

b) $\chi=n^{imp}/n$ e\` la concentrazione di impurezze con una
distribuzione random uniforme.

c) $b/a$, dove $b$ \`e la lunghezza di scattering in onda $s$ tra
particella-impurezza.

Nella prima parte della Tesi si \`e studiato il gas di Bose
diluito, trattando il potenziale esterno random in maniera
perturbativa. In questo regime, utilizzando il modello di
Bogoliubov, si possono ottenere analiticamente l'effetto del
disordine sull'energia dello stato fondomentale, il comportamento
superfuido e la componente di condensato.

Nella seconda parte della Tesi si \`e affrontato il problema
utilizzando il metodo {\em Diffusion Monte Carlo} (DMC). Questo
metodo numerico permette di risolvere esattamente l'equazione di
Schr\"odinger a molti corpi per lo stato fondamentale di sistema di
bosoni. Il DMC \`e stato utilizzato per lo studio del regime a
basso disordine ed i risultati della simulazione mostrano perfetto
accordo con quanto previsto dal modello di Bogoliubov. Il metodo
DMC \`e adatto anche allo studio del regime di forte disordine. In
questo regime abbiamo studiato la relazione tra il componento
superfluido e la condensazione di Bose-Einstein. A bassa
densiti\`a, troviamo che le componenti superfluide e condensate del
sistema sono ugualmente soppresse dal disordine. Tuttavia, per
concentrazioni molto alte di impurezze, troviamo che la frazione di
superfluido $\rho_s/\rho$ diviene significativamente pi\'u piccola
della alla frazione di condensato $N_0/N$.

Di seguito si presenta l'organizzazione della Tesi:

Nel primo capitolo vengono riviste brevemente dapprima la teoria di
campo medio di Gross-Pitaevski e la teoria di Bogoliubov per il gas
di Bose diluito. Viene derivata l'equazione di Gross-Pitaevskii per
il parametro d'ordine e applicata al calcolo dell'energia dello
stato fondamentale del sistema. Le energie dell eccitazioni
elementari sono ottenute considerando piccole oscillazioni del
parametro d'ordine intorno alla soluzione d'equilibrio. Andando
oltre l'approssimazione della teoria di campo medio si \`e discussa
l'Hamiltoniana efficace di Bogoliubov per il gas di Bose diluito.
Utilizzando questo modello sono state calcolate le correzioni
all'energia dello stato fondamentale, dovute alle fluttuazioni
quantistiche e lo spettro di eccitazione. Discutiamo inoltre i
risultati per la frazione di particelle non condensate e per la
matrice densit\`a ad un corpo.

Nel secondo capitolo \`e presentata la teoria del gas di Bose
diluito in presenza di disordine. Utilizzando il modello di
Bogoliubov si sono studiati gli effetti di un debole potenziale
esterno (random), che simula la distribuzione delle impurit\`a.
Sono state calcolate la correzione all'energia dello stato
fondamentale e la soppressione condensato dovuto al disordine ed il
comportamento della matrice densit\`a ad un corpo. La seconda parte
del capitolo \`e dedicata alla definizione microscopica della
densit\`a superfluida. Usando il modello di Bogoliubov abbiamo
studiato l'effetto del campo esterno sulla densit\`a superfluida.
Gli stessi risultati sono ottenuti con un nuovo metodo che fa uso
dell'equazione di Gross-Pitaevskii.

Il terzo Capitolo \`e dedicato al metodo Monte Carlo. La tecnica
DMC viene brevemente descritta e sono discusse le sue principali
caratteristiche. Viene inoltre presentata l'implementazione di una
versione parallela dell'algoritmo. I metodi {\em Diffusion Monte
Carlo} (DMC) e {\em Variational Monte Carlo} (VMC) sono applicati
ad un gas di Bose omogeneo a sfere dure: una specifica funzione di
prova viene costruita e testata. Sono inoltre discusse le tecniche
per il calcolo dell'energia dello stato fondamentale e la matrice
densit\`a ad un corpo. Si deriva inoltre una formula per il calcolo
della densit\`a suprefluida con l'algoritmo DMC e si mostra la sua
indipendenza dalla funzione di prova. I vari tipi di errore
sistematico presenti nel DMC sono studiati nell'applicazione al
modello a sfere dure.

Nell'ultimo capitolo si applica il metodo DMC allo studio del gas
di Bose in presenza di impurezze fisse schematizzate con sfere
dure. Si prova che l'energia dello stato fondamentale di un sistema
diluito $na^3 \ll 1$ nel regime di basso disordine $\chi(b/a)^2 \ll
1$ \`e descritto correttamente dalle predizioni del modello di
Bogoliubov. Si \`e studiata la dipendenza della frazione
suprefluida $\rho_s/\rho$ e della frazione di condensato $N_0/N$
dalla densit\`a $na^3$ e dai parametri di disordine $\chi$ e $b/a$.
Abbiamo trovato che, nel limite di sistemi diluiti con debole
disordine, sia $\rho_s/\rho$ che $N_0/N$ sono in accordo con le
predizioni analitiche. Si \`e verificata l'esistenza di un
comportmento di {\em scaling} in termini del parametro
$R=\chi(b/a)^2$, come predetto dal modello di Bogoliubov, e si \`e
dimostrato che questo \`e valido in un vasto {\em range} di valori
di $R$. L'utilizzo del metodo DMC ha permesso anche di indagare il
regime ad alto disordine. A basse densit\`a e per alti valori di
$R$ abbiamo trovato che il sistema entra in un regime in cui la
superfluidit\`a viene fortemente soppressa, mentre la frazione di
condensato rimane grande. \`E stata calcolata la dipendenza
spaziale della matrice densit\`a ad un corpo e si \`e mostrato
l'accordo con le previsioni analitiche per basse densit\`a. Abbiamo
inoltre mostrato che la transizione quantistica da superfluido ad
isolante \`e assente nel nostro modello di impurezze non
sovrapponibilli

%% file: Introduction.tex
Although proposed by Einstein \cite{Einstein 1924, Einstein 1925}
for an ideal quantum gas a long time ago (1924) Bose-Einstein
condensation (BEC) remained only as a mathematical artifact until
London ``rediscovered'' it in 1938 to explain the superfluidity of
liquid $^4$He \cite{London}. Recently (1995), after many years of
struggle, BEC was observed in alkali vapors in a remarkable series
of experiments \cite{JILA, MIT}. Since that time there has been an
explosion of experimental and theoretical interest worldwide in the
study of dilute Bose gases (for a review see \cite{BEC in trapped gases}).

In the last years great attention has been also devoted to the
investigation of disordered Bose systems. The experimental realizations
of these systems are liquid $^4$He adsorbed in various types of
porous media such as vycor and aerogel. These systems exhibit
many interesting properties, which have not yet been fully
understood theoretically, such as the suppression of superfluidity
\cite{Reppy}, a rich variety of elementary excitations \cite{Kanwal}
and a critical behavior near the phase transition different from
the bulk \cite{Wong}.

In this Thesis we study a Bose gas in the presence of quenched
impurities. This model provides a reasonable description of liquid
$^4$He adsorbed in porous media and can become relevant for Bose
condensed gases in the presence of heavy impurities.

At zero temperature the system is described by the following
parameters:\\
a) $na^3$ (gas parameter) where $n$ is the density of particles $na^3$
and $a$ is the $s$-wave scattering length,\\
b) $\chi = n^{imp}/n$ is the concentration of impurities
with a uniform random distribution,\\
c) $b/a$ where $b$ is the particle-impurity $s$-wave scattering length

In the first part of the Thesis we investigate the dilute Bose gas
by treating the random external potential as a perturbation. In
this regime one can work out analytically, within the Bogoliubov
model, the effect of disorder on the ground-state energy,
superfluid behavior and condensate fraction.

In the second part of the Thesis we approach the problem by
resorting to the Diffusion Monte Carlo (DMC) method. This numerical
method solves exactly the many-body Schr\"odinger equation for the
ground-state of a system of bosons. This method is used for the
investigation of the weak disorder regime and results of the
simulations agree with the predictions of the Bogoliubov model.
Also the DMC method is well suited to study the regime of strong
disorder. In this regime we investigate the relation between
superfluid behavior and Bose-Einstein condensation. At low
densities, we find that the superfluid and condensate components of
the system are equally suppressed by the disorder. However, for the
very large concentration of impurities, we find that the superfluid
fraction $\rho_s/\rho$ becomes significantly smaller than the
condensate fraction $N_0/N$.

The structure of this Thesis is as follows:

In the first chapter the mean-field Gross-Pitaevskii theory and the
beyond mean-field Bogoliubov theory of the dilute Bose gas are
briefly reviewed. The Gross-Pitaevskii equation for the order
parameter is derived and applied to the calculatation of the
ground-state energy of the system. The elementary excitation
energies are obtained by considering the small oscillations of the
order parameter around the equilibrium solution. Beyond mean-field
approximation we discuss the Bogoliubov effective Hamiltonian of a
dilute Bose gas and calculate within this model the excitation
spectrum and corrections to the the ground-state energy arising
from quantum fluctuations. The results for the fraction of
noncondensed particles and the one-body density matrix at zero
temperature are also discussed.

In the second chapter we discuss the theory of a dilute Bose gas in
the presence of disorder. Within the Bogoliubov model we study the
effects of the weak external random potential, modeled by the uniform
random distribution of quenched impurities. The corrections to the
ground-state energy and the condensate depletion due to the
external random potential are calculated, as well as the behavior
of the one-body density matrix. The second part of this chapter is
devoted to the microscopic definition of the superfluid density. By
using the Bogoliubov model we investigate the effect of the
external random field on the superfluid density. The same result is
also obtained in a new alternative way which makes use of the
Gross-Pitaevskii equation.

Chapter Three is devoted to the Quantum Monte Carlo method. The
Diffusion Monte Carlo technique is briefly described and its main
features are discussed. We also discuss the implementation of the
parallel version of the algorithm. The Diffusion Monte Carlo (DMC)
and Variational Monte Carlo (VMC) methods are applied to a
hard-sphere homogeneous Bose gas, and a specific trial
wave-function is constructed and tested. The techniques of
calculating the ground-state energy and the one-body density matrix
are presented. A formula for the calculation of the superfluid
density within the DMC algorithm is derived and proved to be
unbiased by the trial wavefunction. All types of systematic errors
present in the DMC algorithm applied to the hard-sphere model are
investigated.

In the last chapter we apply the DMC method to investigate a Bose
gas in the presence of hard-sphere quenched impurities. We show
that the ground-state energy of a dilute system $na^3 \ll 1$ in the
``weak'' disorder regime $\chi(b/a)^2 \ll 1$ is described correctly
by the prediction of the Bogoliubov model. We study the dependence
of the superfluid fraction $\rho_s/\rho$ and condensate fraction
$N_0/N$ on the density $na^3$ and disorder parameters $\chi$ and
$b/a$. We find that in limit of dilute systems and weak disorder
both $\rho_s/\rho$ and $N_0/N$ are in agreement with analytical
predictions. The existence of scaling in $R=\chi(b/a)^2$, as
predicted by Bogoliubov model, is checked and is shown to be valid
over a large range of $R$. The use of the DMC method enables us to
investigate the regime of strong disorder. At low density and large
values of $R$ we find that the system enters a regime where the
superfluid density is strongly suppressed, whereas the condensate
fraction is still large. The space dependence of the one-body
density matrix is calculated and is shown to agree with analytical
predictions at small densities $na^3$. We show that the
superfluid-insulator quantum transition is absent within our model
of non-overlapping impurities.

%% file: DBGintro.tex
\section{Introduction}
The present chapter is devoted to the theory of homogeneous
dilute Bose gases at zero temperature.

The mean-field theory for the dilute Bose gas is discussed in the
first part of the chapter. We derive the Gross-Pitaevskii equation
for the order parameter and we use it to calculate the ground-state
energy of the system. The small oscillations of the order parameter
around the equilibrium solution provide us with the elementary
excitation energies. The presentation of the material in this
section follows closely the review \cite{BEC in trapped gases}.

In the second part of the chapter we discuss the Bogoliubov model,
which is a theory beyond mean-field and takes into account the
fluctuations of the order parameter. We introduce Bogoliubov
effective Hamiltonian, discuss its diagonalization by means of the
Bogoliubov transformation and we calculate the corrections to the
of the ground state energy arising from the quantum fluctuations.
The excitation spectrum predicted by the Bogoliubov model agrees
with the one obtained from the time-dependent Gross-Pitaevskii
equation. Results for the number of noncondensed particles (quantum
depletion) and the one-body density matrix are also discussed. Much
of the treatment of this part of the chapter parallels closely the
one given in the book \cite{StatPhys II}.

%% file: GPE.tex
\subsection{Dilute Bose gas}

The Hamiltonian of a system of spinless bosons, interacting through the pair
potential $U$ and immersed in the external field $V({\bf r})$ is given,
in second quantization, by

\begin{eqnarray}
\begin{array}{rcl}
\hat H &=&\displaystyle
\int \hat \Psi^{\dagger}({\bf r}) \left(-\frac{\hbar^2}{2m}\triangle + V({\bf r})\right)
\hat\Psi({\bf r})\,{\bf dr}\quad+\\
&&+\quad\displaystyle
\frac{1}{2} \int\!\!\int \hat \Psi^{\dagger}({\bf r_2}) \hat \Psi^{\dagger}({\bf r_1})
\,U(|{\bf r_1 - r_2}|)\,
\hat \Psi({\bf r_1}) \hat \Psi({\bf r_2})\,{\bf dr_1 dr_2},
\end{array}
\label{second quantization}
\end{eqnarray}

here $m$ is the mass of a particle, $\hat\Psi({\bf r})$ and
$\hat\Psi^\dagger({\bf r})$ are the boson field operators that
annihilate and create a particle at the position ${\bf r}$.

If the gas is dilute and cold, then the two-body potential can be
replaced by the pseudopotential $U({\bf r'-r}) = g \delta({\bf
r'-r})$ which is fixed by a single parameter, the $s$-wave
scattering length $a$, through the coupling constant

\begin{eqnarray}
g = \frac{4\pi\hbar^2a}{m}
\label{g}
\end{eqnarray}

The Hamiltonian (\ref{second quantization}) takes the form

\begin{eqnarray}
\hat H =
\int \hat \Psi^{\dagger}({\bf r}) \left(-\frac{\hbar^2}{2m}\triangle + V({\bf r})\right)
\hat\Psi({\bf r})\,{\bf dr} +
\frac{g}{2} \int\!\!\int \hat \Psi^{\dagger}({\bf r}) \hat \Psi^{\dagger}({\bf r})
\hat \Psi({\bf r}) \hat \Psi({\bf r})\,{\bf dr},
\label{second quantization 2}
\end{eqnarray}

The field operator can decomposed as $\hat\Psi({\bf r}) = \sum_{\bf k}
\hat a_{\bf k} \phi_{\bf k}({\bf r})$, where $\phi_{\bf k}({\bf r})$ are
single-particle wavefunctions with quantum number ${\bf k}$. The
bosonic creation and annihilation operators $\hat a_k$ and $\hat
a^\dagger_k$ are defined in Fock space through the relations

\begin{eqnarray}
\hat a^\dagger_{\bf k}~|n_0,\,n_1,\,...,\,n_k,\,...\rangle &=&
\sqrt{n_k+1} |n_0,\,n_1,\,...,\,n_k+1,\,...\rangle,\\
\hat a_{\bf k}~|n_0,\,n_1,\,...,\,n_k,\,...\rangle &=&
~~~\sqrt{n_k}~~|n_0,\,n_1,\,...,\,n_k-1,\,...\rangle,
\end{eqnarray}

where $n_k$ are the eigenvalues of the operator $\hat n_{\bf k} =
\hat a^\dagger_{\bf k} \hat a_{\bf k}$ giving the number of atoms
in the single-particle state ${\bf k}$. The operators $\hat a_{\bf
k}$ and $\hat a^\dagger_{\bf k}$ obey the usual bosonic commutation
rules

\begin{eqnarray}
[\hat a_{\bf k}, \hat a_{\bf k'}^\dagger] = \delta_{\bf k k'},\quad
[\hat a_{\bf k}, \hat a_{\bf k'}] = 0,\quad
[\hat a_{\bf k}^\dagger, \hat a_{\bf k'}^\dagger] = 0.
\label{commutation rules}
\end{eqnarray}

Bose-Einstein condensation occurs when the number of particles in
one particular single-particle state becomes very large $N_{\bf 0}
\gg 1$. In this limit the states with $N_{\bf 0}$ and $N_{\bf 0}
\pm 1 \approx N_{\bf 0}$ correspond to the same physical
configuration and, consequently, the operators $\hat a^\dagger_{\bf
0}$ and $\hat a_{\bf 0}$ can be treated as complex numbers

\begin{eqnarray}
\hat a^\dagger_{\bf 0} = \hat a_{\bf 0} = \sqrt{N_{\bf 0}}
\label{condensate}
\end{eqnarray}

For a uniform gas in a volume $V$ the good single-particle states
correspond to momentum states and BEC occurs in the single-particle
state $\psi_0 = 1 / \sqrt{V}$ having zero momentum. Thus, the field
operator $\hat\Psi_{\bf k}({\bf r})$ can be decomposed in the form
$\hat\Psi_{\bf k}({\bf r}) = \sqrt{N_0/V} + \hat\Psi'({\bf r})$.
The generalization for the case of nonuniform and time-dependent
configurations is given by

\begin{eqnarray}
\hat\Psi_{\bf k}({\bf r},t) = \Phi({\bf r},t) + \hat\Psi'({\bf r},t),
\end{eqnarray}

where the Heisenberg representation for the field operators is
used. Here $\Phi({\bf r},t)$ is a complex function defined as the
expectation value of the field operator
$\Phi({\bf r},t)~=~\langle\hat\Psi({\bf r},t)\rangle$.

The function $\Phi({\bf r},t)$ is a classical field having the
meaning of an order parameter and is often called the {\it
wave-function of the condensate}. The mean-field theory, which
describes the behavior of the classical field $\Phi({\bf r},t)$ and
ignores the fluctuations $\hat\Psi'({\bf r},t)$ is contained in the
Gross-Pitaevskii theory. A more refined approach, which takes into
account the fluctuations of the field operator was proposed by
Bogoliubov.

\subsection{Gross-Pitaevskii equation}
In order to derive the equation for the wavefunction of the
condensate $\Phi({\bf r},t)$ one has to write the time evolution of
the field operator $\hat\Psi({\bf r},t)$ using the Heisenberg
equation

\begin{eqnarray}
i \hbar \frac{\partial}{\partial t} \hat\Psi({\bf r},t)
= [\hat\Psi, \hat H]
\label{Heisenberg equation}
\end{eqnarray}

Substitution of the Hamiltonian (\ref{second quantization 2}) into
(\ref{Heisenberg equation}) gives

\begin{eqnarray}
i \hbar \frac{\partial}{\partial t} \hat\Psi({\bf r},t)
=\left( -\frac{\hbar^2}{2m}\triangle + V({\bf r})
+g |\hat\Psi({\bf r},t)|^2\right) \hat\Psi({\bf r},t)
\end{eqnarray}

We now replace the field operator $\hat\Psi({\bf r},t)$ with the
classical field $\Phi({\bf r},t)$. Then, the following equation for
the order parameter is obtained

\begin{eqnarray}
i \hbar \frac{\partial}{\partial t} \Phi({\bf r},t)
=\left( -\frac{\hbar^2}{2m}\triangle + V({\bf r})
+g|\Phi({\bf r},t)|^2 \right)\Phi({\bf r},t)
\label{Gross-Pitaevskii}
\end{eqnarray}

This equation is called Gross-Pitaevskii equation \cite{Gross 1,
Gross 2, Pitaevskii} and describes the time evolution of the order
parameter.q

\subsection{Ground state energy}

Within the formalism of the mean-field theory it is easy to obtain
the ground state energy from the stationary solution of the
Gross-Pitaevskii equation (\ref{Gross-Pitaevskii}). To this purpose
the condensate wave function should be written as $\Phi({\bf r},t)
= \phi({\bf r}) \exp(-i\mu t/\hbar)$, where $\mu$ is the chemical
potential and the function $\phi({\bf r})$ is real and normalized
to the total number of particles $\int|\phi({\bf r})|^2 {\bf dr} =
N$. Then the Gross-Pitaevskii equation becomes

\begin{eqnarray}
\left(-\frac{\hbar^2}{2m}\triangle+V({\bf r})+g\phi^2({\bf r})\right)\phi({\bf r})
= \mu \phi({\bf r})
\label{nonlinear Schrodinger}
\end{eqnarray}

It has the form of a nonlinear Schr\"odinger equation. In the
absence of interactions ($g = 0$) it reduces to the usual
single-particle Schr\"odinger equation with the Hamiltonian
$-\hbar^2\triangle/2m + V({\bf r})$.

In the uniform case, $V = 0$, $\phi({\bf r})$ is a constant
$\phi({\bf r}) = \sqrt{n}$ and the kinetic term in (\ref{nonlinear
Schrodinger}) disappears. The chemical potential is given by

\begin{eqnarray}
\mu = gn
\label{mu}
\end{eqnarray}

At $T = 0$ the chemical potential is the derivative of the energy
with respect to the number of particles $\mu = \partial E /
\partial N$. Substitution of (\ref{g}) into (\ref{mu}) and simple
integration gives the ground state energy per particle

\begin{eqnarray}
\frac{E}{N} = 4\pi(na^3)\frac{\hbar^2}{2ma^2}
\label{E GP}
\end{eqnarray}

\subsection{Elementary excitations}

In the low-temperature regime, the excited states of the system can
be calculated from the ``classical'' frequencies of the linearized
GP equation. Let us look for solutions in the form of small
oscillations of the order parameter around the stationary value.

\begin{eqnarray}
\Phi({\bf r},t) = e^{-i\mu t/\hbar}
[\phi({\bf r})+u({\bf r})e^{-i\omega t}+v^{\star}({\bf r})e^{i\omega t}]
\label{small oscillations}
\end{eqnarray}

By keeping terms linear in the complex functions $u$ and $v$,
equation (\ref{Gross-Pitaevskii}) becomes

\begin{eqnarray}
\left\{
\begin{array}{rll}
\hbar \omega u({\bf r}) & = &
[H_0-\mu+2g\phi^2({\bf r})] u({\bf r}) + g\phi^2({\bf r})v({\bf r})\\
-\hbar \omega v({\bf r}) & = &
[H_0-\mu+2g\phi^2({\bf r})] v({\bf r}) + g\phi^2({\bf r})u({\bf r})\\
\end{array}
\right.
\end{eqnarray}

where $H_0 = -\hbar^2\triangle/2m + V$.

In a uniform gas, the amplitudes $u$ and $v$ are plane waves and
the resulting dispersion law takes the Bogoliubov form

\begin{eqnarray}
(\hbar \omega)^2 =
\left(\frac{\hbar^2k^2}{2m}\right)
\left(\frac{\hbar^2k^2}{2m} + 2gn\right),
\label{Bogoliubov spectrum}
\end{eqnarray}

where ${\bf k}$ is the wave vector of the excitations and $n =
|\phi|^2$ is the density of the gas. For large momenta the spectrum
coincides with the free-particle energy $\hbar^2 k^2/2m$. At low
momenta equation (\ref{Bogoliubov spectrum}) yields the phonon
dispersion $\omega = ck$, where the sound velocity $c$ is given by
the formula

\begin{eqnarray}
c = \sqrt{\frac{gn}{m}}
\label{c}
\end{eqnarray}

%% file: Bogoliubov.tex
\subsection{Bogoliubov Hamiltonian and elementary excitations
\label{BH & EE}}

In the absence of the external potential $V({\bf r})$, the
Hamiltonian (\ref{second quantization}) can be conveniently
expressed in momentum space

\begin{eqnarray}
\hat H =
\sum \frac{p^2}{2 m}\,\hat a_{\bf p}^\dagger \hat a_{\bf p}
+ \frac{1}{2} \sum \langle{\bf p_1, p_2}|U|{\bf p_1', p_2'}\rangle\,
{\hat a_{\bf p_1}}^\dagger {\hat a_{\bf p_2}}^\dagger \hat a_{\bf p_1'}\hat a_{\bf p_2'}\,
\delta_{\bf p_1+p_2,\,p_1'+p_2'},
\label{Hamiltonian no V}
\end{eqnarray}

where the summation is carried out over all indices that appear
twice. By assuming that the relevant scattering processes involve
particles at low momenta, the matrix elements in the Hamiltonian
(\ref{Hamiltonian no V}) can be replaced by their values at zero
momenta, then

\begin{eqnarray}
\hat H =
\sum\frac{p^2}{2 m}\,\hat a_{\bf p}^\dagger \hat a_{\bf p}
+\frac{U_0}{2V}\sum{\hat a_{\bf p_1}}^\dagger{\hat a_{\bf p_2}}^\dagger \hat a_{\bf p_1'}\hat a_{\bf p_2'}
\label{Hamiltonian no V 2}
\end{eqnarray}

In a dilute gas almost all particles are found in the condensed
state $N~\approx~N_{\bf 0}~=~{\hat a_{\bf 0}}^\dagger \hat a_{\bf 0}$,
then, as it was already discussed above (see eq.
(\ref{condensate})) the operators $\hat a_{\bf 0}$ and ${\hat
a_{\bf 0}}^\dagger$ can be treated as ordinary numbers.

The application of perturbation theory means that the last term in
(\ref{Hamiltonian no V 2}) should be decomposed in powers of the
small quantities ${\hat a_{\bf p}}^\dagger$ and $\hat a_{\bf p}$,
with ${\bf p\ne 0}$. The zeroth term is

\begin{eqnarray}
{\hat a_{0}}^\dagger{\hat a_{0}}^\dagger \hat a_{0}\hat a_{0}
= a_0^4
\label{zero order terms}
\end{eqnarray}

The first order terms are absent because they do not satisfy the
law of momentum conservation. The second order terms are

\begin{eqnarray}
a_{\bf 0}^2 \sum\limits_{\bf p\ne 0} (\hat a_{\bf p}\hat a_{\bf -p}+
{\hat a_{\bf p}}^\dagger{\hat a_{\bf -p}}^\dagger+4{\hat a_{\bf p}}^\dagger \hat a_{\bf p})
\label{second order terms}
\end{eqnarray}

Here the $a_{\bf 0}^2 = N_{\bf 0}$ factor can be substituted with the total
number of particles $N$, although in equation (\ref{zero order
terms}) it is necessary to use the more precise formula

\begin{eqnarray}
a_{\bf 0}^2 + \sum\limits_{\bf p\ne 0} {\hat a_{\bf p}}^\dagger \hat a_{\bf p} = N
\end{eqnarray}

As a result the sum of equations (\ref{zero order terms}) and
(\ref{second order terms}) becomes equal to

\begin{eqnarray}
N^2 + N
\sum\limits_{\bf p\ne 0} (\hat a_{\bf p}\hat a_{\bf -p}+
{\hat a_{\bf p}}^\dagger{\hat a_{\bf -p}}^\dagger+2{\hat a_{\bf p}}^\dagger \hat a_{\bf p})
\end{eqnarray}

and substitution into Hamiltonian (\ref{Hamiltonian no V 2}) gives

\begin{eqnarray}
\hat H
= \frac{N^2}{2V}U_0
+ \sum\limits_{\bf p}\frac{p^2}{2m} {\hat a_{\bf p}}^\dagger\hat a_{\bf p}
+ \frac{N}{2V} U_0
\sum\limits_{\bf p\ne 0} (\hat a_{\bf p}\hat a_{\bf -p}+
{\hat a_{\bf p}}^\dagger{\hat a_{\bf -p}}^\dagger+2{\hat a_{\bf p}}^\dagger\hat a_{\bf p})
\label{H1}
\end{eqnarray}

The matrix element $U_0$ has to be expressed in terms of
the scattering length $a$. In the second order terms this
can be done using the first Born approximation
$U_0 = 4\pi\hbar^2a/m$, although in the zeroth order term
one should use the second Born approximation for collisions of
two particles from the condensate

\begin{eqnarray}
U_0 = \frac{4\pi\hbar^2a}{m}
\left(1+\frac{4\pi\hbar^2a}{V}\sum\limits_{\bf p\ne 0}\frac{1}{p^2}\right)
\end{eqnarray}

or by introducing the speed of sound (see equation (\ref{c}))

\begin{eqnarray}
c = \sqrt{\frac{4\pi \hbar^2 a N}{m^2 V}}
\end{eqnarray}

one obtains

\begin{eqnarray}
U_0 = \frac{Vmc^2}{N}\left(1+\frac{1}{N}\sum\limits_{\bf p\ne 0}
\biggl(\frac{mc}{p}\biggr)^2\right)
\end{eqnarray}

Substitution of this formula into (\ref{H1}) gives

\begin{eqnarray}
\begin{array}{lcr}
\hat H &\quad=\quad&\displaystyle
\frac{N}{2} mc^2
\left(1+\frac{1}{N}\sum\limits_{\bf p\ne 0}\biggl(\frac{mc}{p}\biggr)^2\right)
+ \sum\limits_{\bf p}\frac{p^2}{2m}\,{\hat a_{\bf p}}^\dagger\hat a_{\bf p}
\quad+\\&&
\displaystyle +\quad\frac{mc^2}{2}\sum\limits_{\bf p\ne 0} (\hat a_{\bf p}\hat a_{\bf -p}+
{\hat a_{\bf p}}^\dagger{\hat a_{\bf -p}}^\dagger+2{\hat a_{\bf p}}^\dagger \hat a_{\bf p})
\label{H2}
\end{array}
\end{eqnarray}

In order to calculate the energy levels of the system
one has to diagonalize the Hamiltonian (\ref{H2}).
This can be accomplished by using the {\it Bogoliubov canonical transformation}
of the field operators \cite{Bogoliubov}.
The operators $\hat a_{\bf p}^\dagger$ and $\hat a_{\bf p}$ should be
expressed as a linear superposition of the qusiparticle operators
$\hat b_{\bf p}^\dagger$ and $\hat b_{\bf p}$

\begin{eqnarray}
\left\{
\begin{array}{rcl}
\hat a_{\bf p}&=&u_p \hat b_{\bf p} + v_p \hat b_{\bf -p}^\dagger\\
\hat a_{\bf p}^\dagger&=&u_p \hat b_{\bf p}^\dagger + v_p \hat b_{\bf -p}\\
\end{array}
\right.
\label{Bogoliubov transformation}
\end{eqnarray}

which have to satisfy the same commutation rules as the operators
$\hat a_{\bf p}^\dagger,\hat a_{\bf p}$ (see eq.(\ref{commutation rules}))

\begin{eqnarray}
[\hat b_{\bf p}, \hat b_{\bf p'}^\dagger] = \delta_{\bf pp'},\quad
[\hat b_{\bf p}, \hat b_{\bf p'}] = 0,\quad
[\hat b_{\bf p}^\dagger, \hat b_{\bf p'}^\dagger] = 0.
\label{commutation b}
\end{eqnarray}

From the commutation rules (\ref{commutation b}) one can show that
the coefficients must satisfy the condition $u_p^2  - v_p^2 = 1$.
The transformation (\ref{Bogoliubov transformation}) can be rewritten as

\begin{eqnarray}
\left\{
\begin{array}{rcl}
\displaystyle\hat a_{\bf p}&=&\displaystyle\frac{\hat b_{\bf p} + L_p \hat b_{\bf -p}^\dagger}{\sqrt{1-L_p^2}}\\
\displaystyle\hat a_{\bf p}^\dagger&=&\displaystyle\frac{\hat b_{\bf p}^\dagger + L_p \hat b_{\bf-p}}{\sqrt{1-L_p^2}}
\end{array}
\right.
\label{linear}
\end{eqnarray}

Let us substitute (\ref{linear}) into the Hamiltonian (\ref{H1})
and set to zero the coefficient of the term proportional to
$\hat b_{\bf p}\hat b_{\bf -p}$. This gives an equation for $L_p$

\begin{eqnarray}
\label{bkb_p}
L_p^2 + 2 \frac{\frac{p^2}{2m}+ mc^2}{mc^2}L_p +1 = 0,
\end{eqnarray}

which has two solutions

\begin{eqnarray}
L_p = \frac{1}{mc^2}
\left(-\frac{p^2}{2m} - mc^2
\pm \sqrt{\left(\frac{p^2}{2m}\right)^2+(pc)^2}\,\right)
\label{L_p}
\end{eqnarray}

The solution with negative sign is unphysical, because the
$1-L_p^2$ term in the square root in (\ref{linear}) becomes
negative. Thus, the solution is

\begin{eqnarray}
L_p = \frac{1}{mc^2}
\left(E(p) - \frac{p^2}{2m} - mc^2\right),
\label{Lp}
\end{eqnarray}

where $E(p)$ stands for

\begin{eqnarray}
E(p) = \sqrt{\left(\frac{p^2}{2m}\right)^2+(pc)^2}
\label{Ep}
\end{eqnarray}

The condition that the coefficient of the term proportional to
$\hat b_{\bf p}^\dagger \hat b_{\bf -p}^\dagger$ be zero gives the same
equation (\ref{bkb_p}). Thus, if condition (\ref{Lp}) is satisfied,
the Hamiltonian has been diagonalized and has the form

\begin{eqnarray}
\hat H
= E_0 + \sum\limits_{\bf p}E(p)~\hat b_{\bf p}^\dagger \hat b_{\bf p},
\label{H diag}
\end{eqnarray}

where

\begin{eqnarray}
E_0 = \frac{N}{2} mc^2
+ \frac{1}{2} \sum\limits_{\bf p\ne 0}
\left[E(p) - \frac{p^2}{2m} -
mc^2\left(1 - \biggl(\frac{mc}{p}\biggr)^2\right)\right]
\label{E_0}
\end{eqnarray}

From the Hamiltonian (\ref{H diag}) and the commutation rules
(\ref{commutation b}) one can identify $\hat b_{\bf p}$ and $\hat
b_{\bf p}^\dagger$ as the creation and annihilation operators of
quasiparticles with energy $E(p)$. The ground state energy is given
by $E_0$ which is the energy of the ``vacuum'' of quasiparticles
$\hat H |0\rangle = E_0 |0\rangle$, where the ``vacuum'' state is
defined as $\hat b_{\bf p}|0\rangle = 0$ for any value of ${\bf p
\ne 0}$ The excited states are given by $|{\bf p}\rangle = \hat
b_{\bf p}|0\rangle$ and have energy $E(p)$ and momentum ${\bf p}$.

It is interesting to note that the spectrum (\ref{Ep}) of the
elementary excitations was already obtained in (\ref{Bogoliubov
spectrum}) from the Gross-Pitaevskii equations by considering small
oscillations of the order parameter around the stationary solution.

\subsection{Ground state energy}

In equation (\ref{E_0}) the discrete summation over momenta in a
volume $V$ should be replaced by integration over
$V{\bf dp}/(2\pi\hbar)^3$. The result is

\begin{eqnarray}
\frac{E_0}{N} = \frac{mc^2}{2}
\left(1+\frac{128}{15\sqrt{\pi}} (na^3)^{1/2}\right)
\end{eqnarray}

The first term of this expression gives the mean-field energy and
coincides with the result obtained from the Gross-Pitaevskii
equation (\ref{E GP}). The second term gives the correction to the
ground state energy arising from the zero-point motion of the quasiparticles

\begin{eqnarray}
\frac{E_0}{N} = \frac{\hbar^2}{2ma^2}\left(4\pi na^3+\frac{512\sqrt{\pi}}{15}(na^3)^{3/2}\right)
\label{E Bogolibov}
\end{eqnarray}

The result is valid if the system is dilute, i.e. if the gas
parameter is small $na^3 \ll 1$.

\subsection{Quantum depletion of the condensate}

Another interesting result that can be obtained from Bogoliubov
theory is the momentum distribution of the particles. The number of
particles with momentum ${\bf p}$ is given by
$N_{\bf p} = {\hat a_{\bf p}}^\dagger \hat a_{\bf p}$
or using the transformation (\ref{linear}) it is given by

\begin{eqnarray}
N_{\bf p} = \frac{n_{\bf p}+L_p^2(n_{\bf p}+1)}{1-L_p^2},
\label{Np}
\end{eqnarray}

here $n_{\bf p}= {\hat b_{\bf p}}^\dagger\hat b_{\bf p}$ is the number of elementary excitations,
which satisfy the usual Bose distribution $n_p = (\exp\{E(p)/k_B T\}-1)^{-1}$.
At zero temperature such excitations are absent and (\ref{Np}) simplifies to

\begin{eqnarray}
N_p = \frac{(mc^2)^2}{2E(p)\left[E(p)+\frac{p^2}{2m} + mc^2\right]}
\label{N_p}
\end{eqnarray}

As $p \to 0$ the momentum distribution diverges as $N_p \to mc/2p$.
The number of atoms in the condensate can be obtained by taking
the difference between $N$ and the number of non-condensed atoms.

\begin{eqnarray}
N_0 = N - \sum\limits_{\bf p \ne 0}N_p =
N -\frac{V}{(2\pi\hbar)^3}\int N_p {\bf dp}
\end{eqnarray}

The integration can be carried out and gives the result

\begin{eqnarray}
N_0 = N \left(1 - \frac{8}{3\sqrt{\pi}} (na^3)^{1/2}\right)
\label{quantum depletion}
\end{eqnarray}

Due to interactions particles are pushed out of the condensate and
a fraction of particles with nonzero momenta is present even at
zero temperature. This phenomenon is called
{\it quantum depletion of the condensate}

Also result (\ref{quantum depletion}) is valid in the dilute regime
$na^3 \ll 1$ in which the quantum depletion is small and Bogoliubov
theory applies.

\subsection{One-body density matrix \label{Bogoliubov OBDM}}

For a homogeneous system the one body density matrix is
defined as the Fourier transform of the momentum distribution (\ref{N_p})

\begin{eqnarray}
\rho({\bf r}) =
\frac{N_0}{V}+\int N_{\bf p} e^{i{\bf pr}/\hbar} \frac{\bf dp}{(2\pi\hbar)^3},
\label{rho}
\end{eqnarray}

where the contribution of the condensate has been
extracted from the integral. After angular integration one gets

\begin{eqnarray}
\rho(r) =
\frac{N_0}{V} +\rho^{(1)}(r)
= \frac{N_0}{V}+
\frac{1}{2\pi^2r} \int\limits_{0}^{\infty} N_p \sin\left(\frac{pr}{\hbar}\right)
\frac{p dp}{\hbar^2}
\label{rho 2}
\end{eqnarray}

At $T = 0$ the momentum distribution $N_p$ is given by (\ref{N_p}).
By introducing the dimensionless variable $\xi = p / mc$, one
obtains the following result for the coordinate dependent part of
the one-body density matrix

\begin{eqnarray}
\rho^{(1)}(r) =
n\frac{4}{\pi x}
\int\limits_0^\infty
\frac{\sin(\sqrt{4\pi na^3}x\xi) d\xi}
{\sqrt{4+\xi^2}(\xi\sqrt{4+\xi^2}+\xi^2+2)},
\label{OBDM 1}
\end{eqnarray}

where $r = ax$.

For $r = 0$, $\rho^{(1)}(0) = \displaystyle\frac{8}{3\sqrt{\pi}} (na^3)^{3/2}$
and the one-body density matrix coincides with the total density
$\rho(0) = n$. For $r \gg r_0$ (where $r_0 = \hbar/mc =
a/\sqrt{8\pi na^3}$ is the healing length) $\rho^{(1)}(r) =
\sqrt{na}/(2\pi\sqrt{\pi}r^2)$
(see derivation and comments in Appendix \ref{OBDM expansion}).
Thus the asymptotic value of the one-body density matrix
coincides with the condensate density
$\lim\limits_{r\to\infty} \rho(r) = N_{0}/V$.

\begin{figure}
\includegraphics[width=\textwidth]{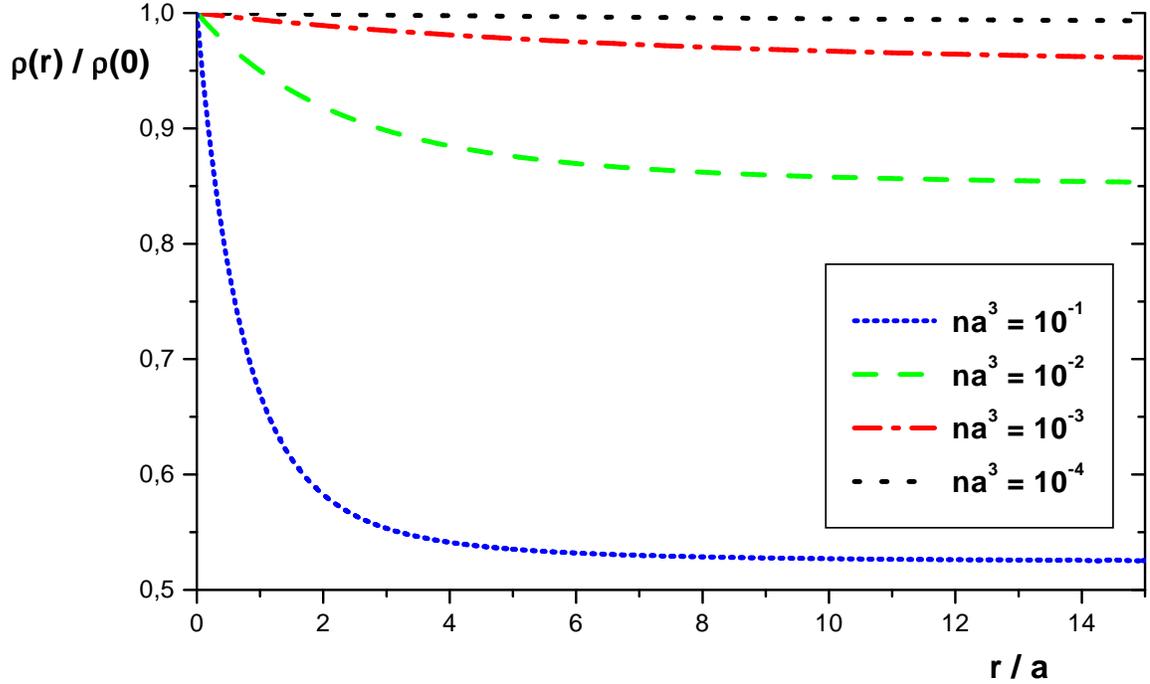}
\caption{One-body density matrix}
\label{OBDMpure}
\end{figure}

For arbitrary values of $r$ the integral (\ref{OBDM 1}) can be
calculated numerically. Results for different values of $na^3$
are shown in Fig.~{\ref{OBDMpure}}

%% file: DBGDintro.tex
In this chapter we discuss the theory of a dilute Bose gas in the
presence of disorder.

Within the Bogoliubov model we study the effects of a weak external
random potential, which is modeled by a uniform random distribution
of quenched impurities. We calculate the corrections to the ground
state energy and to the condensate fraction due to the external
random potential. We also investigate the behavior of the one-body
density matrix.

The second part of this chapter is devoted to the microscopic
definition of the superfluid density. By using the
Bogoliubov model we study the effect of the external
field on the superfluid density. This result is also obtained in
a new alternative way which makes use of the Gross-Pitaevskii equation.

%% file: disorder.tex
\subsection{Random external potential \label{Random external potential}}

As a simple model for disorder we use the potential produced by a
uniform random distribution of quenched impurities. The random
external potential is then given by

\begin{eqnarray}
V({\bf r}) = \sum\limits_{i=1}^{N_{imp}} v_{imp}({\bf r-r_i}),
\label{Vv}
\end{eqnarray}

where $N_{imp}$ is the number of impurities
present in the volume $V$
located at the fixed positions ${\bf r_i},~i = \overline{1,N_{imp}}$
and $v_{imp}({\bf r})$ is a two-body potential which describes the
particle-impurity interaction.

This model of disorder is particularly convenient for two reasons:
\begin{itemize}
\item it can be easily treated analytically within the Bogoliubov
theory of a dilute Bose gas
\item can be easily implemented in a numerical simulation.
\end{itemize}

If the gas of impurities is dilute, as it is the case in the
``weak'' disorder regime which is of interest here, the
particle-impurity interaction potential can be replaced by a
pseudopotential

\begin{eqnarray}
V({\bf r}) = \sum\limits_{i=1}^{N_{imp}} g_{imp} \delta ({\bf r-r_i}),
\label{V}
\end{eqnarray}

The coupling constant $g_{imp}$ is defined by the $s$-wave scattering
length $b$ of the particle-impurity collision process

\begin{eqnarray}
g_{imp} = \frac{2\pi \hbar^2 b}{m},
\label{gimp}
\end{eqnarray}

here $m$ is the mass of the scattering particle, since the mass of
the impurity is taken to be infinite (quenched impurities). In this
case the particle-impurity reduced mass is twice as large as the
corresponding particle-particle reduced mass. This explains the
factor two difference between (\ref{g}) and (\ref{gimp}).

The important quantities which describe the statistical properties
of the random external potential are the mean value

\begin{eqnarray}
\overline V_0 = \left\langle\frac{1}{V} \int V({\bf r})\,{\bf dr}\right\rangle,
\end{eqnarray}

and the correlation function $\langle V_{\bf p}V_{\bf-p}\rangle$,
where $V_{\bf p}$ denotes the Fourier component

\begin{eqnarray}
V_{\bf p} = \frac{1}{V} \int e^{-i{\bf pr}/\hbar}V({\bf r})\,{\bf dr}
\end{eqnarray}

Here $\langle ...\rangle$ means average over disorder
configurations.

For our random external potential (\ref{V}) the correlation
function can be rewritten as

\begin{eqnarray}
\nonumber
\langle V_{\bf p}V_{\bf -p}\rangle =
\left<\frac{1}{V^2}\int\!\!\int e^{-i{\bf p(r_1-r_2)}/\hbar}
\sum\limits_i g_{imp} \delta({\bf r_1 - r_i})
\sum\limits_j g_{imp} \delta({\bf r_2 - r_j})
\,{\bf dr_1 dr_2}\right> =\\
= \frac{g_{imp}^2}{V^2} \sum\limits_i \sum\limits_j \langle e^{-i{\bf p(r_1 - r_i})/\hbar}\rangle
\qquad\qquad\qquad
\end{eqnarray}

By assuming that the impurities have a uniform distribution
the mean value of the random potential is given by

\begin{eqnarray}
\overline V_0 = n_{imp} g_{imp}
= \frac{1}{2} mc^2 \chi\left(\frac{b}{a}\right),
\label{Vo}
\end{eqnarray}

while $\langle V_{\bf p}\rangle = 0$ for ${\bf p \ne 0}$. It can be
easily shown that the correlation function becomes

\begin{eqnarray}
\langle V_{\bf p}V_{\bf -p}\rangle  =
\frac{N_{imp} g_{imp}^2}{V^2} =
\frac{1}{4} \frac{(mc^2)^2}{nV} \chi\left(\frac{b}{a}\right)^2,
\label{corr}
\end{eqnarray}

where $n_{imp} = N_{imp}/V$ is the density of impurities and $\chi
= N_{imp}/N = n_{imp}/n$ is the concentration of impurities. Eq.
(\ref{corr}) implies that the external potential is treated as a
short correlated white noise in momentum space with amplitude
proportional to $g_{imp}$.

The independent parameters that describe the properties of
the system are the following

\begin{eqnarray}
\begin{tabular}{|p{2.5 cm}|p{7cm}|}
\hline
$na^3$ & gas parameter\\
\hline
$b/a$  & relative size of the impurity\\
\hline
$\chi = n^{imp}/n$ & concentration of the impurities\\
\hline
\end{tabular}
\label{parameters}
\end{eqnarray}

%% file: BogoliubovDis.tex
\subsection{Diagonalization of the Hamiltonian \label{DH}}

Let us rewrite Hamiltonian (\ref{second quantization}) in terms
of the creation and annihilation operators $\hat a_{\bf p}$ and
$\hat a_{\bf p}^\dagger$ in momentum representation

\begin{eqnarray}
\begin{array}{lcc}
\hat H &=&\displaystyle
\sum \frac{p^2}{2 m} \hat a_{\bf p}^\dagger\hat a_{\bf p}
+\frac{1}{2}\sum\langle{\bf p_1, p_2}|U|{\bf p_1', p_2'}\rangle\,
{\hat a_{\bf p_1}}^\dagger {\hat a_{\bf p_2}}^\dagger \hat a_{\bf p_1'}\hat a_{\bf p_2'}
\,\delta_{\bf p_1+p_2,\,p_1'+p_2'}\quad+\\
&&\displaystyle
+\quad\sum\langle{\bf p}|V|{\bf p'}\rangle\,\hat a_{\bf p}^\dagger \hat a_{\bf p'}
\label{Hamiltonian}
\end{array}
\end{eqnarray}

where we have included the external potential $V$. We use the
Bogoliubov prescription $\hat a^\dagger_0 = \hat a_0 = \sqrt{N_0}$
and we consider $\hat a_{\bf p}, {\bf p \ne 0}$ as small perturbations.
To second order in $\hat a^\dagger_{\bf p}$ for $\hat a_{\bf p}$
the external potential term can be written as

\begin{eqnarray}
\sum \langle{\bf p}|V|{\bf p'}\rangle\,\hat a_{\bf p}^\dagger \hat a_{\bf p'}
= \sum V_{\bf p-p'}\,\hat a_{\bf p}^\dagger \hat a_{\bf p'}
\approx N_0 \overline{V_0} + \sqrt{N_0}
\sum(\hat a_{\bf p}^\dagger V_{\bf p} + \hat a_p V_{\bf -p}),
\end{eqnarray}

The term $N_0 \overline V_0$ must be calculated in the second Born
approximation in order to obtain an expression which is correct up
to second order in the particle-impurity scattering amplitude.

\begin{eqnarray}
\overline{V_0} =
\frac{mc^2}{2}\chi\left(\frac{b}{a}\right)
+\frac{mc^2}{2N}\chi\left(\frac{b}{a}\right)^2
\sum\limits_{\bf p\ne 0} \left(\frac{mc}{p}\right)^2
\end{eqnarray}

The part of the Hamiltonian which is independent of the external
potential can be diagonalized by the Bogoliubov transformation
(\ref{Bogoliubov transformation}). The Hamiltonian takes the form

\begin{eqnarray}
\begin{array}{lcr}
\hat H
&=&\displaystyle
N \frac{mc^2}{2}\left(1 + \chi\frac{b}{a} \right)+
\frac{1}{2} \sum\limits_{{\bf p \ne 0}}
\left[E(p) - \frac{p^2}{2m}
- mc^2\left\{1-\biggl(\frac{mc}{p}\biggr)^2
-\chi\biggl(\frac{b}{a}\biggr)^2\biggl(\frac{mc}{p}\biggr)^2
\right\}\right]
+\\
&&\displaystyle
+\quad\sum\limits_{{\bf p \ne 0}}E(p)\,\hat b_{\bf p}^\dagger \hat b_{\bf p}
+ \sqrt{N_0}\sum\limits_{{\bf p \ne 0}}
\left(\frac{\hat b_{\bf p} + L_p \hat b_{\bf -p}^\dagger}{\sqrt{1-L_p^2}}V_{\bf p}
+\frac{\hat b_{\bf p}^\dagger + L_p \hat b_{\bf -p}}{\sqrt{1-L_p^2}}V_{\bf -p}\right)
\qquad\qquad
\end{array}
\end{eqnarray}

where $E_p$ and $L_p$ are defined by (\ref{Ep}) and (\ref{Lp}) respectively.


The linear term in the quasiparticle operators can be eliminated by
means of the following transformation (analogous transformation, but for
a different model of the disorder was introduced in \cite{Huang})

\begin{eqnarray}
\left\{
\begin{array}{ccl}
\hat b_{\bf p}&=& \hat c_{\bf p} + Z_p V_{\bf p}\\
\hat b_{\bf p}^\dagger&=&\hat c_{\bf p}^\dagger + Z_p V_{\bf -p}\\
\end{array}
\label{Tranformation}
\right.
\end{eqnarray}

with $Z_p$ defined by

\begin{eqnarray}
Z_p = -\sqrt{\frac{1+L_p}{1-L_p}} \frac{\sqrt{N_0}}{E(p)}
= -\sqrt{\frac{p^2}{2m E(p)}} \frac{\sqrt{N_0}}{E(p)}
\end{eqnarray}

The transformation (\ref{Tranformation}) does not change the commutation
rules and the new quasiparticle operators $\hat c_{\bf p}$, $\hat c_{\bf p}^\dagger$
satisfy the usual bosonic commutation relations.

Finally, the Hamiltonian takes the form

\begin{eqnarray}
\begin{array}{lc}
\hat H&\displaystyle
=\qquad\qquad
N\,\frac{mc^2}{2}\left(1 + \chi\frac{b}{a} \right)\quad+\quad
\sum\limits_{\bf p \ne 0} E(p)\,\hat c_{\bf p}^\dagger\hat c_{\bf p}
\quad+\\&\displaystyle+
\frac{1}{2}\sum\limits_{\bf p \ne 0}
\left[E(p) - \frac{p^2}{2m}
- mc^2\left\{1-\left(\frac{mc}{p}\right)^2 \left(
1+\chi\biggl(\frac{b}{a}\biggr)^2\right)
\right\}
-2N_0\frac{p^2}{2m} \frac{\langle V_{\bf p} V_{\bf -p}\rangle }{E(p)^2}\right]
\end{array}
\label{H}
\end{eqnarray}

The creation and annihilation particle operators $\hat
a_{\bf p}^\dagger$, $\hat a_{\bf p}$ are obtained from the corresponding
quisiparticle operators $\hat c_{\bf p}^\dagger$, $\hat c_{\bf p}$ in the
following way

\begin{eqnarray}
\left\{
\begin{array}{ccccc}
\displaystyle\hat a_{\bf p}&=&
\displaystyle\frac{\hat c_{\bf p} + L_p \hat c_{\bf -p}^\dagger}{\sqrt{1-L_p^2}}
-\sqrt{N_0}\frac{1+L_p}{1-L_p}\frac{V_{\bf p}}{E_p}&=&
\displaystyle\frac{\hat c_{\bf p} + L_p \hat c_{\bf -p}^\dagger}{\sqrt{1-L_p^2}}
-\sqrt{N_0}\frac{p^2}{2m} \frac{V_{\bf p}}{E^2(p)}\\
\displaystyle\hat a_{\bf p}^\dagger&=&
\displaystyle\frac{\hat c_{\bf p}^\dagger + L_p \hat c_{\bf -p}}{\sqrt{1-L_p^2}}
-\sqrt{N_0}\frac{1+L_p}{1-L_p}\frac{V_{\bf -p}}{E_p}&=&
\displaystyle\frac{\hat c_{\bf p}^\dagger + L_p \hat c_{\bf -p}}{\sqrt{1-L_p^2}}
-\sqrt{N_0}\frac{p^2}{2m} \frac{V_{\bf -p}}{E^2(p)}\\
\end{array}
\right.
\label{Bogoliubov transformation 2}
\end{eqnarray}

\subsection{Ground state energy}

By inserting the results (\ref{Vo}) and (\ref{corr}) for $\overline
V_0$ and $\langle V_{\bf p} V_{\bf -p}\rangle$respectively into the
Hamiltonian (\ref{H}) and after integration one gets the following
result for the ground state energy

\begin{eqnarray}
\frac{E_0}{N} = \frac{\hbar^2}{2ma^2} \left\{
4\pi na^3 \left(1+\chi\frac{b}{a}\right)
+  (na^3)^{3/2}
\left(\frac{512\sqrt{\pi}}{15} + 16 \pi^{3/2}
\chi\biggl(\frac{b}{a}\biggr)^2 \right)
\right\}
\label{E}
\end{eqnarray}

The first term in the above equation simply gives the mean field energy
(see eq.(\ref{E GP})) with the correction due to the presence of
the random external potential. This becomes clearer if we rewrite this
term in terms of the coupling constant $E_{MF} = 1/2\,(gn+g_{imp}n_{imp})$.
The beyond mean-field correction to the ground state energy is
given by

\begin{eqnarray}
\frac{E_0}{N} - \frac{E_{MF}}{N}
=\frac{\hbar^2}{2ma^2} (na^3)^{3/2}
\left(\frac{512\sqrt{\pi}}{15} + 16 \pi^{3/2}
\chi\left(\frac{b}{a}\right)^2 \right)
\label{E-EMF}
\end{eqnarray}

The correction due to disorder in the above result is proportional to

\begin{eqnarray}
R = \chi \left(\frac{b}{a}\right)^2,
\label{R}
\end{eqnarray}

which as we will see represents an important parameter to describe
the effect of disorder.

\subsection{Quantum depletion of the condensate}

It is easy to obtain the particle momentum distribution
$\langle N_{\bf p}\rangle  = \langle \hat a^\dagger_{\bf p} \hat a_{\bf p}\rangle$
from the operator transformation (\ref{Bogoliubov transformation 2}).

\begin{eqnarray}
\langle N_\p\rangle  =
\frac{\langle n_\p\rangle + L_p^2 (\langle n_\p\rangle +1)}{1-L_p^2} +
\left(\frac{p^2}{2m}\right)^2
\frac{\langle V_\p V_{\bf -p}\rangle }{E^4(p)}N_0,
\label{Np 2}
\end{eqnarray}

where $\langle n_\p\rangle $ is the number of excitations with momentum $\p$.
At zero temperature such excitations are absent and the distribution
(\ref{Np 2}) takes the form

\begin{eqnarray}
\langle N_\p\rangle  =
\nonumber
\frac{L_p^2}{1-L_p^2} +
\left(\frac{p^2}{2m}\right)^2
\frac{\langle V_\p V_{\bf -p}\rangle }{E^4(p)} = \qquad\qquad\\
= \frac{(mc^2)^2}{2E(p)(E(p)+ \frac{p^2}{2m}+ mc^2)}
+ \frac{\chi\left(\frac{b}{a}\right)^2}{\left(4+\left(\frac{p}{mc}\right)^2 \right)^2}
\label{Np Dis}
\end{eqnarray}

The first term of the above result corresponds to the momentum
distribution in the absence of disorder. The second term gives the
contribution due to disorder. The presence of disorder induces an
extra quantum depletion of the condensate arising from the
particle-impurity interaction. By integrating over momentum one
obtains the following result for the condensate fraction

\begin{eqnarray}
\frac{N_0}{N} =
1 - \frac{8}{3\sqrt{\pi}} (na^3)^{1/2}
- \frac{\sqrt{\pi}}{2} (na^3)^{1/2} \chi \left(\frac{b}{a}\right)^2
\label{condensate depletion}
\end{eqnarray}

It is interesting to note, that the effect of disorder again is
described by the combination of parameters $R = \chi(b/a)^2$, as it
was found for the energy (\ref{E}).

\subsection{One body density matrix}

The one body density matrix is given by the Fourier transform of
the momentum distribution (see (\ref{rho})). There are three
contributions to the one-body density matrix coming from
condensate, particle-particle interaction effects and
particle-impurity interaction effects:

\begin{eqnarray}
\rho(r) = \frac{N_0}{V} + \rho^{(1)}(r) + \rho^{(2)}(r)
\end{eqnarray}

The result for $\rho^{(1)}(r)$ has been given in (\ref{OBDM 1}).
Here we calculate the contribution due to disorder

\begin{eqnarray}
\rho^{(2)}(r)
= \frac{2\chi\left(\frac{b}{a}\right)^2}{\pi V x}
\int\limits_0^\infty
\frac{\sin(\sqrt{4\pi na^3 x \xi})}{(4+ \xi^2)^2}\xi d\xi
= \frac{\sqrt{\pi}}{2}\sqrt{na^3}\,
\chi\left(\frac{b}{a}\right)^2
\exp\left(-\frac{r}{\sqrt{2}r_0}\right)n,
\label{correlation}
\label{OBDM 2}
\end{eqnarray}

where $r_0 = a/\sqrt{8\pi na^3}$ is the healing length.

Notice that at $r=0$ the value of the integral equals to the
density of particles which are scattered out of the condensate due
to the presence of the external field.

%% file: SD.tex
\subsection{Connection between $\rho_s/\rho$ and the transverse current-current
response function}

\begin{figure}
\center
\resizebox{7 cm}{3 cm}{\includegraphics{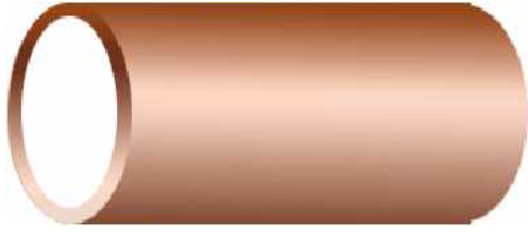}}
\caption{Long tube filled with superfluid}
\label{Tube}
\end{figure}

One of the striking properties of superfluids is the ability to
flow without friction. This fact allows us to define the
normal fluid density $\rho_n$ as the fraction of liquid which is
carried along by the walls if they are set in motion. For example
\cite{Baym, Huang 2},
consider the liquid inside a long tube (see. Fig~\ref{Tube}), which
was at rest at time $t = -\infty$ and was then adiabatically
accelerated up to time $t = 0$ (for instance with the exponential
law $v(t) = v \exp(\varepsilon t)$ with infinitesimal $\varepsilon >
0$). The normal component $\rho_n$ can be defined through the
momentum density $\langle\vec j\rangle$ at $t = 0$

\begin{eqnarray}
\langle\vec j \rangle = \rho_n \vec v
\end{eqnarray}

and the superfluid density as the difference between the total
density $\rho$ and $\rho_n$

\begin{eqnarray}
\rho_s = \rho - \rho_n
\end{eqnarray}

The effect of the perturbation caused by the moving walls is given
by the energy

\begin{eqnarray}
V_{walls} = - \int \vec j(x) \vec v(x,t) d^3x,
\end{eqnarray}

where $\vec j(x)$ is the momentum density and $\vec v(x,t)$ is the
external velocity field. The linear response is given by the Kubo
formula

\begin{eqnarray}
\langle j_i(x,t) \rangle =
\int\limits_{-\infty}^{\infty}ds \int d^3y~\chi_{ij}(x,t;y,s) v_j(y,s),
\label{Kubo}
\end{eqnarray}

where

\begin{eqnarray}
\chi_{ij}(x,t;y,s) = i\theta(t-s)\,\langle [j_i(x,t), j_j(y,s)]\rangle
\end{eqnarray}

For uniform systems the response function $\chi_{ij}$ depends
only on the difference of its arguments $\chi_{ij} = \chi_{ij}(x-y,t-s)$.
The {\it static susceptibility} is defined as

\begin{eqnarray}
\chi_{ij}(x) =
\int\limits_{-\infty}^{0}\chi_{ij}(x,t) e^{\varepsilon t} dt
\end{eqnarray}

or in terms of Fourier components

\begin{eqnarray}
\chi_{ij}(k) =
\int\limits_{-\infty}^{\infty} \frac{\chi_{ij}(k,\omega)}{\omega+i\varepsilon}
\frac{d\omega}{2\pi}
\end{eqnarray}

At time $t = 0$ the linear response function satisfies the equation

\begin{eqnarray}
\langle j_i(k)\rangle = \chi_{ij}(k) v_j(k),
\qquad i,j = \{x,y,z\}
\end{eqnarray}

Since $\chi_{ij}(k)$ is a second rank tensor, it can be decomposed
into the sum of longitudinal and transverse components

\begin{eqnarray}
\chi_{ij}(k) = \frac{k_i k_j}{k^2} \chi_L(k) +
\left(\delta_{ij} - \frac{k_i k_j}{k^2}\right) \chi_T(k),
\label{chi_ij}
\end{eqnarray}

Let us consider first the transverse response. Due to the
rotational invariance of $\chi_T(k)$ it is enough to examine the
response in an arbitrary direction, for example $\chi_{zz}$, that is the
momentum response $j_z({\bf k})$ due to an imposed velocity field in the
$z$ direction $v_z({\bf k})$.

\begin{figure}
\center
\includegraphics[width=.5\textwidth]{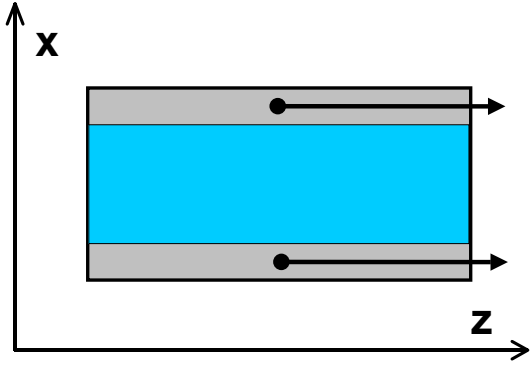}
\caption{Illustration of the transverse response. Only the normal
component is dragged in $z$ direction}
\label{TubeTransverse}
\label{TubeNormal}
\end{figure}

Suppose first that the velocity field is created by dragging the
walls of an indefinitely long pipe (see Fig.~\ref{TubeNormal}) and
the cross section of the pipe tends to infinity. This arrangement
corresponds to the limiting procedure $k_z \to 0$, followed by $k_x
\to 0$, $k_y \to 0$. Then, the part of the system responding to the
shear force is defined as the normal fluid. Carrying out the
limiting procedure $k^2 = k_x^2 + k_y^2 + k_z^2 \to 0$ gives

\begin{eqnarray}
\rho_n =
\lim\limits_{k_x \to 0\atop  k_y \to 0}
\lim\limits_{k_z \to 0}
\chi_{ij}(k)
=\lim\limits_{k \to 0} \chi_T(k)
\label{rho_n}
\end{eqnarray}

\begin{figure}
\center
\includegraphics[width=.5\textwidth]{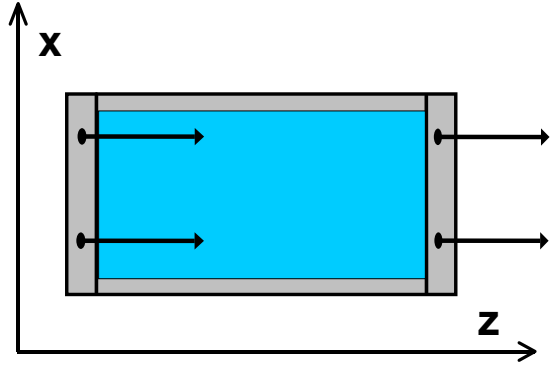}
\caption{Illustration to the longitudinal response. Both superfluid
and normal components are pushed in the $z$ direction}
\label{TubeLongitudinal}
\label{TubeTotal}
\end{figure}

Next suppose that the pipe of infinite radius is constrained by two
plates, normal to the $z$ axis, with separation between the plates
approaching infinity, as illustrated in Fig.~\ref{TubeTotal}. In
this case the entire fluid $\rho = \rho_s +\rho_n$ responds to the
external probe. This arrangement corresponds to the limiting
procedure $k_x \to 0$, $k_y\to 0$, followed by $k_z \to 0$. The
result of the limiting procedure is

\begin{eqnarray}
\rho =
\lim\limits_{k_z \to 0}
\lim\limits_{k_x \to 0\atop k_y \to 0}
\chi_{ij}(k)
=\lim\limits_{k \to 0} \chi_L(k)
\label{rho_total}
\end{eqnarray}

\subsection{Superfluid fraction $\rho_s/\rho$ in the presence of disorder}

Let us apply the theory developed in the previous section to a
dilute Bose system. As we will see, the system without disorder is
fully superfluid at zero temperature while the presence of
impurities create a depletion of the superfluid density.

According to (\ref{chi_ij}) and (\ref{rho_n}) the normal component
is given by the limit of the transverse current-current response
function. Let us consider, for example, $\chi_T = \chi_{z}$, i.e.
$z$ response and in the $z$ direction (see
Fig.~\ref{TubeTransverse}). First one has to take the limit $k_z\to
0$ and after $k_x\to 0$ and $k_y\to 0$. This can be accomplished by
considering ${\bf k} =(k,0,0)$ and letting $k$ decrease toward
zero.

The $k$-component of the current operator $\hat j$ in second
quantization is given by the formula

\begin{eqnarray}
\hat j_{\bf k} = \frac{\hbar}{\sqrt{V}}
\sum\limits_{\bf q} \left({\bf q}+\frac{\bf k}{2}\right)
\hat a^\dagger_{\bf q}\hat a_{\bf q+k},
\label{j}
\end{eqnarray}

where in the absence of disorder the particle creation and
annihilation operators can be expressed in terms of quasiparticle
operators by means of the Bogoliubov transformation (\ref{linear}).
Once the current (\ref{j}) is calculated, the transverse response
function can be obtained by averaging the commutator

\begin{eqnarray}
\chi_T({\bf k},t) = -i\Theta(t)\,
\langle[j^z_{\bf k}(t), j^z_{\bf -k}(0)]\rangle
\label{chi}
\end{eqnarray}

By taking the Fourier transform

\begin{eqnarray}
\chi_T({\bf k},\omega)=\int\limits_{-\infty}^{+\infty}\exp(i\omega t)\chi_T({\bf k},t)\,dt
\end{eqnarray}

the limiting procedure

\begin{eqnarray}
\rho_n = \lim\limits_{k\to 0} \lim\limits_{\omega \to 0} \chi_T(k,\omega)
\end{eqnarray}

yields the density of the normal fluid.

It is easy to check that in the absence of disorder $\chi_T({\bf
k}, \omega)$ goes to zero as $\bf k \to 0$ and $\omega \to 0$ that
the commutator (\ref{chi}) of $\chi_{zz}$ goes to zero in the limit
$k\to 0$. This corresponds to the fact that a homogeneous dilute
Bose gas is completely superfluid at $T = 0$.

A useful check consists in the calculation of the longitudinal
response (see Fig.~\ref{TubeLongitudinal}). This means taking first
the limit $k_x \to 0$, $k_y \to 0$ in $\chi_{zz}({\bf k})$. We
consider ${\bf k} = (0,0,k)$ and then let $k \to 0$. It is easy to
calculate $\chi_{zz} $ in Bogloliubov approximation.

For the $z$ component of the current operator one has

\begin{eqnarray}
\hat j^z_{\bf k} = \frac{\hbar{\bf k}\sqrt{N_0}}{\sqrt{V}}
\frac{(\hat a_{\bf k}+\hat a_{\bf -k}^\dagger)}{2}
+\frac{\hbar}{\sqrt{V}}
\sum\limits_{\bf q\ne 0}
\left({\bf q}+\frac{\bf k}{2}\right) \hat a^\dagger_{\bf q} \hat a_{\bf q+k},
\approx
\frac{\hbar{\bf k}\sqrt{N_0}}{\sqrt{V}}
\frac{(\hat a_{\bf k}+\hat a_{\bf -k}^\dagger)}{2}
\end{eqnarray}

Within this level of accuracy the limit (\ref{rho_total}) of the
longitudinal component is given by

\begin{eqnarray}
\lim\limits_{\bf k\to 0} \lim\limits_{\omega \to 0}
\chi_L({\bf k},\omega) = n_0m \approx nm
\end{eqnarray}

Let us now study the system in the presence of the random external
potential.Starting from the transformation (\ref{Bogoliubov
transformation 2}) between the particles operators $\hat a_{\bf k}$,
$\hat a^\dagger_{\bf k}$ and the corresponding quasiparticle
operators $\hat c_{\bf k}$, $\hat c^\dagger_{\bf k}$ one can write
the contribution to the current proportional to the external
potential (there is no need to consider the contribution independent
of the external potential, because as calculated before it is equal to
zero).

\begin{eqnarray}
\hat j_0^{z}(t)
= \sum\limits_{\bf q} \frac{\hbar q_z Z_q}{\sqrt{V}} \sqrt{\frac{1-L_q}{1+L_q}}
(\hat c^\dagger_{\bf q} V_{\bf -q}-\hat c V_{\bf q}) + ...
\end{eqnarray}

Notice that in order to make calculations simpler we take $k = 0$
from the very begging. Result (\ref{rho_n}) is independent of the
order of the two limits.

The response function is then given by

\begin{eqnarray}
\chi_T(0, t) = -i \Theta(t)
\frac{1}{V}\sum\limits_{\bf q}\hbar^2q^2_x |Z_q|^2 \langle V_{\bf q} V_{\bf -q}\rangle
\frac{1-L_q}{1+L_q}
(e^{-i\omega_{\bf q}t}-e^{i\omega_{\bf q}t})
\end{eqnarray}

its Fourier transformation being equal to

\begin{eqnarray}
\chi_T(0, w) =
\frac{1}{V}\sum\limits_{\bf q}\hbar^2 q^2_x |Z_q|^2 \langle V_{\bf q}V_{\bf-q}\rangle
\frac{1-L_q}{1+L_q}
\left(\frac{1}{\omega-\omega_q}-\frac{1}{\omega+\omega_q}\right),
\end{eqnarray}

and, finally, setting $\omega = 0$ one obtains

\begin{eqnarray}
\rho_n
=\chi_T(0, 0) =
\frac{4m\hbar}{3V}
\sum\limits_q q^2_x |Z_q|^2 \langle V_q V_{-q}\rangle
= \frac{2\sqrt{\pi}}{3}(na^3)^{1/2}
\chi\left(\frac{b}{a}\right)^2 nm
\label{SD Bogoliubov}
\end{eqnarray}

This result gives the depletion of the superfluid density due to
the presence of impurities.

%% file: GPsd.tex
\subsection{calculation of the superfluid fraction from GPE \label{GP SD}}

In this section the superfluid fraction $\rho_s/\rho$ will be
obtained directly from Gross-Pitaevskii equation in a perturbative
manner. This derivation is new and the result coincides with the one
obtained from the Bogoliubov model presented in the previous
section.

The Gross-Pitaevskii equation (\ref{Gross-Pitaevskii}) for the
condensate wavefunction in the absence of external field takes the form

\begin{eqnarray}
i\hbar\frac{\partial }{\partial t} \Psi({\bf r}, t) =
\left(-\frac{\hbar^2}{2m} \triangle + g|\Psi({\bf r}, t)|^2\right)
\Psi({\bf r}, t)
\label{GP}
\end{eqnarray}

Let us add a moving impurity that creates an external field
$V_{ext}({\bf r}, t) = g_{imp} \delta({\bf r - V}t)$ and let us
treat it as a perturbation to the solution $\Psi_0({\bf r}, t)$ of
the equation (\ref{GP}), i.e.

\begin{eqnarray}
\Psi({\bf r}, t) = \left[\Psi_0({\bf r}) + \delta\Psi({\bf r}, t)\right]
\exp\left(-i \frac{\mu t}{\hbar} \right),
\label{expansion}
\end{eqnarray}

Then, substitution of (\ref{expansion}) into (\ref{GP}) gives

\begin{eqnarray}
i\hbar\frac{\partial }{\partial t} \delta \Psi =
\left(-\frac{\hbar^2}{2m} \triangle - \mu\right) \delta \Psi
+ 2 g|\Psi_0|^2 \delta \Psi
+ g \Psi_0^2 \delta \Psi^\star
+ g_{imp} \delta ({\bf r - V}t) \Psi_0
\label{first approx}
\end{eqnarray}

The perturbation follows the moving impurity, so
$\delta \Psi$ is a function of ${\bf r - V}t$.

Let us introduce the new variable ${\bf r' = r - V}t$. 
It means that the coordinate derivative can be related to time derivative

\begin{eqnarray}
\frac{\partial}{\partial t} \delta \Psi({\bf r - V}t)& =&
-{\bf V \nabla} \delta \Psi({\bf r - V}t), \label{dt2nabla}\\
\triangle {\bf r'} &=& \triangle {\bf r}
\end{eqnarray}

and instead of (\ref{first approx}) one has

\begin{eqnarray}
\left( i\hbar {\bf V \nabla} - \frac{\hbar^2}{2m} \triangle
- \mu + 2 g|\Psi_0|^2 \right) \delta \Psi
+ g \Psi_0({\bf r})^2 \delta \Psi^\star
+ g_{imp} \delta ({\bf r}) \Psi_0 = 0,
\end{eqnarray}

(from now on, the subscript over ${\bf r}$ will be dropped)

Taking the Fourier transform of this equation and treating $\Psi_0$
as a real constant (i.e. the solution for the uniform case $g\Psi_0^2 = \mu$)
one obtains

\begin{eqnarray}
\left( -\hbar {\bf k V} + \frac{\hbar^2k^2}{2m}
- \mu + 2 \mu \right) \delta \Psi_k
+ \mu\delta (\Psi_{-k})^\star + g_{imp} \Psi_0 = 0,
\label{Fourier}
\end{eqnarray}

where we used the property of Fourier components
$\delta (\Psi^\star)_{k} = \delta (\Psi_{-k})^\star$.

The substitution of $(-{\bf k})$ in the equation complex
conjugate of (\ref{Fourier}) gives

\begin{eqnarray}
\left( \hbar {\bf k V} + \frac{\hbar^2k^2}{2m}
- \mu + 2 \mu \right) (\delta \Psi_{-k})^\star
+ \mu\delta \Psi_{k} + g_{imp} \Psi_0 = 0
\label{Fourier conjugate}
\end{eqnarray}

The solution for the system of linear equations
(\ref{Fourier} - \ref{Fourier conjugate}) is given by

\begin{eqnarray}
\delta \Psi_{k}
= -\frac{g_{imp} \left(\hbar{\bf k V} + \frac{\hbar^2 k^2}{2m}\right)\Psi_0}
{\frac{\hbar^2 k^2}{2m} \left(\frac{\hbar^2 k^2}{2m}+2\mu\right)
-(\hbar{\bf k V})^2}
\label{dPsi}
\end{eqnarray}

The energy $E' = E-\mu N$ has the minimum at fixed $\mu$ for the
ground state function $\Psi_0$. It means that $E'$ does not have
terms linear in $\delta \Psi_k$ and $\delta \Psi_k^\star$, so
$E'= E^{(0)}+E^{(2)}+g_{imp} (\Psi_0^\star\delta\Psi(0)+\Psi_0\delta\Psi^\star(0))$.
Here the last term comes from the linear expansion of the energy
$\int |\Psi({\bf r})|^2 g_{imp} \delta({\bf r}){\bf dr}$. The term
$E^{(2)}$ being quadratic in $\delta \Psi_k$ and $\delta
\Psi_k^\star$ satisfies the Euler identity:

\begin{eqnarray}
2 E^{(2)} =
\int \left[
\delta \Psi({\bf r}) \frac{\delta E^{(2)}}{\delta(\delta \Psi({\bf r}))}
+\delta \Psi^\star({\bf r}) \frac{\delta E^{(2)}}{\delta(\delta
\Psi^\star({\bf r}))}
\right]{\bf dr}
\end{eqnarray}

which using the variational equation

\begin{eqnarray}
i \hbar \frac{\partial\delta(\delta \Psi)}{\partial t} =
\frac{\delta E^{(2)}}{\delta(\delta\Psi^\star)}
+ g_{imp}\Psi_0\delta({\bf r})
\end{eqnarray}

can be rewritten as

\begin{eqnarray}
\begin{array}{rcc}
E^{(2)}= E^{(2)}_1 + E^{(2)}_2&=&
\displaystyle\frac{i\hbar}{2} \int \left[
\delta \Psi^\star({\bf r}) \frac{\partial\delta \Psi({\bf r})}{\partial t}
-\frac{\partial\delta \Psi^\star({\bf r})}{\partial t}\delta \Psi({\bf r})
\right]{\bf dr}\quad-\\
&&\displaystyle-\quad\frac{g_{imp} }{2} (\Psi_0^\star\delta\Psi(0)+\Psi_0\delta\Psi^\star(0))
\end{array}
\end{eqnarray}

To start with, let us Fourier transform the first term. Exchanging
time derivatives with gradients by the rule (\ref{dt2nabla}) one
obtains

\begin{eqnarray}
E^{(2)}_1 = \int \hbar{\bf k V} |\delta \Psi_k|^2 \frac{\bf dk}{(2\pi)^3}
\end{eqnarray}

The terms of interest are the ones that are quadratic in the
velocity ${\bf V}$. It means that the term $(\hbar{\bf kV})^2$ in
the denominator of (\ref{dPsi}) can be neglected and $|\delta
\Psi_k|^2$ turns out to be

\begin{eqnarray}
|\delta \Psi_{k}|^2 = \frac{g_{imp} ^2 |\Psi_0|^2
\left[\left(\frac{\hbar^2 k^2}{2m}\right)^2 +
2\frac{\hbar^2 k^2}{2m} \hbar{\bf k V} \right]}
{\left[\frac{\hbar^2 k^2}{2m}
\left( \frac{\hbar^2 k^2}{2m}+2 \mu\right)\right]^2}
\label{dPsi2}
\end{eqnarray}

The energy does not have terms linear in ${\bf V}$, because all
terms independent of ${\bf V}$ in (\ref{dPsi2}) are even in ${\bf
k}$, so multiplied by ${\bf k}$ and integrated over momentum space
they provide zero contribution to the energy. The only term that is
left is the following

\begin{eqnarray}
E^{(2)}_1 =
2g_{imp} ^2|\Psi_0|^2
\int \frac{(\hbar{\bf kV})^2}
{\frac{\hbar^2 k^2}{2m}
\left( \frac{\hbar^2 k^2}{2m}+2 \mu\right)^2}
\frac{\bf dk}{(2\pi)^3}
\end{eqnarray}

For the calculation of $\delta\Psi(0)$ one should consider $\delta
\Psi_k$ taking into account that $\hbar {\bf kV} \ll \mu$ and then
integrate it over momentum space

\begin{eqnarray}
\begin{array}{rcl}
\delta \Psi_{k}&
\approx&\displaystyle
\frac{(\hbar{\bf kV})^2}{\frac{\hbar^2 k^2}{2m}\left(\frac{\hbar^2 k^2}{2m}+2\mu\right)^2}
g_{imp} \Psi_0\quad-\\
&&\displaystyle-\quad\left\{
\frac{\hbar{\bf kV} \left[\frac{\hbar^2 k^2}{2m}\left(\frac{\hbar^2 k^2}{2m}+2\mu\right)\right]
+(\hbar{\bf kV})^3}
{\left[\frac{\hbar^2 k^2}{2m}\left(\frac{\hbar^2 k^2}{2m}+2\mu\right)\right]^2}
+\frac{1}{\frac{\hbar^2 k^2}{2m}+2\mu}
\right\}
g_{imp} \Psi_0
\end{array}
\end{eqnarray}

The integral of the second term over momentum space is equal to
zero. The third term is diverging and needs the renormalization of
$g_{imp}$ (as discussed in sections \ref{BH & EE} and \ref{DH}) in
order to be calculated correctly. However, its correction does not
depend on ${\bf V}$ and will be omitted.

The energy is defined by the following integral

\begin{eqnarray}
\nonumber
\delta E = E^{(2)}_1 +  E^{(2)}_2 +
g_{imp} (\Psi_0^\star\delta\Psi(0)+\Psi_0\delta\Psi^\star(0)) =\qquad\qquad\\
=\left(2g_{imp} ^2|\Psi_0|^2 - \left(g_{imp} -\frac{g_{imp} }{2}\right)
\left(\Psi_0^\star\Psi_0+\Psi_0\Psi^\star_0\right)\right)
\int \frac{(\hbar{\bf kV})^2 d{\bf k}}
{\frac{\hbar^2 k^2}{2m}
\left( \frac{\hbar^2 k^2}{2m}+2 \mu\right)^2}
\frac{d{\bf k}}{(2\pi)^3}
\end{eqnarray}

In the integral $(\bf{kV})^2 d{\bf k}$ can be replaced by $1/3~k^2
V^2 4\pi k^2 dk$ due to the equivalence of different directions.
Then the integral can be easily calculated if one recall the
following integral identity

\begin{eqnarray}
\int \frac{x^2dx}{(x^2+a^2)^2} = -\frac{x}{2(x^2+a^2)^2}
+\frac{1}{2a} \arctg\frac{x}{a}
\end{eqnarray}

The result is the following

\begin{eqnarray}
\delta E = \frac{m^{5/2}g_{imp} ^2|\Psi_0|^2}{3\pi\hbar^3\sqrt{\mu}}\frac{V^2}{2},
\label{dE}
\end{eqnarray}

where

\begin{eqnarray}
g = \frac{4\pi\hbar^2a}{m},\quad
g_{imp}  = \frac{2\pi\hbar^2b}{m},\quad
|\Psi_0|^2 = n
\nonumber
\end{eqnarray}

The term in front of $V^2/2$ in (\ref{dE}) can be interpreted as an
effective mass $m^\star$ of the particles which follow the external
perturbation, i.e. the normal (and not superfluid) component of the
fluid. Then normal fraction can then be easily obtained which, as
anticipated, coincides with result (\ref{SD Bogoliubov}).

\begin{eqnarray}
\frac{\rho_n}{\rho} = \frac{m^\star}{m} \chi =
\frac{2\sqrt{\pi}}{3} (na^3)^{1/2} \chi\left(\frac{b}{a}\right)^2
\label{SD}
\end{eqnarray}

Let us compare the results for the superfluid density (\ref{SD})
and the condensate fraction (\ref{condensate depletion}). It is
interesting to note that in both formulae the effect of disorder
enters as $\sqrt{na^3} R$, where $R=\chi (b/a)^2$ is the universal
scaling parameter which already entered the result (\ref{E}) for
the energy. This means that systems with different disorder
concentration $\chi$ and size of the impurities $b/a$, but same $R$
experience the same effect due to disorder.

Another interesting result is that disorder is more efficient (by a
factor $4/3$) in depleting the superfluid density than the
condensate. Taking into account that even pure systems ($R = 0$)
exhibit a nonzero quantum depletion due to particle interactions,
one infers that at the critical amount of disorder $R_c =
16/\pi\approx 5.1$ the depletion of the superfluid density becomes
larger than the depletion of the condensate fraction.

Huang and Meng \cite{Huang}, who first derived results
(\ref{condensate depletion}) and (\ref{SD Bogoliubov}) even if for
a different model of disorder, have used them at $T = 0$ to predict
two distinct transitions as a function of the amount of disorder :
first a superfluid-insulator transition where $\rho_s = 0$ followed
by a Bose-Einstein transition where $N_0 = 0$. These authors also
argue that the intermediate phase corresponding to $\rho_s = 0$ and
$N_0 \ne 0$ should be identified with a Bose-glass phase. However,
in \cite{disorder} it is stressed that results (\ref{condensate
depletion}) and (\ref{SD Bogoliubov}) are valid in the weak
disorder regime and cannot be applied if the depletion due to
disorder is large.

The range of validity of results (\ref{condensate depletion}) and
(\ref{SD}) will be investigated in detail in section
\ref{SD CF res} using Monte Carlo techniques.

%% file: DMC.tex
\subsection{Introduction}

Monte Carlo methods are very powerful tools for the investigation
of quantum many body systems (for a review see, for example,
\cite{Ceperley}).

The simplest of the quantum Monte-Carlo methods is the {\it
variational} method (VMC). The idea of this method is to use an
approximate wavefunction $\psi_T$ for the system (trial
wavefunction) and then to sample the probability distribution
$p(r)~=~|\psi(r)|^2$ and calculate averages of physical quantities over
this distribution. The average of the local energy $E_L =
\psi^{-1}_T H \psi_T$ gives an upper bound to the ground-state
energy. In this method one must make a good guess for the trial
wavefunction, and there is no regular way for doing it and further
improving it. In VMC the closer is the trial wavefunction to the
stationary eigenfunction the smaller is the energy variance $\langle
H^2\rangle-\langle H\rangle^2$. In usual applications the trial
wavefunction depends on the particle coordinates and on some
external parameters $\psi_T = \psi_T(r_1,...,r_N,a,b,...)$.
By minimizing the variational energy with respect to the external
parameters one can optimize the wavefunction within the given
class of wavefunctions considered.

The Diffusion Monte Carlo method (DMC) can be successfully applied
to the investigation of boson systems at low temperatures. It is
based on solving the Schr\"odinger equation in imaginary time and
allows us to calculate the exact (in statistical sense)
value of the ground state energy. The DMC method will be
extensively discussed in the next sections.

The Path Integral Monte Carlo (PIMC) is based on carrying out
discretized Feynman integral in the imaginary time which allows to
calculate the density matrix of the system. The main advantage
of this method is that it works at finite temperatures and one has
access to the study of thermodynamic properties such as the
critical behavior in the proximity of a phase transition
(\cite{2D Ceperley}, \cite{PI sd}).

In this study we use DMC method because we are interested in the ground
state properties of the system.

\subsection{Schr\"odinger equation}

The wavefunction of the system satisfies the Schr\"odinger equation

\begin{eqnarray}
i\hbar \frac{\partial}{\partial \tau} \varphi ({\bf R},\tau) =
\hat H \varphi ({\bf R},\tau),
\end{eqnarray}

where ${\bf R} = (\vec r_1, \vec r_2, ...)$ denotes the particle
coordinates. This equation can be rewritten in imaginary time $t =
i\tau / \hbar$.

\begin{eqnarray}
-\frac{\partial}{\partial t} \varphi ({\bf R},t) =
(\hat H -E) \varphi ({\bf R},t),
\label{shifted Schrodinger}
\end{eqnarray}

where $E$ is an energy shift whose meaning will become clearer later.

The formal solution of this equation is

\begin{eqnarray}
\psi ({\bf R},t) = e^{-(\hat H - E)t} \psi ({\bf R},0)
\label{Schrodinger}
\end{eqnarray}

This solution can be expanded in eigenstate functions of the
Hamiltonian $\hat H\phi_n~=~E_n\phi_n$, $E_0 < E_1 < ...$

\begin{eqnarray}
\psi ({\bf R},t) =
\sum\limits_n c_n \phi_n({\bf R},t) =
\sum\limits_n c_n \phi_n({\bf R},0) e^{-(E_n - E)t}
\label{sum}
\end{eqnarray}

The amplitudes of the components change with time, either
increasing or decreasing depending on the sign of $(E_n-E)$. At
large times the term that corresponds to the projection on the
ground state dominates the sum. In other words all excited states
decay exponentially fast and only contribution from ground state
survives

\begin{eqnarray}
\psi ({\bf R},t) \to c_0 \phi_0({\bf R},0) e^{-(E_0 - E)t}
\qquad \mbox{if } t \to \infty
\label{psi limit}
\end{eqnarray}

In the long time limit the wavefunction remains finite only when
$E$ is equal to $E_0$. This provides a method to obtain the ground
state energy by adjusting the parameter $E$ in a way that the norm
of $\psi({\bf R},t)$ is constant.

Let us consider system of $N$ particles, introducing the
Hamiltonian through a pair-wise potential

\begin{eqnarray}
\hat H =
-\frac{\hbar^2}{2m} \sum\limits_{i=1} \Delta_i
+ \sum\limits_{i<j}^N V(|\vec r_i -\vec r_j|),
\end{eqnarray}

and the Schr\"odinger equation reads

\begin{eqnarray}
-\frac{\partial}{\partial t} \psi ({\bf R},t) =
-D \sum\limits_{i=1}^N \Delta_i \psi ({\bf R},t)
+ V({\bf R}) \psi ({\bf R},t) - E \psi ({\bf R},t),
\end{eqnarray}

where the following notation is used: $D = \hbar^2 /2m$ and $V({\bf
R}) = \sum\limits_{i<j} V(|\vec r_i - \vec r_j|)$. In principle,
any external field which is independent of the particle momenta and
is a function only of the particle coordinates can be included into $V({\bf R})$
without any harm to the reasoning.

Better efficiency is achieved if the importance sampling is used.
In the DMC method this means that one has to solve the
Schr\"odinger equation for the modified wavefunction
\footnote[1]{One of the reasons for using the product of
wavefunctions as the probability distribution instead of sampling
$\psi$ is that the average over the latter is ill defined $\langle
A\rangle = \int A\psi\,dR/\int\psi\,dR$, on the contrary the
average over the product of wavefunctions has the meaning of the
mixed estimator $\langle A\rangle = \int \psi_T A \psi\,dR/
\int\psi_T\psi\,dR$}

\begin{eqnarray}
f({\bf R},t) = \psi_T ({\bf R},t) \psi ({\bf R},t)
\label{f}
\end{eqnarray}

Here $\psi_T({\bf R},t)$ is the trial wavefunction which
approximates the true wavefunction $\psi({\bf R},t)$ of the system.
The distribution function $f$ satisfies the following equation

\begin{eqnarray}
-\frac{\partial}{\partial t} f ({\bf R},t) =
-D \sum\limits_{i=1}^N \Delta_i f ({\bf R},t)
+ D \vec \nabla (\vec F f({\bf R},t))
+ (E_L({\bf R}) - E) f({\bf R},t),
\end{eqnarray}

here $E_L$ denotes the {\it local energy} which is the average of
the Hamiltonian with respect to trial wavefunction

\begin{eqnarray}
E_L({\bf R}) =
\frac{\psi_T^*({\bf R}) \hat H \psi_T({\bf R})}{\psi_T^*({\bf R})\psi_T({\bf R})}
\label{local energy}
\end{eqnarray}

and $\vec F$ is the {\it drift force} which is proportional to the
gradient of the trial wavefunction and consequently always points
in the direction where $\psi_T$ increases

\begin{eqnarray}
\vec F = \frac{2}{\psi_T({\bf R})} \vec \nabla \psi_T({\bf R})
\end{eqnarray}

\subsection{Green's function}

The formal solution of the Schr\"odinger equation written in
coordinate space is given by

\begin{eqnarray}
\langle{\bf R}|f(t)\rangle =
\sum\limits_{{\bf R'}} \langle{\bf R}|e^{-(\hat H-E)t}| {\bf R'}\rangle
\langle{\bf R'}|f(0)\rangle,
\end{eqnarray}

or, expressed in terms of the Green's function
$G({\bf R},{\bf R'},t) = \langle{\bf R}|e^{-(\hat H-E)t}|{\bf R'}\rangle$,
the above equation reads

\begin{eqnarray}
f({\bf R},t) = \int G({\bf R},{\bf R'},t) f({\bf R'},0)\,{\bf dR'}
\label{Green}
\end{eqnarray}

In other words, the differential Schr\"odinger equation
(\ref{Schrodinger}) corresponds to the integral equation
(\ref{Green}), which can be integrated with help of Monte
Carlo methods. Although the Green's function $G({\bf R'},{\bf R},t)$ is not
known, it can be approximated for small values of the argument $t$,
and then equation (\ref{Green}) can be solved step by step

\begin{eqnarray}
f({\bf R},t + \triangle t) =
\int G({\bf R},{\bf R'},\triangle t) f({\bf R'},t)\,{\bf dR'}
\label{Green approx}
\end{eqnarray}

For further convenience let us split the Hamiltonian into three
operators

\begin{eqnarray}
\label{H summands}
\hat H = \hat H_1 + \hat H_2 + \hat H_3,
\end{eqnarray}

where

\begin{eqnarray}
\begin{array}{lcl}
\hat H_1&=&-D \Delta,\\
\hat H_2&=&D((\vec \nabla \vec F) + \vec F \vec \nabla)),\\
\hat H_3&=&E_L({\bf R}) - E
\end{array}
\end{eqnarray}

and let us introduce the corresponding Green's functions

\begin{eqnarray}
G_i({\bf R},{\bf R'},t) = \langle{\bf R}|e^{-\hat H_i t}|{\bf R'}\rangle,
\qquad i = 1,2,3
\end{eqnarray}

The exponential operator can be approximated as (the error comes
from the noncommutativity of the $\hat H_i$'s, $i = 1,2,3$)

\begin{eqnarray}
e^{-\hat H t} =
e^{-\hat H_1 t}e^{-\hat H_2 t}e^{-\hat H_3 t}+O(t^2)
\label{t2 approximation}
\end{eqnarray}

This formula, rewritten in coordinate representation, gives approximation
for the Green's function

\begin{eqnarray}
\nonumber
G({\bf R},{\bf R'},t) = \int\int
G_1({\bf R},{\bf R_1},t) G_2({\bf R_1},{\bf R_2},t) G_3({\bf R_2},{\bf R'},t)\,
{\bf dR}_1 {\bf dR}_2
\end{eqnarray}

To obtain the three Green's functions one must solve the
differential equations

\begin{eqnarray}
\left\{
{\begin{array}{rcll}
\displaystyle -\frac{\partial}{\partial t} G({\bf R},{\bf R'},t) &=&
\displaystyle H_i\,G({\bf R},{\bf R'},t),
&\displaystyle i = 1,2,3\\
\displaystyle G({\bf R},{\bf R'},0)& =&\displaystyle\delta({\bf R}-{\bf R'})&\\
\end{array}}
\right.
\label{G}
\end{eqnarray}

The equation for the kinetic term has the form

\begin{eqnarray}
-\frac{\partial G_1({\bf R},{\bf R'},t)}{\partial t} =
-D\triangle G_1({\bf R},{\bf R'},t)
\end{eqnarray}

This is the diffusion equation with diffusion constant $D = \hbar^2/2m$
and its solution is a Gaussian

\begin{eqnarray}
\label{Green's function 1}
G_1({\bf R},{\bf R'},t) =
(4 \pi D t)^{-3N/2} \exp\left(-\frac{({\bf R}-{\bf R'})^2}{4Dt}\right)
\end{eqnarray}

The equation for the drift force term is

\begin{eqnarray}
-\frac{\partial G_2({\bf R},{\bf R'},t)}{\partial t} =
-D \vec \nabla (\vec F G_2({\bf R},{\bf R'},t))
\end{eqnarray}

and its solution is

\begin{eqnarray}
\label{Green's function 2}
G_2({\bf R},{\bf R'},t) = \delta({\bf R}-{\bf R}(t)),
\end{eqnarray}

here ${\bf R}(t)$ is the solution of the classical equation of motion

\begin{eqnarray}
\left\{
{\begin{array}{rcl}
\displaystyle\frac{{\bf dR}(t)}{dt}&=&\displaystyle D F({\bf R}(t)),\\
\displaystyle{\bf R}(0)&=&\displaystyle{\bf R'}
\end{array}}
\right.
\label{importance sampling}
\end{eqnarray}

The last equation from (\ref{G}) has trivial solution, which
describes the {\it rate term}

\begin{eqnarray}
\label{Green's function 3}
G_3({\bf R},{\bf R'},t) = \exp((E-E_L({\bf R}))t) \delta({\bf R}-{\bf R'})
\end{eqnarray}

\subsection{DMC algorithm}

If the wavefunction of the system $f({\bf R}, t)$ is
real and positive, as it happens in case of
ground state of a bose system, it can be treated as population
density distribution\footnote[1]{
The formula (\ref{Walkers}) should be understood in the statistical sense,
the average of any value $A$ over the l.h.s.and r.h.s distributions
are equal to each other in the limit when size of the population
$N_W$ tends to infinity
$\int A({\bf R}) f({\bf R},t)\,{\bf dR} = \lim\limits_{N_W\to\infty} \int A({\bf R})
\sum\limits_{i=1}^{N_W} C \delta ({\bf R - R_i}(t))\,{\bf dR}$}

\begin{eqnarray}
f({\bf R}, t) = \sum\limits_{i=1}^{N_W} C \delta ({\bf R - R_i}(t)),
\label{Walkers}
\end{eqnarray}

here $C$ is a positive constant, ${\bf R_i}(t)$ are coordinates of
a population element (so called {\it walker}) in $3N$-dimensional
configuration space, $f({\bf R}, t)\,{\bf dR}$ gives the probability
to find a walker at time $t$ in vicinity ${\bf dR}$ of point ${\bf R}$.

Let us now interpret the action of the each of the three terms of
the Hamiltonian (\ref{H summands}) on the population distribution
or, being the same, the action of the corresponding Green's
functions (\ref{Green's function 1}, \ref{Green's function 2},
\ref{Green's function 3}). In terms of Markov Chains the Green's
function is the $G({\bf R},{\bf R'},t)$ is the transition matrix
which determines the evolution of the distribution (see
eq.(\ref{Green approx})).

The first term means {\it diffusion} of each of the walkers in
configuration space

\begin{eqnarray}
{\bf R^{(1)}}(t+\triangle t) = {\bf R}(t) + {\bf \chi},
\label{R1}
\end{eqnarray}

here ${\bf \chi}$ is a random value from a gaussian distribution
$\exp(-\chi^2/(4 D \triangle t))$.

The second term describes the action of the drift force, which
guides the walkers to places in the configuration space, where the
trial wavefunction is maximal. This is the way how importance
sampling acts in this algorithm.

\begin{eqnarray}
{\bf R^{(2)}}(t+\triangle t) = {\bf R}(t) + \triangle t D F({\bf R})
\label{R2}
\end{eqnarray}

The corresponding Green's functions of these two steps
(\ref{Green's function 1} - \ref{Green's function 2}) are
normalized to one $\int G({\bf x,x'},t)\,{\bf dx} = 1$. The normalization of
wavefunction $f$ is then conserved meaning that the number of
walkers remains constant.

The third term is the {\it branching term}

\begin{eqnarray}
f^{(3)}({\bf R}, t+ \triangle t) =
\exp\left(-(E_L({\bf R})-E) \triangle t \right)
f({\bf R}, t)
\label{third step}
\label{branching}
\end{eqnarray}

Here the corresponding Green's function (\ref{Green's function 3})
is no longer normalized and, when the quantity in the exponent in
is negative (i.e. large values of local energy), then the density
of population decreases and vice-versa.

\subsection{Parallel DMC algorithm}

The simulation of a homogeneous infinite system is done by
repeating periodically in space the ``simulation box'' with side $L$.
Such a substitution leads to correlation in space for distances $r
> L/2$. It means that the one-body density matrix, the pair
distribution functions, can be calculated only for $r<L/2$. In some
cases it is important to have information about the large-scale
properties of the system. For example this happens in the region
close to the phase transition, where the correlation length is
large. In our case we need to find the asymptotic value of the
one-body density matrix and as a result it is necessary to use
large $L$. The problem is that doubling the number of particles $N$
at a constant density $n = N/L^3$ enlarges the size of the box only
$\sqrt[3]{2} = 1.26$ times, while the time of the calculation
scales quadratically. This makes the calculation heavy and one of
the possible way out consists in using the parallel computations.

The Diffusion Monte Carlo algorithm can be parallelized in a
natural way. In the algorithm the wavefunction $f$ is treated as
the density of the probability distribution of walkers (see
(\ref{Walkers})). The walkers explore the coordinate space moving
according to (\ref{R1}, \ref{R2}), and then some of the walkers are
removed or added during the branching process (\ref{branching}).
The key point is that the walkers move absolutely independently,
and consequently can be evaluated on different processors
independently.

Here is the list of corrections to the serial calculation:

\begin{enumerate}
\item walkers move in space according to (\ref{R1}, \ref{R2}), done in parallel
\item the local energy of each walker is calculated, done in parallel
\item other quantities different from energy are calculated (if necessary), done in parallel
\item averages over all walkers are calculated, the regeneration coefficient (\ref{branching}) is
calculated for each walker and walkers are redistributed among the
processors to keep the load constant
\end{enumerate}

The explained above algorithm of numerical simulations produces
very high productivity, because all heavy calculations are done in
the parallel regime and only few, like the the balance load, are done in
the non-parallel way.Although one also should consider the loss of
time due to communication and data transfer between the processes,
but amount of this time is negligible in comparison to the length
of the calculation.

This parallel algorithm is called {\it distributed-system method},
another way to build the program is described by the {\it
replicated-system method\cite{parallel}}, which has $5\%$ higher
performance.

Monte Carlo methods are very robust to numerical errors and this
makes their usage very advantageous. For example, the MC run on a
cluster of computers can survive even if one of the processes is
switched off during the computation. This will lead only to a
decrease in statistics at this step and different loading
distribution at the next one. Many other methods, like the direct
solution of the differential equation, will break down in a case of
such an event.

%% file: HBG.tex
In this section we apply the DMC algorithm to the study of a
homogeneous Bose gas modeled by hard spheres.

\subsection{Trial Wavefunction \label{Trial Wavefunction}}

The trial wavefunction $\psi_T$ should be chosen as close to the
true wave function $\psi$ of the system as possible. If we were
able to approximate the system wavefunction with satisfying
accuracy, then the sampling over the corresponding distribution
(for example with the help of Metropolis algorithm) would give us
all properties of the system. However, the problem is that very
often it is impossible to find the wavefunction of the system using
analytical methods. Here enters the Diffusion Monte Carlo method,
which compensates our lack of knowledge and corrects the trial
wavefunction provided that the projection of the trial wavefunction
on the true system wavefunction of the system differs from zero.

As the use of the trial wavefunction lies at the heart of the
method, it has to be expressed in a way that is fast to calculate
or it has to be tabulated.

The common way to construct many-body wavefunctions is to use the Jastrow
function consisting of the product of an uncorrelated state and
a correlation factor, which is a product of two-body wavefunctions.

\begin{eqnarray}
\psi_T = \prod\limits_i \phi(\vec r_i) \prod\limits_{i<j} g(\vec r_i,\vec r_j)
\end{eqnarray}

The one-body term describes the effect of an external field and is
absent in the case of a homogeneous system. For a homogeneous
system the trial wavefunction can be written in general as

\begin{eqnarray}
\psi_T(r_1,...,r_N) =
\prod\limits_{i=1\atop j=1}^N g(|\vec r_i-\vec r_j|)
\label{prod}
\end{eqnarray}

Since we are mainly interested to dilute system a possible way to
obtain the pair function $g(r)$ is through the solution of the
two-body Schr\"odinger equation.

\begin{eqnarray}
\left(-\frac{\hbar^2}{2\mu}\triangle+V(\vec r\,) \right)g = {\cal E} g,
\end{eqnarray}

here $\mu = m_1 m_2 / (m_1+m_2) = 2m$ is the reduced mass and $r$
is the interparticle distance. Let's search for the $l = 0$
solution in spherical coordinates

\begin{eqnarray}
-\frac{\hbar^2}{2\mu} \left(g''+\frac{2}{r}g'\right) + V(r) g = {\cal E} g
\end{eqnarray}

The particles are modeled by hard spheres of diameter $a$ and the
interaction potential is

\begin{eqnarray}
V({\bf r}) =
\left\{
{\begin{array}{ll}
+\infty,&|r| \le a\\
0,&|r| > a\\
\end{array}}
\right.
\end{eqnarray}

The dimensionless Schr\"odinger equation is obtained by expressing
all distances in units of $a$ and energy in units of $\hbar^2/(2ma^2)$

\begin{eqnarray}
\left\{
{\begin{array}{ll}
\displaystyle g(x) = 0,&|x| \le 1\\
\displaystyle g''+\frac{2}{x}g' - 2E g = 0,& |x| > 1 \\
\end{array}}
\right.
\end{eqnarray}

So, it is necessary to solve the differential equation

\begin{eqnarray}
\left\{
{\begin{array}{l}
\displaystyle g''+\frac{2}{x}g'- 2E g = 0\\
\displaystyle g(1) = 0\\
\end{array}}
\right.
\label{DE1}
\end{eqnarray}

The solution of equation (\ref{DE1}) is
$g(x) = A\sin\left(\sqrt{2E}(x-1)\right)\Bigl/x\Bigr.$, where $A$ is an arbitrary constant.

In dilute systems for small interparticle distance $r$ the
correlation factor is well approximated by the function $g(r)$,
i.e. by the wavefunction of a pair of particles in vacuum. At large
distances the pair wavefunction should be constant, which
corresponds to uncorrelated particles.

Taking these facts into account let us introduce the trial function
in the following way \cite{DMC Bose Gas}

\begin{eqnarray}
g(x) =
\left\{
{\begin{array}{ll}
\displaystyle \frac{A}{x} \sin(\sqrt{2E}(x-1)),& |x| \le R\\
\displaystyle 1- B \exp\left(-\frac{x}{\alpha}\,\right),& |x| > R\\
\end{array}}
\right.
\label{f_PP}
\end{eqnarray}

This function has to be smooth at the matching point $R$, i.e.

1) the $f(x)$ must be continuous

\begin{eqnarray}
\frac{A}{R} \sin(\sqrt{2E}(R-1)) = 1 - B \exp\left(-\frac{R}{\alpha}\,\right)
\end{eqnarray}

2) its derivative $f'(x)$ must be continuous

\begin{eqnarray}
-\frac{A}{R^2} \sin(\sqrt{E}(R-1))
+\frac{A\sqrt{2E}}{R} \cos(\sqrt{2E}(R-1)) = \frac{B}{\alpha} \exp\left(-\frac{R}{\alpha}\,\right)
\end{eqnarray}

3) the local energy $E_{L}(x) = \psi_T^{-1}\hat H\psi_T$ must be
continuous

\begin{eqnarray}
2E =
\frac{\left(\displaystyle\frac{1}{\alpha^2}-\frac{2}{R\alpha}\right)
B \exp\left(-\displaystyle\frac{R}{\alpha}\,\right)
}{1-B \exp\left(-\displaystyle\frac{R}{\alpha}\,\right)}
\end{eqnarray}

The solution of this system is

\begin{eqnarray}
\left\{
{\begin{array}{l}
A =\displaystyle\frac{R}{\sin(u(1-1/R))}\frac{\xi^2-2\xi}{\xi^2-2\xi+u^2},\\
B =\displaystyle\frac{u^2 \exp(\xi)}{\xi^2-2\xi+ u^2},
\end{array}}
\right.
\label{f_PP 1}
\end{eqnarray}

where we used the notation $u = \sqrt{2E}R$ and $\xi = R/\alpha$.
The value of $\xi$ is obtained from the equation

\begin{eqnarray}
1-\frac{1}{R} = \frac{1}{u} \arctg\frac{u(\xi-2)}{u^2+\xi-2}
\label{f_PP 2}
\end{eqnarray}

There are three conditions for the determination of five unknown
parameters, consequently two parameters are left free. The usual
way to define them is minimize the variational energy in
Variational Monte Carlo which yields an optimized trial
wavefunction.

\subsection{Comparison between VMC and DMC methods}

\begin{figure}
\includegraphics[width=\textwidth]{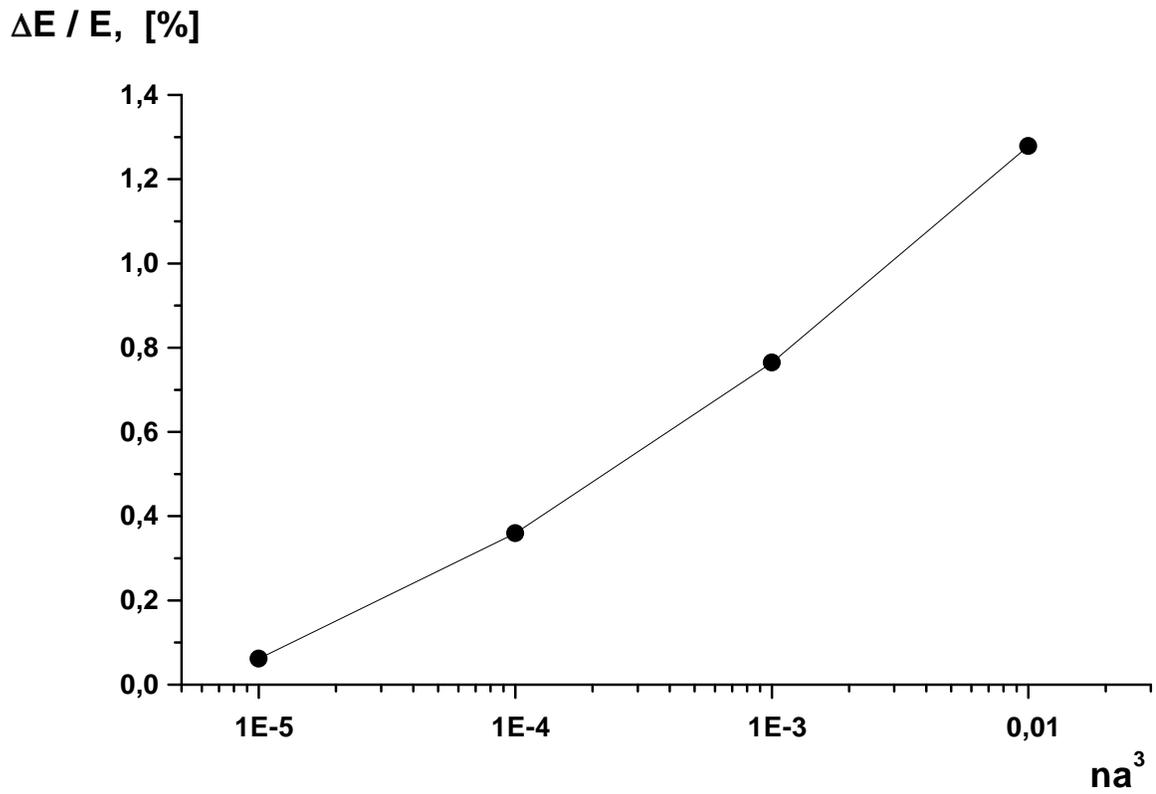}
\caption{Energy comparison between VMC and DMC calculations
$\triangle{E} = E^{VMC}-E^{DMC}$
at different densities $na^3$}
\label{Ecompare}
\end{figure}

The Jastrow trial-wavefunction obtained in
the previous section givers a very good approximation for the ground
state wavefunction of dilute Bose gases.
One can appreciate this by comparing the VMC and
DMC energies (Fig.~\ref{Ecompare}).
On this plot the relative error $\triangle E/E$ of the energy
estimate is presented as a function of the density.
The coincidence of DMC and VMC results is very good
in the density region $n<10^{-4}$ and remains quite good
for higher densities.

The energy obtained in VMC is always larger than the ground state
energy (see Fig.~\ref{Ecompare}). In other words, variational
simulations always give an {\it upper bound} to the ground state
energy.

\begin{eqnarray}
E^{VMC} \ge E
\end{eqnarray}


%% file: values.tex
\section{Outputs of the calculation}

The advantage of Diffusion Monte Carlo with respect to usual
Variational Monte Carlo methods is that it provides the possibility
to calculate pure estimators, which have no bias due to the choice
of the trial wave function. This is true for such important
quantities as the energy, radial distribution function and
superfluid density. For other quantities such as the one body
density matrix the DMC method provides mixed estimator which, used
in correction with outputs of VMC calculations, allows one to
reduce the bias of the trial wavefunction.

\subsection{Energy}

The energy is a direct output of the DMC algorithm. In fact, the
population of walkers is stable only if the energy shift $E$ is
equal to the value of ground state energy.

The ground state energy can be expressed as an integral ratio

\begin{eqnarray}
E_0 = \frac{\int\psi_T({\bf R})\hat H \phi_0({\bf R}){\bf dR}}
{\int\psi_T({\bf R})\phi_0({\bf R})\,{\bf dR}},
\label{Eo}
\end{eqnarray}

where $\phi_0({\bf R})$ is the ground state
eigenfunction of the Hamiltonian $\hat H\phi_0({\bf R})~=~E_0\phi_0({\bf R})$.
By multiplying and dividing the integrand in the numerator
by $\psi_{T}({\bf R})$, the formula (\ref{Eo}) can be rewritten as

\begin{eqnarray}
E_0 = \frac{\int\psi_T^{-1}({\bf R})\hat H\psi_T({\bf R})\psi_T({\bf R})\phi_0({\bf R})\,{\bf dR}}
{\int\psi_T({\bf R})\phi_0({\bf R})\,{\bf dR}}
\end{eqnarray}

The average of the Hamiltonian over
the trial wavefunction is the local energy (see definition (\ref{local energy})).
Since in the large time limit the distribution function
$f$ is proportional to the product of the
trial and the ground-state wavefunctions (see eq. (\ref{psi limit}))

\begin{eqnarray}
\lim\limits_{t \to \infty} f({\bf R},t) = c_0 \psi_T({\bf R}) \phi_0({\bf R},t)
\label{f limit}
\end{eqnarray}

The calculation of the mean local energy of the walkers provides
the value of the ground state energy\footnote[1]{In real DMC
simulations the upper limit of the integrals is truncated by $L/2$.
The ``tail'' energy, which is small and is typically much less than
$1\%$ of the total energy can be approximated by the formula
$E_{tail} = n \int_{L/2}^{\infty}E_L(r) 4\pi r^2 dr$. The idea is
that at large distances, where the integral is evaluated, one can
safely assume uniform distribution of particles.}

\begin{eqnarray}
E_0 = \frac{\int E_L({\bf R})f({\bf R}){\bf dR}}{\int f({\bf R}){\bf dR}}
= \frac{1}{N_W} \sum\limits_{i=1}^{N_W} E_L({\bf R_i})
\end{eqnarray}

\subsection{Superfluid density \label{superfluid density}}

\newcommand{\V}{{\upsilon}}

The normal and superfluid fractions of a liquid can be obtained by
measuring the momenta of inertia of a rotating bucket. Consider a
liquid which is inserted between two cylindrical walls of radii $R$
and $R+d$. If $d \ll R$ then the system can be described as moving
between two planes. Let us denote by $E_{\V}$ the ground state
energy of the system in equilibrium with the walls which move with
velocity $\V$ and $E_0$ the ground state energy of the system at
rest. The difference between the energies $E_{\V}$ and $E_0$ is due
to the superfluid component, which remains immobile in contrast to
the normal component which is carried along by the moving walls.
Thus, the superfluid fraction $\rho_s/\rho$ can be defined as

\begin{eqnarray}
\frac{N m \V^2}{2} \frac{\rho_s}{\rho} = E_{\V} - E_0
\end{eqnarray}

Let us introduce the wave-functions $f_0$ and $f_{\V}$ related to
the wave-functions of the system in the reference frames at rest
and in motion.

\begin{eqnarray}
\label{f_0}
f_0({\bf R}, t) &=& \psi_T({\bf R}) \phi_0({\bf R}, t),\\
f_{\V}({\bf R}, t) &=& \psi_T({\bf R}) \phi_{\V}({\bf R}, t)
\label{f_V}
\end{eqnarray}

These wavefunctions satisfy the Schr\"odinger equation with the
following Hamiltonians

\begin{eqnarray}
\hat H_0 = \frac{1}{2m} \sum\limits_i (-\imath \hbar \nabla_i)^2 + V({\bf R})
\end{eqnarray}

for the reference frame at rest and

\begin{eqnarray}
\hat H_{\V} = \frac{1}{2m} \sum\limits_i
(-\imath \hbar \nabla_i -m \vec{\V} )^2 + V({\bf R})
\end{eqnarray}

for the reference frame at moving with velocity $\V$.

In the reference frame at rest one has

\begin{eqnarray}
-\frac{\partial}{\partial t} f_0({\bf R}, t) =
-\frac{\hbar^2}{2m} \sum\limits_{i=1}^{N}
\biggl[\triangle_i f_0({\bf R},t)
-\nabla_i \Bigl(\vec F f_0({\bf R},t)\Bigr)\biggr]
+ (E_L({\bf R}) - E_0) f_0({\bf R}, t)
\label{Schr 0}
\end{eqnarray}

The Schr\"odinger equation in the moving frame is instead

\begin{eqnarray}
\begin{array}{ll}
\displaystyle-\frac{\partial}{\partial t} f_{\V}({\bf R}, t)=&
\displaystyle-\frac{\hbar^2}{2m} \sum\limits_{i=1}^{N}
\biggl[\triangle_i f_{\V}({\bf R},t)
-\nabla_i \Bigl(\vec F f_{\V}({\bf R},t)\Bigr)\biggr]
+ (E_L({\bf R})-E_{\V}) f_{\V}({\bf R}, t) +\\
&\displaystyle+\quad\frac{N m \V^2}{2}
f_{\V}({\bf R}, t) + \sum\limits_{i=1}^{N} i\hbar \vec{\V}\,\nabla_i
f_{\V}({\bf R}, t)
- \frac{i\hbar}{2} \vec{\mathstrut \V} \vec{\mathstrut F} f_{\V} ({\bf R}, t)
\end{array}
\label{Schr V}
\end{eqnarray}

Looking at (\ref{Schr 0}) and (\ref{Schr V}) it is easy to write
the Bloch equations for the Green's functions in the rest frame
$G_0({\bf R}, {\bf R'}, t)$ and in the moving frame $G_{\V}({\bf R},
{\bf R'}, t)$

\begin{eqnarray}
\label{Go}
-\frac{\partial}{\partial t} G_0({\bf R}, {\bf R'}, t) =
\left(-\frac{\hbar^2}{2m} \sum\limits_{i=1}^{N}
\Bigl[\triangle_i-(\nabla_i\vec F) - \vec F\nabla_i\Bigr]
+ E_L({\bf R}) - E_0\right) G_0({\bf R}, {\bf R'}, t)
\end{eqnarray}

and

\begin{eqnarray}
\begin{array}{lc}
\label{Gv}
\displaystyle-\frac{\partial}{\partial t} G_{\V}({\bf R}, {\bf R'}, t) =&
\left(-\frac{\hbar^2}{2m} \sum\limits_{i=1}^{N}
\Bigl[\triangle_i-(\nabla_i\vec F) - \vec F\nabla_i\Bigr]
+ E_L({\bf R}) - E_0 + \frac{N m \V^2}{2}
\right) G_{\V}({\bf R}, {\bf R'}, t) +\\
\displaystyle
&+\quad \left(\sum\limits_{i=1}^{N} i\hbar \vec{\V} \nabla_i
- \frac{i\hbar}{2} \vec{\V} \vec F\right)
G_{\V}({\bf R}, {\bf R'}, t)
\end{array}
\end{eqnarray}

In general the wavefunction $\psi({\bf R}, t)$ of the system satisfies the
Schr\"odinger equation (\ref{shifted Schrodinger}) and its
evolution in time is described by

\begin{eqnarray}
\psi({\bf R},t) = e^{-(\hat H - E)t} \psi ({\bf R},0)
\label{evolution psi}
\end{eqnarray}

The wavefunctions $f$ evolves in time as

\begin{eqnarray}
f({\bf R},t) = e^{-At} f({\bf R},0)
\label{evolution f}
\end{eqnarray}

so, substitution of (\ref{f_0}) or (\ref{f_V}) into (\ref{evolution f})
gives

\begin{eqnarray}
\psi_T({\bf R})\psi({\bf R},t) = e^{-At}\psi_T({\bf R})\psi({\bf R},0)
\label{evolution psi_T psi}
\end{eqnarray}

Combining together (\ref{evolution psi}) and (\ref{evolution psi_T
psi}) one has

\begin{eqnarray}
e^{-At}
= \psi_T({\bf R}) e^{-(\hat H - E)t} \psi_T^{-1}({\bf R})
= B e^{-(\hat H - E)t} B^{-1},
\label{exp(-At)}
\end{eqnarray}

where the operator $B$ is defined as
$B|\psi\rangle = \sum_{\bf R} \psi({\bf R})\,
\langle {\bf R}|\psi\rangle~|{\bf R}\rangle$.

Let us calculate the trace of the Green's function. From
(\ref{exp(-At)}) it follows that the trace $T$ of the Green's
function is equal to

\begin{equation}
T = \int\!G_0({\bf R}, {\bf R}, t)\,{\bf dR}
= \int\langle {\bf R}| e^{-tA}| {\bf R} \rangle\,{\bf dR}
= \int\langle {\bf R}| B e^{-t(\hat H - E) B^{-1}}| {\bf R} \rangle\,{\bf dR}
\end{equation}

Here it is possible to use the permutation property of the trace
$\tr(AB) = \tr(BA)$

\begin{equation}
T = \int \langle {\bf R}|e^{-t(\hat H - E)}| {\bf R} \rangle\,{\bf dR}
\label{T}
\end{equation}

This formula means that the trace of the Green's function is
unaffected by the presence of the trial wavefunction $\psi_T$.

\begin{equation}
T = \sum\limits_{k,l} \int
\langle{\bf R}|\phi_k\rangle\,\langle \phi_k|e^{-t(\hat H - E)}|\phi_l\rangle
\,\langle\phi_l|{\bf R}\rangle\,{\bf dR}
= \sum\limits_k e^{-(E_k-E)t}
\end{equation}

After long enough time of evolution the traces of the Green's function
$G_0$ is fixed by the ground state energy

\begin{equation}
\int G_0({\bf R}, {\bf R}, t)\,{\bf dR}\to e^{-E_0t},
\quad t\to\infty,
\label{trace}
\end{equation}

Approximation (\ref{trace}) is valid for times $t$ such that

\begin{eqnarray}
t \gg 1 / E_0
\label{time}
\end{eqnarray}

Analogously, the trace of $G_{\V} ({\bf R, R},t)$ is fixed
by the ground state energy $E_{\V}$ in the moving frame

\begin{eqnarray}
\int G_{\V}({\bf R, R},t)\,{\bf dR}\to e^{-t E_{\V}}\quad t\to\infty,
\label{trace 2}
\end{eqnarray}

The Green's function has to comply with periodic boundary conditions,
i.e. it must remain the same if one of the arguments is shifted
by the period $\vec L$

\begin{eqnarray}
\label{boundary conditions}
G_0(\vec r_1, ..., \vec r_i + \vec L, ..., \vec r_N,~{\bf R'},~t) &=&
G_0(\vec r_1, ..., \vec r_i, ..., \vec r_N,~{\bf R'},~t), \\
\label{boundary conditions 2}
G_{\V}(\vec r_1, ..., \vec r_i + \vec L, ..., \vec r_N,~{\bf R'},~t) &=&
G_{\V}(\vec r_1, ..., \vec r_i, ..., \vec r_N,~{\bf R'},~t)
\end{eqnarray}

Let us define a new Green's function $\tilde G({\bf R}, {\bf R'}, t)$
in such a way that

\begin{eqnarray}
G_{\V}({\bf R, R'},t) =
exp\left(i\frac{m}{\hbar} \vec{\V} \sum\limits_i (\vec r_i - \vec r_i\mathstrut')\right)
\tilde G({\bf R, R'},t)
\end{eqnarray}

The Green's function $\tilde G({\bf R, R'}, t)$ satisfies the same
Bloch equation (\ref{Go}) as $G_0({\bf R}, {\bf R'}, t)$, but the
boundary conditions differ from (\ref{boundary conditions},
\ref{boundary conditions 2}) by the presence of a phase factor

\begin{eqnarray}
\tilde G(\vec r_1, ..., {\vec r_i+\vec L}, ..., \vec r_N,~{\bf R'},~t) =
exp\left(-i\frac{m}{\hbar} \vec{\mathstrut \V} \vec{\mathstrut  L}\right)
\tilde G(\vec r_1, ..., \vec r_i, ..., \vec r_N,~{\bf R'},~t),
\end{eqnarray}

Results (\ref{trace}) and (\ref{trace 2}) give the following
relation

\begin{eqnarray}
\frac{\int \tilde G({\bf R, R},t)\,{\bf dR}}{\int G_0({\bf R, R},t)\,{\bf dR}}=
\frac{\int G_\V({\bf R, R},t)\,{\bf dR}}{\int G_0({\bf R, R},t)\,{\bf dR}} \approx
\frac{ e^{-t E_{\V}}}{e^{-t E_0}}
\end{eqnarray}

By assuming that $t (E_{\V} - E_0) \ll 1$ one gets

\begin{eqnarray}
\frac{ e^{-t E_{\V}}}{e^{-t E_0}} \approx 1 - t(E_{\V} - E_0)
\end{eqnarray}

The ratio of the traces is related to the energy difference

\begin{eqnarray}
\frac{\int\tilde G({\bf R, R},t)\,{\bf dR}}{\int G_0({\bf R, R},t)\,{\bf dR}} =
1 - t(E_\V - E_0)
\end{eqnarray}

The Green's function $\tilde G$ coincides with $G_0$ apart when the
boundary conditions are invoked. Let us introduce the {\it winding
number} $W$ \cite{Pollock and Ceperley},
which counts how many times the boundary conditions
were used during the time evaluation

\begin{eqnarray}
 1 - t(E_\V - E_0) =
\frac{\int |f({\bf R},t)|^2 e^{-i\frac{m}{\hbar}\vec{\V}\,W\vec L}\,{\bf dR}}
{\int |f({\bf R},t)|^2\,{\bf dR}}
\end{eqnarray}

In the case of slowly moving walls, i.e. when
$\vec{\V}\,\frac{m}{\hbar} W\vec L \ll 1$, the exponential
can be expanded in a Taylor series

\begin{eqnarray}
e^{-i \frac{m}{\hbar}\vec{\V}\,W\vec L} \approx
1 -i\frac{m}{\hbar}\vec{\V}\,W\vec L -
\frac{m^2}{\hbar^2} (\vec{\V}\,W\vec L)^2
\end{eqnarray}

Let us define $W$ through the distance the particles have gone
during the time $t$

\begin{eqnarray}
W\vec L = \sum\limits_{i = 1}^N\Bigl(\vec r_i(t) - \vec r_i(0)\Bigr)
\end{eqnarray}

The average value of the linear term is equal to zero and the final
result is

\begin{eqnarray}
\frac{\rho_s}{\rho} =
\frac{2m}{\hbar^2} \frac{1}{6 N}
\lim\limits_{t \to \infty}
\frac{1}{t}
\frac{\int |f({\bf R},t)|^2 (WL)^2\,{\bf dR}}
{\int |f({\bf R},t)|^2\,{\bf dR}}
\label{sd limit}
\end{eqnarray}

An interpretation of this result is that the superfluid fraction is
equal to the ratio between the diffusion constant $D_{\V}$ of the
center of the mass of the system and the free diffusion constant
$D_0$\footnote[1]{
It is necessary to note that here the ``diffusion'' occurs in
imaginary time and it has nothing to do with diffusion in real space.}

\begin{eqnarray}
\frac{\rho_s}{\rho} = \frac{D_{\V}}{D_0},
\end{eqnarray}

where the diffusion constants are defined as

\begin{eqnarray}
D_0 &=& \frac{\hbar^2}{2m} ,\\
D_{\V} &=& \lim\limits_{t \to \infty}
\frac{N}{6t}
\frac{\int f({\bf R}, t)\Bigl(\vec R_{CM}(t)-\vec R_{CM}(0)\Bigr)^2
\,{\bf dR}} {\int f({\bf R},t)\,{\bf dR}},
\end{eqnarray}

where the center of the mass of the system is

\begin{eqnarray}
\vec R_{CM}(t) = \frac{1}{N} \sum\limits_{i=1}^N \vec r_i(t)
\end{eqnarray}

By calculating the ratio $D_\V/D_0$ as a function of time one finds
that this ratio starts from $1$ at small time step, decreases and
finally reaches a constant plateau, In practice the best way of
finding the asymptotic value is to fit the ration $D_\V/D_0$ with
the function $C_0+C_1(1-exp(-C_2t))/t$, where $C_0$, $C_1$, $C_2$
are fitting parameters \cite{2D DIS}.

It is worth to remind that the calculation of $\rho_s/\rho$ is
independent of the choice of the trial wave-function and similarly
to the calculation of energy the superfluid fraction is a pure
estimator.

\subsection{One body density matrix and condensate fraction}

The one body density matrix (OBDM) of a homogeneous system described by the
many body wavefunction $\psi(r_1, ..., r_N)$ is defined as follows

\begin{eqnarray}
\rho(|\vec r~'-\vec r~''|) =
N \frac{\int...\int \phi_0^*(\vec r\,', \vec r_2, ..., \vec r_N) \phi_0 (\vec r\,'', \vec r_2, ..., \vec r_N)\,d\vec r_2 ...d\vec r_N}
{\int...\int |\phi_0(\vec r_1, ..., \vec r_N)|^2\,d\vec r_1 ... d\vec r_N}.
\end{eqnarray}

Since in DMC calculation one can not sample the ground-state
probability distribution $\phi_0^2$, instead one samples the mixed
probability $\psi_T\phi_0$ and one can calculate the mixed
one-body density matrix

\begin{eqnarray}
\rho_{mixed} (r) =
N \frac{\int...\int \psi^*_T(\vec r\,''+\vec r, \vec r_2,..., \vec r_N) \phi_0 (\vec r\,'', \vec r_2, ..., \vec r_N)\,d\vec r_2 ... d\vec r_N}
{\int...\int \psi^*_T(\vec r_1, ...,\vec r_N)\phi_0(\vec r_1, ...,\vec r_N)\,d\vec r_1 ... d\vec r_N},
\end{eqnarray}

This formula can be further developed

\begin{eqnarray}
\begin{array}{l}
\displaystyle
\rho_{mixed}(r) =
N \frac{\int...\int \psi^*_T(\vec r_1+\vec r,\vec r_2,...,\vec r_N)\phi_0(\vec r_1, \vec r_2, ..., \vec r_N)
\delta (\vec r_1 - \vec r\,'')d\vec r_1 ... d\vec r_N}
{\int...\int \psi^*_T(\vec r_1, ...,\vec r_N)\phi_0(\vec r_1, ...,\vec r_N)d\vec r_1 ... d\vec r_N} = \\
=
\displaystyle
n\frac{\int...\int
[\psi^*_T(\vec r_1+\vec r, \vec r_2,..., \vec r_N)(\psi^*_T(\vec r_1,\vec r_2,...,\vec r_N))^{-1}]
f(\vec r_1, ..., \vec r_N)d\vec r_1 ... d\vec r_N}
{\int...\int f(\vec r_1, ..., \vec r_N)d\vec r_1 ... d\vec r_N},
\quad t\to\infty
\end{array}
\label{OBDM mixed}
\end{eqnarray}

where we have used the asymptotic formula (\ref{f limit}). If the
trial wavefunction is chosen as a product of pair functions (see
eq. (\ref{prod}) then using the notation $\mu (|\vec r_i -\vec r_j|) = \ln
g(|\vec r_i -\vec r_j|)$) one has

\begin{eqnarray}
\psi_T(\vec r_1, ...,\vec r_N) = \prod\limits_{i<j} e^{\mu(|\vec r_i-\vec r_j|)}
\end{eqnarray}

Then, the ratio of trial wavefunction appearing in (\ref{OBDM mixed})
becomes

\begin{eqnarray}
\nonumber
\frac{\psi_T(\vec r_1+\vec r, ...,\vec r_N)}{\psi_T(\vec r_1, ...,\vec r_N)} =
\prod\limits_{j>1} exp\left(\mu(|\vec r_1+\vec r-\vec r_j|)-\mu(|\vec r_1-\vec r_j|)\right) =\\
= exp\left(\sum\limits_{j>1} \mu(|\vec r_1+\vec r-\vec r_j|)-\mu(|\vec r_1-\vec r_j|)\right).
\end{eqnarray}

In order to gain better statistics one can average over all particles

\begin{eqnarray}
\nonumber
\frac{1}{N} \sum\limits_{i=1}^N
\frac{\psi_T(\vec r_1, ..., \vec r_i+\vec r, ...,\vec r_N)}{\psi_T(\vec r_1, ...,\vec r_N)}
= \frac{1}{N} \sum\limits_{i=1}^N
exp\left(\sum\limits_{j \neq i}^N \mu(|\vec r_i+\vec r-\vec r_j|)-\mu(|\vec r_i-\vec r_j|)\right)
\end{eqnarray}

The asymptotic limit of the OBDM gives the condensate density

\begin{eqnarray}
\lim\limits_{r\to\infty} \rho(r) = \frac{N_0}{V}
\end{eqnarray}

and the condensate fraction is obtained by the calculating the
asymptotic ratio

\begin{eqnarray}
\lim\limits_{r\to\infty} \frac{\rho(r)}{\rho} = \frac{N_0}{N}
\end{eqnarray}

\subsection{Extrapolation technique from mixed and variational estimators
\label{Extrapolation technique}}

The one body density matrix (\ref{OBDM mixed}) corresponds to a
mixed estimator, when the averaging of the variable $A$ is
done in an asymmetric way $\langle\phi_0|\hat A|\psi_T\rangle$.
If the trial wavefunction is close to the true ground-state
wavefunction $\phi_0$ one can estimate the ground-state average
$\langle\phi_0|\hat A|\psi_0\rangle$ by using the following
technique.

Let us denote the difference between the trial wave function and
ground-state wave function as $\delta \psi$

\begin{eqnarray}
\phi_0 = \psi_T + \delta \psi
\end{eqnarray}

Then the ground-state average can be written as

\begin{eqnarray}
\langle\phi_0|\hat A|\phi_0\rangle =
\langle\psi_T|\hat A|\psi_T\rangle + 2\langle\phi_0|\hat A|\delta\psi\rangle
+ \langle\delta\psi|\hat A|\delta\psi\rangle
\end{eqnarray}

If $\delta \psi$ is small the second order term
$\langle\delta\psi|\hat A|\delta\psi\rangle$ can be neglected.
After substitution
$\langle\phi_0|\hat A|\delta\psi\rangle =
\langle\psi_T|\hat A|\phi_0\rangle -
\langle\psi_T|\hat A|\psi_T\rangle$
the extrapolation formula becomes

\begin{eqnarray}
\langle A\rangle =
\langle\phi_0|\hat A|\phi_0\rangle =
2\langle\phi_0|\hat A|\psi_T\rangle - \langle\psi_T|\hat A|\psi_T\rangle
\label{extrapolation}
\end{eqnarray}

%% file: errors.tex
\section{Systematic errors}
\subsection{Population of walkers}

A key ingredient of the DMC algorithm is the branching process
(\ref{branching}). The algorithm is exact, i.e. gives the exact
ground-state energy of the system, in the limit of an infinite
population of walkers. However it is necessary to understand how
many walkers have to be used in practice to estimate the energy
with a given value of accuracy \cite{population effects}.

\begin{figure}
\includegraphics[width=\textwidth]{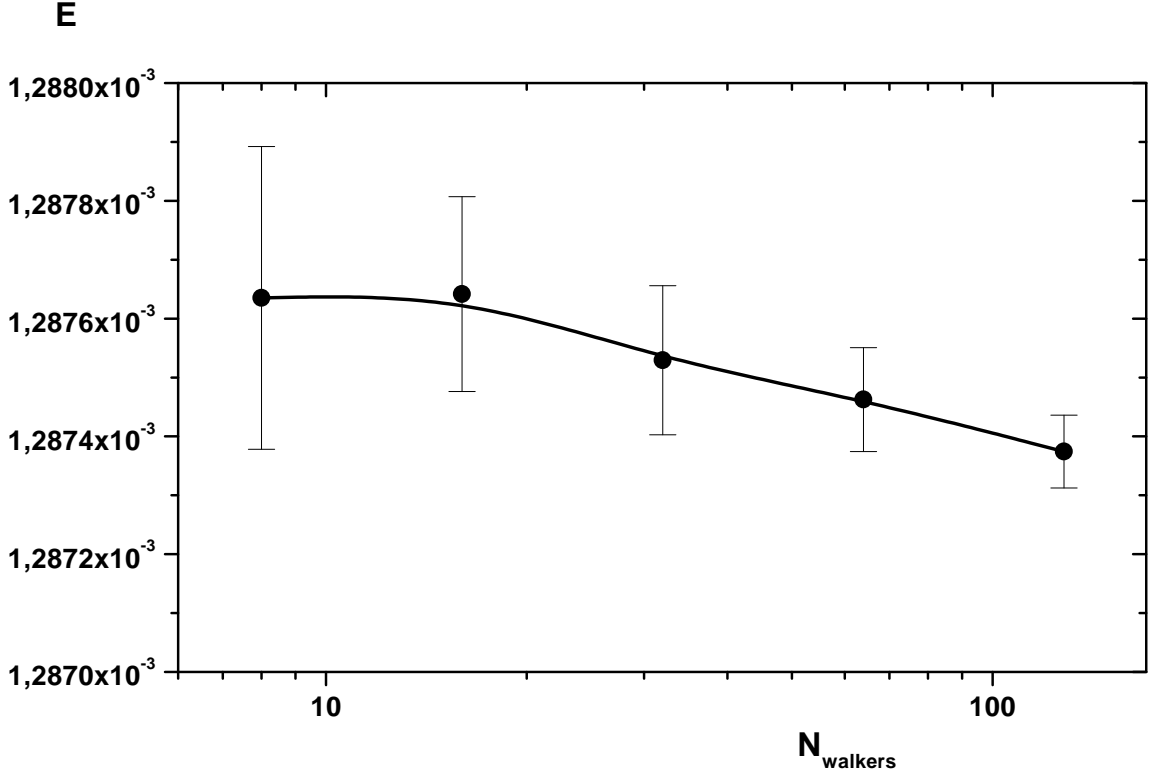}
\caption{Hard-spheres at $na^3 = 10^{-4}$. Dependence of
the energy on the size of the population of walkers.}
\label{Enwalk}
\end{figure}

In Fig.~\ref{Enwalk} we plot the energy for a given value of the
density $na^3 = 10^{-4}$ as a function of the mean number of
walkers. In all simulations carried out in this work we have used
about one hundred walkers.

\subsection{Time step\label{timestep}}

The approximation (\ref{t2 approximation}) of the Green's function
has first order accuracy in the timestep. High order approximations
can be used. One of the possibilities to gain second order accuracy
is to use the formula

\begin{eqnarray}
e^{-\hat H t} =
e^{-\hat H_3 t/2} e^{-\hat H_2 t/2} e^{-\hat H_1 t}
e^{-\hat H_2 t/2}  e^{-\hat H_3 t/2} + O(t^3)
\label{t3 approximation}
\end{eqnarray}

\begin{figure}
\includegraphics[width=0.75\textwidth]{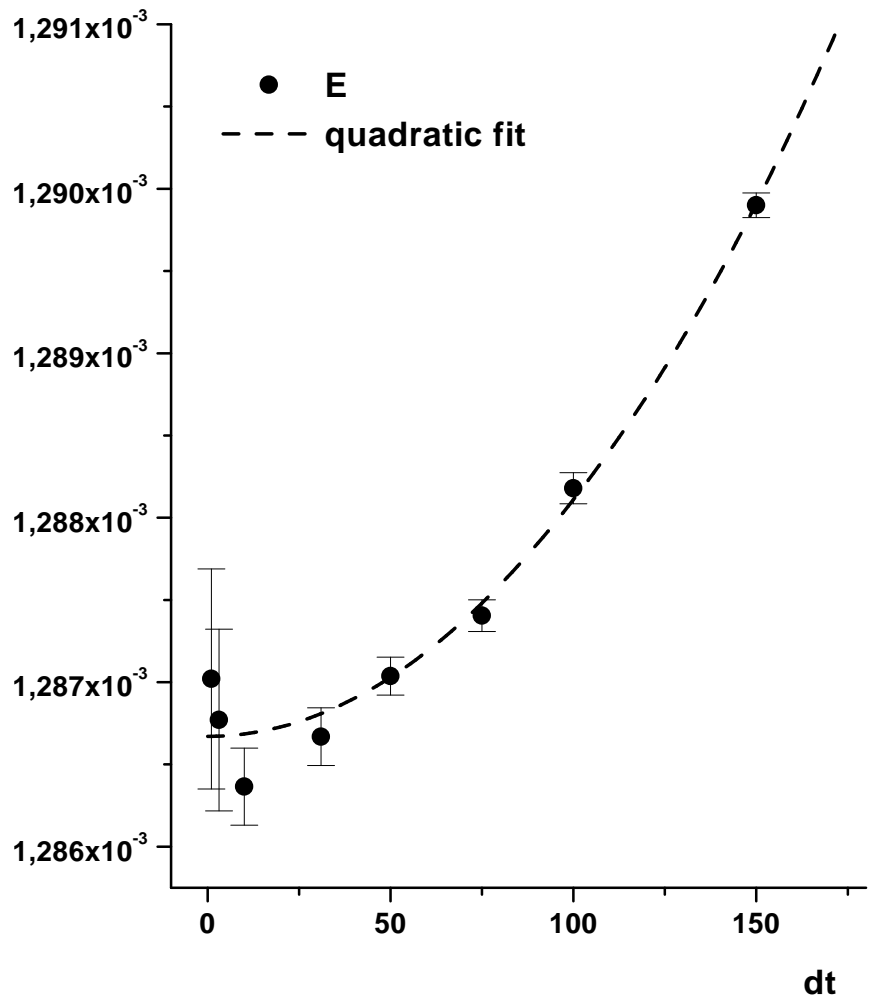}
\caption{Hard spheres at $na^3 = 10^{-4}$.
Dependence of the energy on the time step.}
\label{Etimestep}
\end{figure}

The result for the energy in the DMC algorithm depends on the value
of the timestep used. The exact ground-state energy is obtained by
extrapolating the results to the zero timestep. Approximation
(\ref{t3 approximation}) for the evaluation operator leads to a
quadratic dependence of the energy on the timestep. The result of
such a calculation is presented in Fig.~\ref{Etimestep}. In this
respect the use of a quadratic algorithm, such as (\ref{t3
approximation}), is preferable because for small timestep the
results are less sensitive to the choice of the timestep and with a
judicious choice one does not need to extrapolate.

On one side the timestep has to be small so that the approximation
in the Green's function is good, on the other side the larger is
the timestep the faster the phase space is explored and less number
of iterations are needed for the same statistical accuracy.

\subsection{Finite size errors \label{Finite size errors}}

The standard way to simulate infinite systems is to use a finite
box with periodic boundary conditions. As a result, if the size of
the box is not large enough, one can have large errors due to
finite size effects.
These  effects are very important in the proximity of a
phase transition.

\begin{figure}
\includegraphics[width=\textwidth]{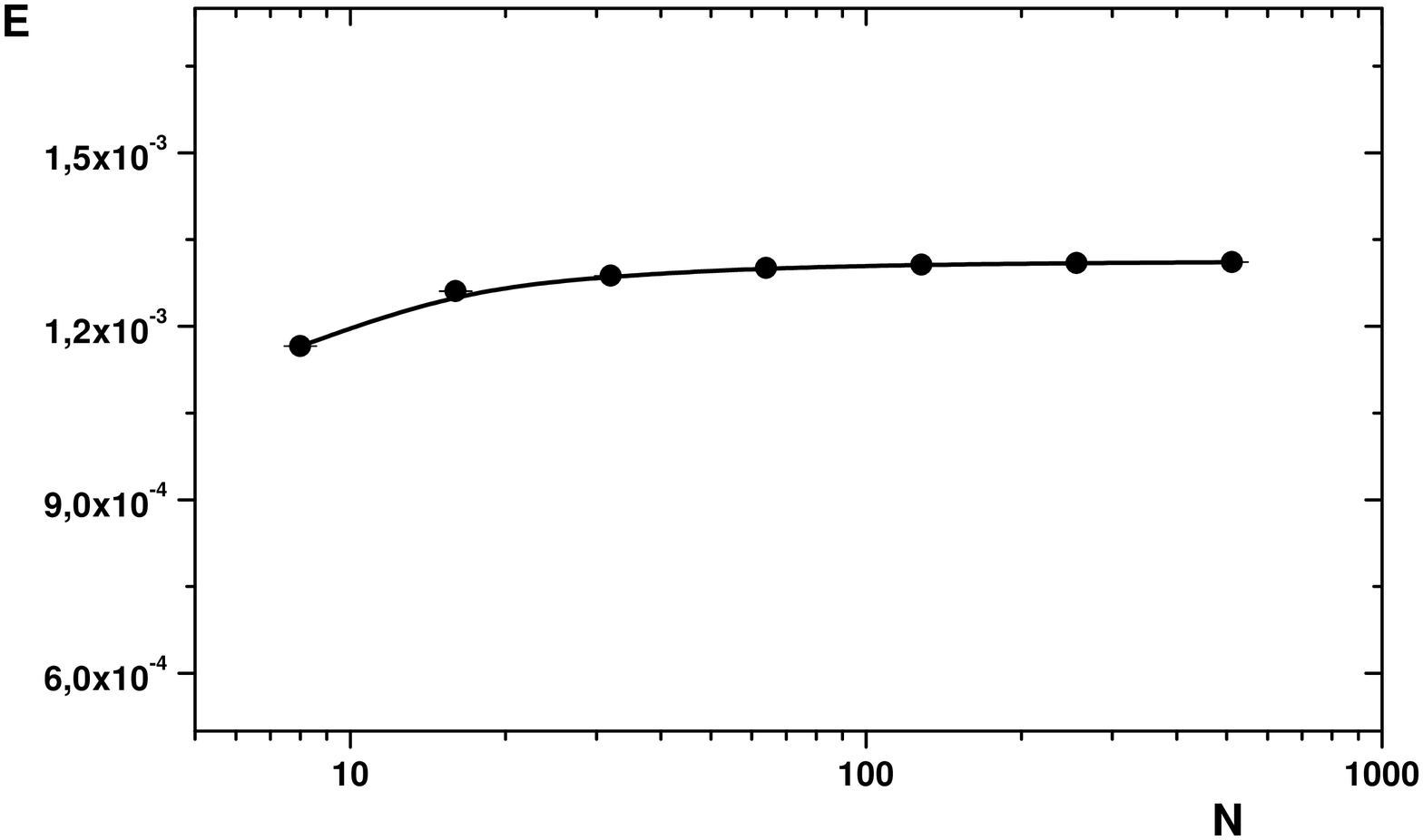}
\caption{Dependence of the energy $E$ measured in DMC simulation
on the size of the system $N$}
\label{Esize}
\end{figure}

In our simulations finite size effects are well under control, as
is evident from Fig.\ref{Esize}, and few tens of particles are
enough to properly simulate the system in the thermodynamic limit.

\subsection{Other sources of errors}

Another type of error comes from the bias due to the trial wave
function. For the quantities measured as pure estimators no effect
was found. Also in the case of other estimators, such as the OBDM,
the use of an optimized trial wavefunction yields reasonably small
differences between mixed and variational estimators (less than
$10\%$). In this case we believe that the extrapolated technique
of section \ref{Extrapolation technique} gives reliable results.

There is another source of error which arises from the hard-sphere
model. In the DMC method the particles propagate according to
diffusion (\ref{R1}) and drift (\ref{R2}) moves. Due to finite
timestep it can happen during the simulation that some particle
overlap with an impurity or another particle. This walker should
not contribute to the calculation of averages because it does not
satisfy the boundary conditions of the problem. One possibility is
to artificially eliminate the walker or to redo the last diffusion
jump so that it will choose another trajectory. Both of the choices
bring in a systematic error. It is hard to investigate the effect
of this error because it comes always mixed up with the timestep
errors (\ref{timestep}) and it is impossible to separate two
effects. It is clear however that the percentage of the restarted
or removed walkers must be kept small. For example in our
calculations this percentage was always smaller than few percent.

%% file: intro.tex
The problem of bosons in the presence of disorder has generated
much theoretical and experimental interest.

The superfluid fraction in liquid $^4$He has been measured for
different types of adsorbing porous media. In vycor, which has
small ($\approx 70$ \AA) pores and porosity of $30\%$ the superfluid
transition is considerably suppressed, but exhibits the same
critical exponent as in the bulk \cite{Reppy}. In contrast, in
aerogel, which is characterized by larger pores with a broad
distribution of sizes and porosity $85-99.5\%$, the superfluid
transition is changed by only a few milli-kelvins while the
critical exponents are quite different from the bulk 
\cite{Wong, Critical exponents}.
Some experimental studies have also investigated, by measuring the
dynamic structure factor, the nature of the elementary excitations
in these systems 
\cite{Aerogel 2}
and the role played by the condensate fraction
\cite{Dynamics of liquid $^4$He in Vycor}.

Theoretical studies of these effects have been proposed, mostly
concerning models on a lattice. Many of the theoretical works
address the problem of the superfluid-insulator transition and the
critical behavior near the phase transition \cite{Fisher, Kanwal,
Ma}. The boson localization and the structure of the Bose-glass
phase have also been investigated.

Quantum Monte Carlo techniques have been used for numerical
simulations of the disordered Bose systems at low temperatures.
Most of them concern systems on a lattice using the Bose-Hubbard or
equivalent models. These studies have been carried out in 1D
\cite{1D}, 2D \cite{2D Ceperley, 2D Zhang,
Superfluid-Insulator Transition in Disordered Boson Systems}
and 3D \cite{Ying-Hong Li, Moon} both at zero and finite temperatures. The structure of
the phase diagram has been investigated and the properties of the
superfluid, Mott insulator and Bose-glass phase have been
addressed.

There are very few simulations of disordered boson systems in the
continuum. In ref. \cite{Boninsegni} the effect of impurities on
the excitation spectrum in liqiud $^4$He is investigated using
PIMC. The same technique is applied to study the effect of disorder
on the superfluid transition in a Bose gas \cite{Gordillo}.

We apply DMC to study a hard-sphere Bose gas at zero temperature in
the presence of hard-sphere quenched impurities. Hard sphere
quenched impurities are easy are easy to implement in a numerical
simulation and provide a reasonable model for liquid $^4$He in the
porous media. Another possible physical realization of this model
is given by trapped gases in the presence of heavy impurities.

The free parameters in our simulations are:

\begin{itemize}
\item the density of the particles $na^3$, where $a$ is the
diameter of the hard-sphere particle scattering length,
\item the concentration of impurities $\chi = N^{imp}/N$ fixed
by the ratio of the number of impurities to the number of particles
used in the simulation,
\item the ratio $b/a$, where $b = R^{imp}+a/2$ with $R^{imp}$
radius of the hard-sphere impurity.
\end{itemize}

The same parameters (\ref{parameters}) were used in the
perturbative analysis discussed previously (see section \ref{Random
external potential}).

The goals of our study are:

\begin{itemize}
\item recover the ``weak'' disorder regime
where the results of perturbation theory apply,
\item verify the scaling behavior of the effects due to
disorder in terms of the single parameter $R = \chi\,(b/a)^2$ as
predicted from the Bogoliubov model
\item understand if one can realize situations where the
superfluid density is smaller than the condensate fraction
\item investigate if within our model there exists a quantum
phase transition for strong disorder
\end{itemize}

%% file: trial.tex
The trial wavefunction for the pure system was constructed in
chapter \ref{Trial Wavefunction}. In this section the same approach
will be extended to systems in the presence of quenched impurities.

The wave function of the system is chosen as the product of
one-body and two-body wavefunctions.

\begin{eqnarray}
\psi_T(\vec r_1,...,\vec r_N) =
\prod\limits_{i=\overline{1,N} \atop j=\overline{1,N}} f_{PP}(|\vec r_i-\vec r_j|)
\prod\limits_{i=\overline{1,N} \atop j=\overline{1,N}_{imp}} f_{PI}(|\vec r_i-\vec r^{~imp}_j|),
\label{disorder wf}
\end{eqnarray}

where $f_{PP}$ stands for the particle-particle wavefunction,
which has already been obtained in section \ref{Trial Wavefunction}
and is defined by (\ref{f_PP}), (\ref{f_PP 1}) and (\ref{f_PP 2}).
In eq. (\ref{disorder wf}) $f_{PI}$ describes the effect of the
impurities on each particle.

To construct $f_{PI}$ we use a similar procedure as for $f_{PP}$,
i.e. we solve the particle-impurity Schr\"odinger equation

\begin{eqnarray}
\left(-\frac{\hbar^2}{2m}\triangle+V_{PI}(\vec r\,) \right)f = {\cal E} f,
\end{eqnarray}

where the reduced particle-impurity mass is equal to the mass of a
particle, because the quenched impurity is infinitely massive Let
us look for the symmetric solution in spherical coordinates

\begin{eqnarray}
-\frac{\hbar^2}{2m} \left(f''+\frac{2}{r}f'\right) + V_{PI}(r)f = {\cal E}f
\end{eqnarray}

The particle is modeled by a hard sphere of diameter $a$ and the
impurity by a hard sphere of diameter $2b-a$. The particle-impurity
interaction potential is

\begin{eqnarray}
V_{PI}(r) =
\left\{
{\begin{array}{ll}
+\infty,&|r| \le b\\
0,&|r| > b\\
\end{array}}
\right.
\end{eqnarray}

where $b$ is the particle-particle $s$-wave scattering length

The dimensionless Schr\"odinger equation has form (lengths in units
of $a$ and energies in units of $\hbar^2/(2ma^2)$)

\begin{eqnarray}
\left\{
{\begin{array}{ll}
\displaystyle f(x) = 0,& |x| \le b/a\\
\displaystyle f''+\frac{2}{x}f' - E f = 0,&|x| > b/a\\
\end{array}}
\right.
\end{eqnarray}

and the differential equation which has to be solved is

\begin{eqnarray}
\left\{
{\begin{array}{l}
\displaystyle f''+\frac{2}{x}f' - E f = 0\\
f\left(\frac{b}{a}\right) = 0
\end{array}}
\right.
\label{DE2}
\end{eqnarray}

The solution is $f(x) = A\sin(\sqrt{E}(x-b/a))\,/x$, with
$A$ being an arbitrary constant.

Let us construct the particle-impurity wave function $f_{PI}$ in
the same way as it was done for the particle-particle wavefunction.
We introduce a matching point $R_{PI}$ and choose

\begin{eqnarray}
f_{PI}(x) =
\left\{
{\begin{array}{ll}
\displaystyle \frac{A}{x} \sin\left(\sqrt{E}\biggl(x-\frac{b}{a}\biggr)\right),& |x| \le R_{PI}\\
\displaystyle 1-B \exp\left(-\frac{x}{\alpha}\right),& |x| > R_{PI}\\
\end{array}}
\right.
\end{eqnarray}

The function $f_{PI}$ must be smooth at the matching point. The
request of continuity for $f_{PI}$, its derivative $f_{PI}'$ and
the local energy
$E_{L}(R) = -(f_{PI}''(R)-2f_{PI}'(R)\,/R)\,/\,f_{PI}(R)$ is fulfilled

\begin{eqnarray}
\left\{
{\begin{array}{lll}
A&=&\displaystyle\frac{R_{PI}}{\sin(u(1-a/bR))}\frac{\xi^2-2\xi}{\xi^2-2\xi+ u^2},\\
B&=&\displaystyle\frac{u^2 \exp(\xi)}{\xi^2-2\xi+ u^2},\\
u&=&\displaystyle\sqrt{E}R_{PI},\\
\xi&=&\displaystyle R_{PI}/\alpha,\\
\displaystyle 1-\frac{a}{bR}&=&\displaystyle\frac{1}{u}\arctg\frac{u(\xi-2)}{u^2+\xi-2}
\end{array}}
\right.
\end{eqnarray}

%% file: disorder2.tex
\subsection{Distribution of impurities}

\begin{figure}
\includegraphics[width=\textwidth]{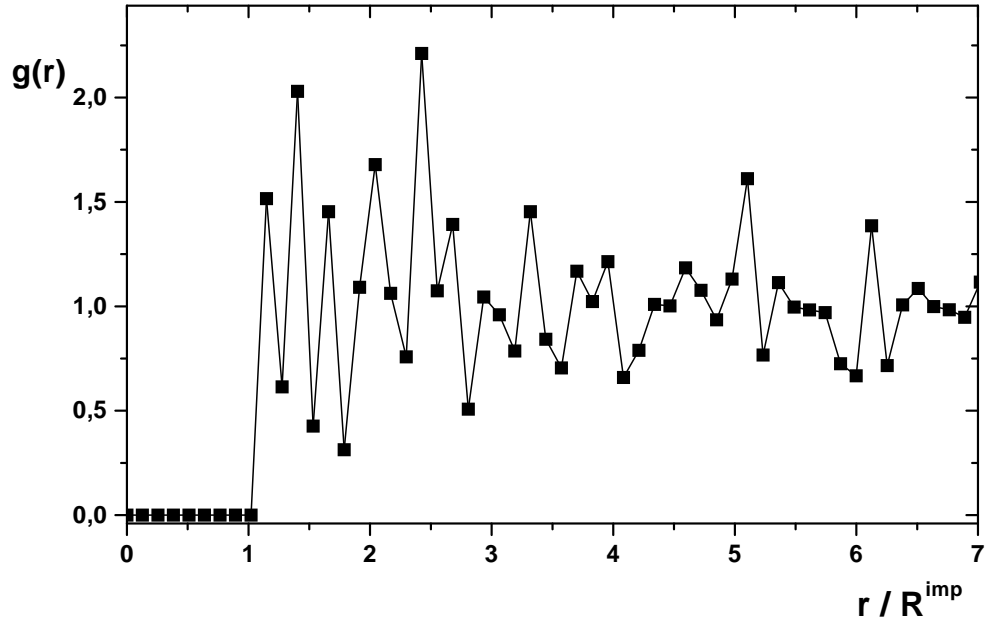}
\caption{Radial distribution function of the impurities for a given
one configuration of disorder}
\label{Disorder g(r)}
\end{figure}

The positions of the quenched impurities in the simulation box are
fixed at the beginning of the simulation run following a uniform
random distribution. We also require that to impurities do not
overlap. To achieve this we throw the impurities at random within
the simulation box and then rethrow all overlapping impurities
until the required configuration is obtained. In
Fig.~\ref{Disorder g(r)} we sow the radial distribution function
of a typical configuration of impurities.

\subsection{Dependence on the number of disorder realizations}

Once the relevant physical quantities (energy, superfluid and
condensate fraction) have been calculated for a given configuration
of disorder, the simulation is repeated for different
configurations and finally the average over disorder
configurations is taken.

\begin{figure}
\includegraphics[width=\textwidth]{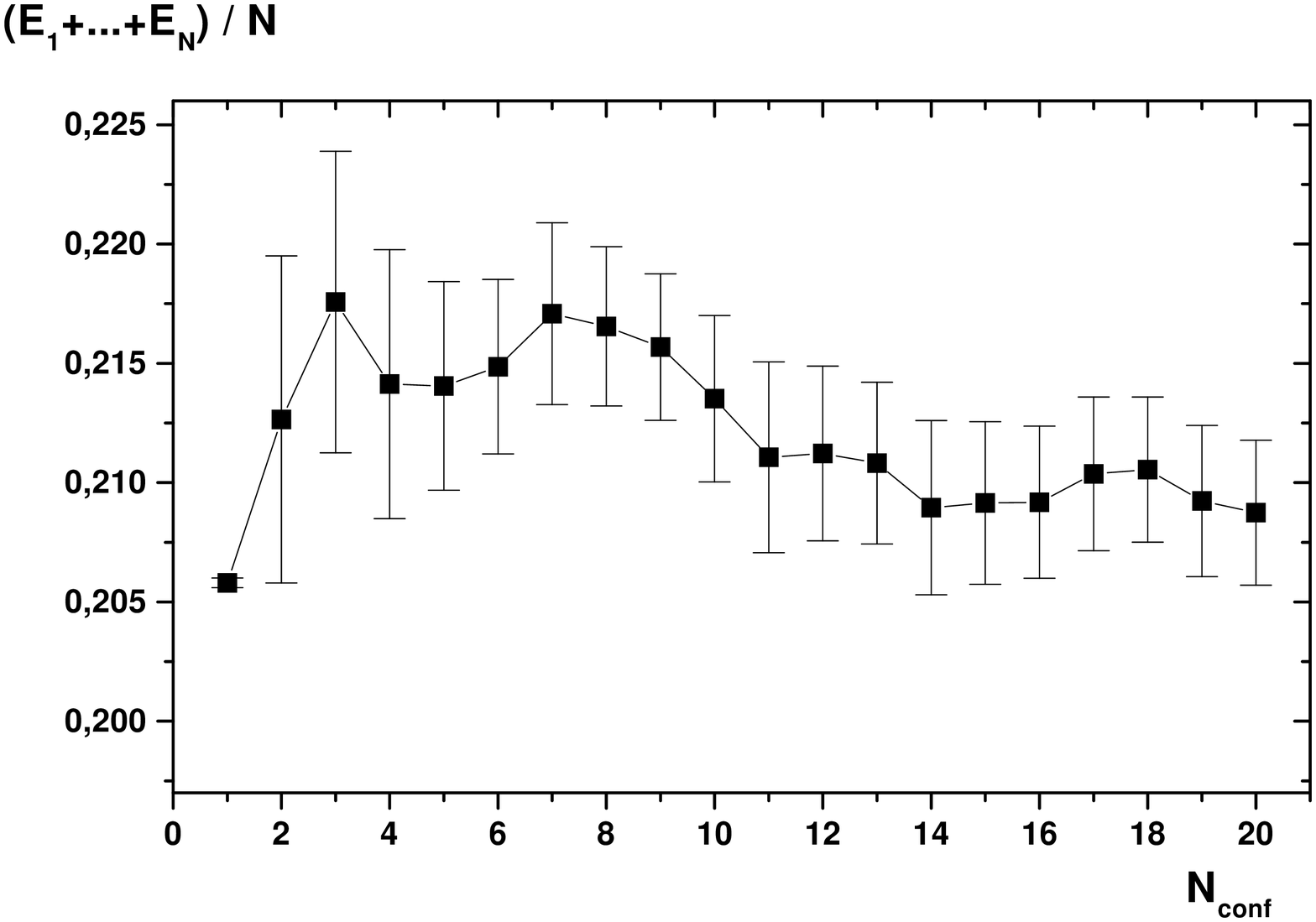}
\caption{The average of the energy over disorder as a function of the
bumber of different disorder realizations. The error bars show the
variance of the averaging.}
\label{Econf}
\end{figure}

As the computation for a given disorder configuration is heavy and
it takes a lot of time to complete the runs with different
configurations, it is very important to understand the behavior of
the statistical error due to the average over realizations. The
results for the energy are shown in Fig.~\ref{Econf}.

Fortunately, energy and the other quantities converge very rapidly
to the mean value. Due to the self-averaging 5 or 6 disorder
configurations are enough for most of the cases.

%% file: results.tex
\subsection{Ground-state energy}

The energy of a dilute Bose gas in the presence of impurities
is given by result (\ref{E}). It is derived under the assumptions that
the gas parameter is small $na^3 \ll 1$ and the external field is
weak.

Using the DMC algorithm we investigate the dependence of the
ground-state energy both on the density $na^3$ and on the strength
of the disorder $R = \chi (b/a)^2$.
 The main
contribution to the energy comes from the mean-field term
(see result (\ref{E GP}) obtained from the Gross-Pitaevskii equation).
In order to better understand the results for the energy
it is useful to subtract the mean-field term
$E_{MN}$ from the total energy $E$.

\begin{figure}
\includegraphics[width=\textwidth]{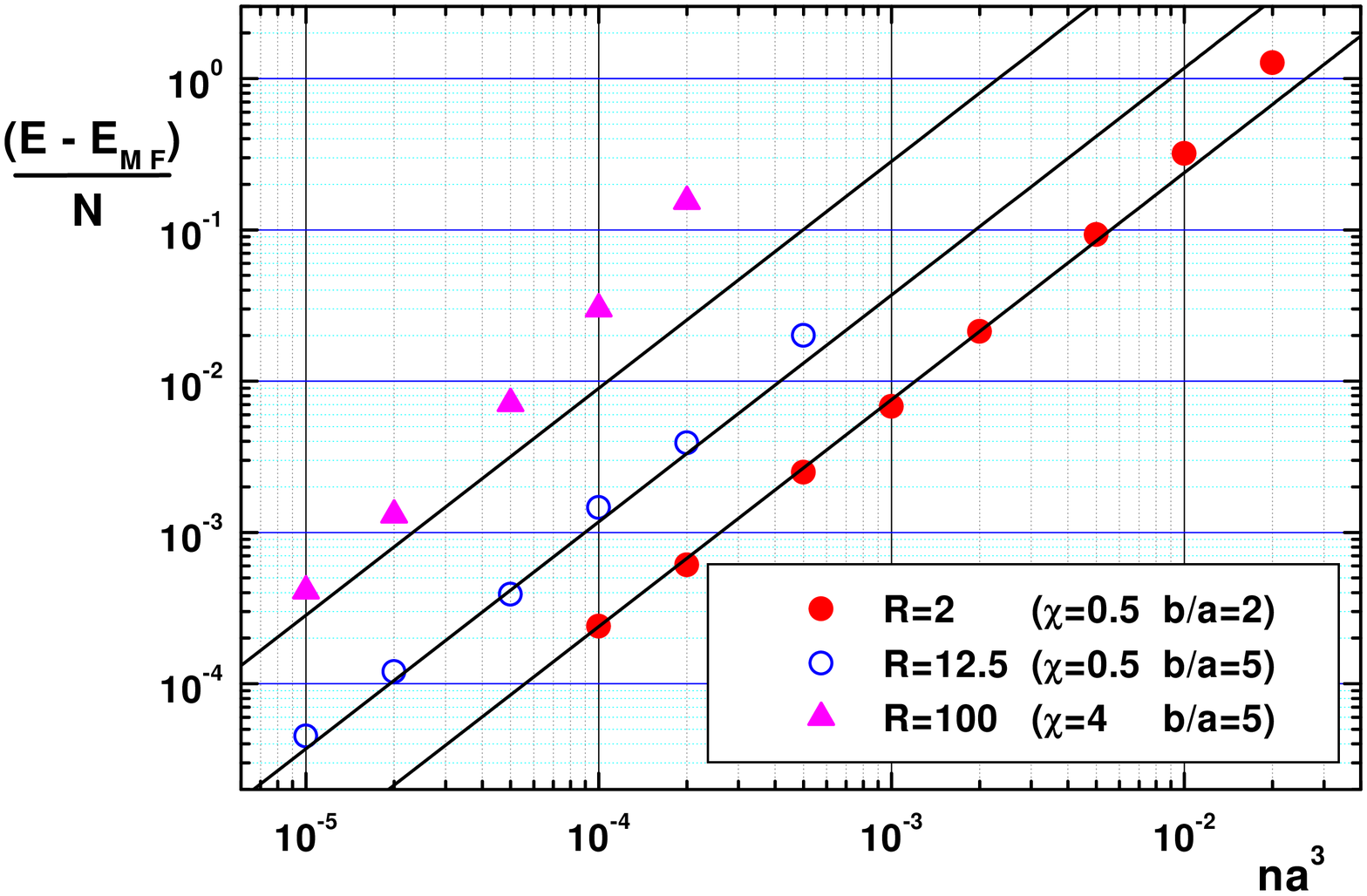}
\caption{Beyond mean-field energy per particle $E - E_{MF}$ as
a function of density $na^3$ for different strengths of disorder
$R$. The solid lines correspond to the analytical prediction
(\ref{E})}
\label{Figure6}
\end{figure}

Let us first consider the energy dependence on $R$. For weak
disorder ($R = 2$) the predictions well agree with the results of
DMC simulations. By increasing $R$ while keeping $na^3$ fixed one
sees deviation from analytical prediction.

The Bogoliubov model is valid if the gas parameter $na^3$ is small.
Fig.~\ref{Figure6} shows that the values of the gas parameter where
the theoretical prediction holds depends on the strength of
disorder. For weak disorder ($R = 2$) agreement is found up to very
high densities $na^3 \approx 10^{-2}$. By increasing the amount of
disorder deviations appear for smaller values of $na^3$. For $R =
12.5$ numerical and analytical results coincide up to $na^3 \approx
5\cdot 10^{-4}$ and for very strong disorder $R = 100$ no agreement
is found at densities $na^3 > 10^{-5}$.

\subsection{Superfluid density and condensate fraction \label{SD CF res}}

The predictions of Bogoliubov theory for the condensate and
superfluid fraction are given by (\ref{condensate depletion}) and
(\ref{SD}). As already mentioned in section \ref{GP SD} a very
interesting consequence of these results is that for any value of
$na^3$ and $R>5.1$ the superfluid fraction $\rho_s/\rho$ is less
than the condensate fraction $N_0/N$.

\begin{figure}
\includegraphics[width=\textwidth]{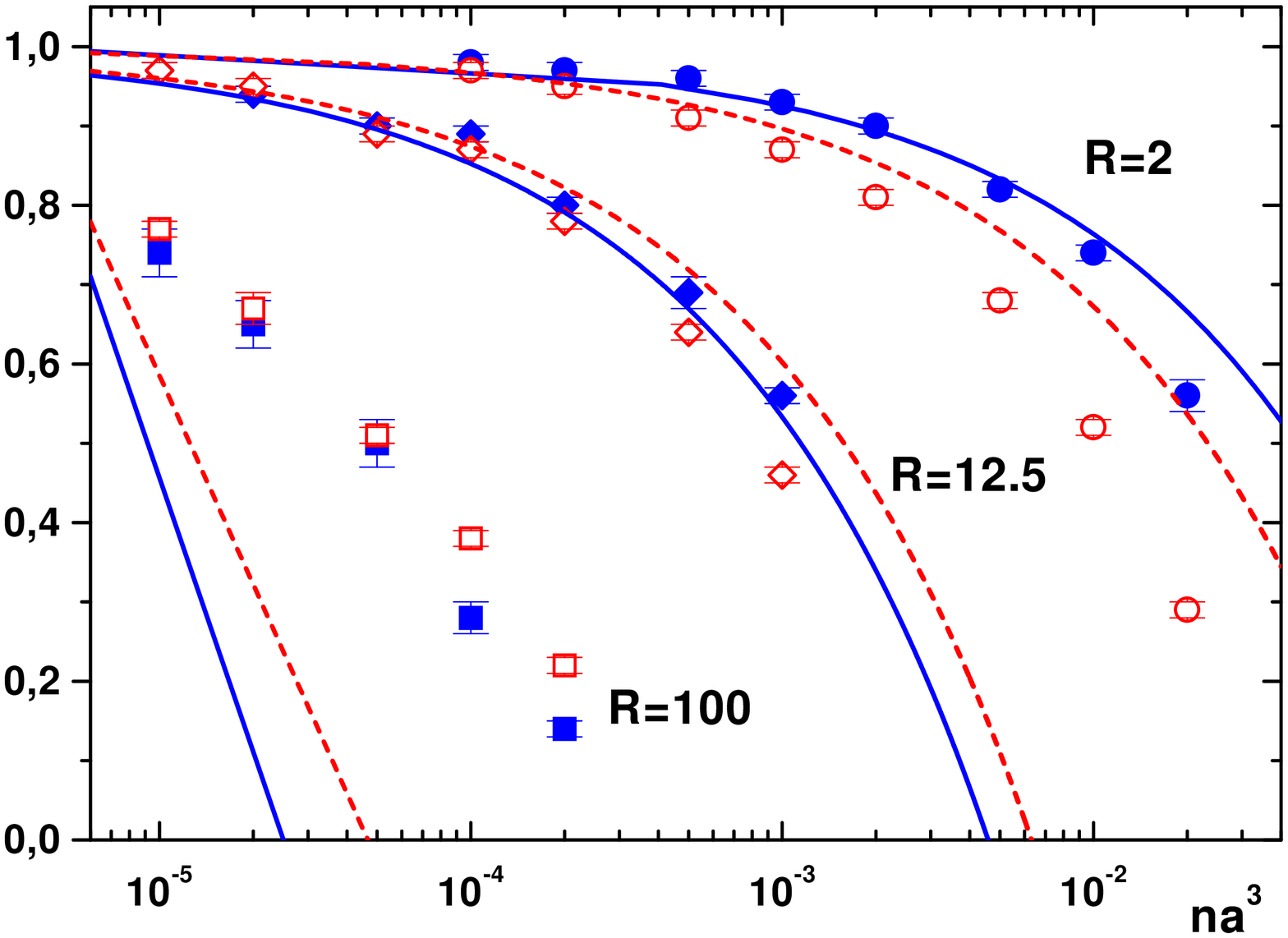}
\caption{Condensate fraction $N_0/N$ (open symbols) and superfluid fraction
$\rho_s/\rho$ (solid symbols) as a functions of density $na^3$, for $R
= 2$, $R = 12.5$, $R = 100$. The solid curve is the Bogoliubov
prediction for the superfluid fraction [Eq.(\ref{SD})], the dashed curve for the
condensate fraction [Eq.(\ref{condensate depletion})].}
\label{Figure5}
\end{figure}

We have investigated the dependence of the superfluid and
condensate fraction on the density $na^3$ and the strength of
disorder $R$. The results of the DMC simulations are presented in
Fig.~\ref{Figure5}. At low density the DMC results always confirm
the predictions of Bogoliubov model, but the region of validity of
the model depends on the strength of disorder. If disorder is weak
($R = 2$) the superfluid fraction is described correctly up to
density $na^3 \approx 5\cdot 10^{-3}$ while the condensate fraction
starts to deviate much earlier. By increasing disorder we find
agreement over a smaller range in density. For $R = 12.5$ the
superfluid fraction agrees with the Bogoliubov prediction only up
to $na^3 \approx 5\cdot 10^{-4}$ and the condensate fraction only
up to $na^3 \approx 2\cdot 10^{-4}$. The strength of disorder is
here larger than the critical value $R_c=5.1$ and Bogoliubov model
predicts $\rho_s/\rho < N_0/N$. Our results show that the
condensate fraction decreases faster than predicted and we do not
see this phenomenon. In the presence of very strong disorder $R =
100$ we find no quantitative agreement for $na^3>10^{-5}$, at large
densities, however, we find $\rho_s/\rho < N_0/N$.

\begin{figure}
\includegraphics[width=\textwidth]{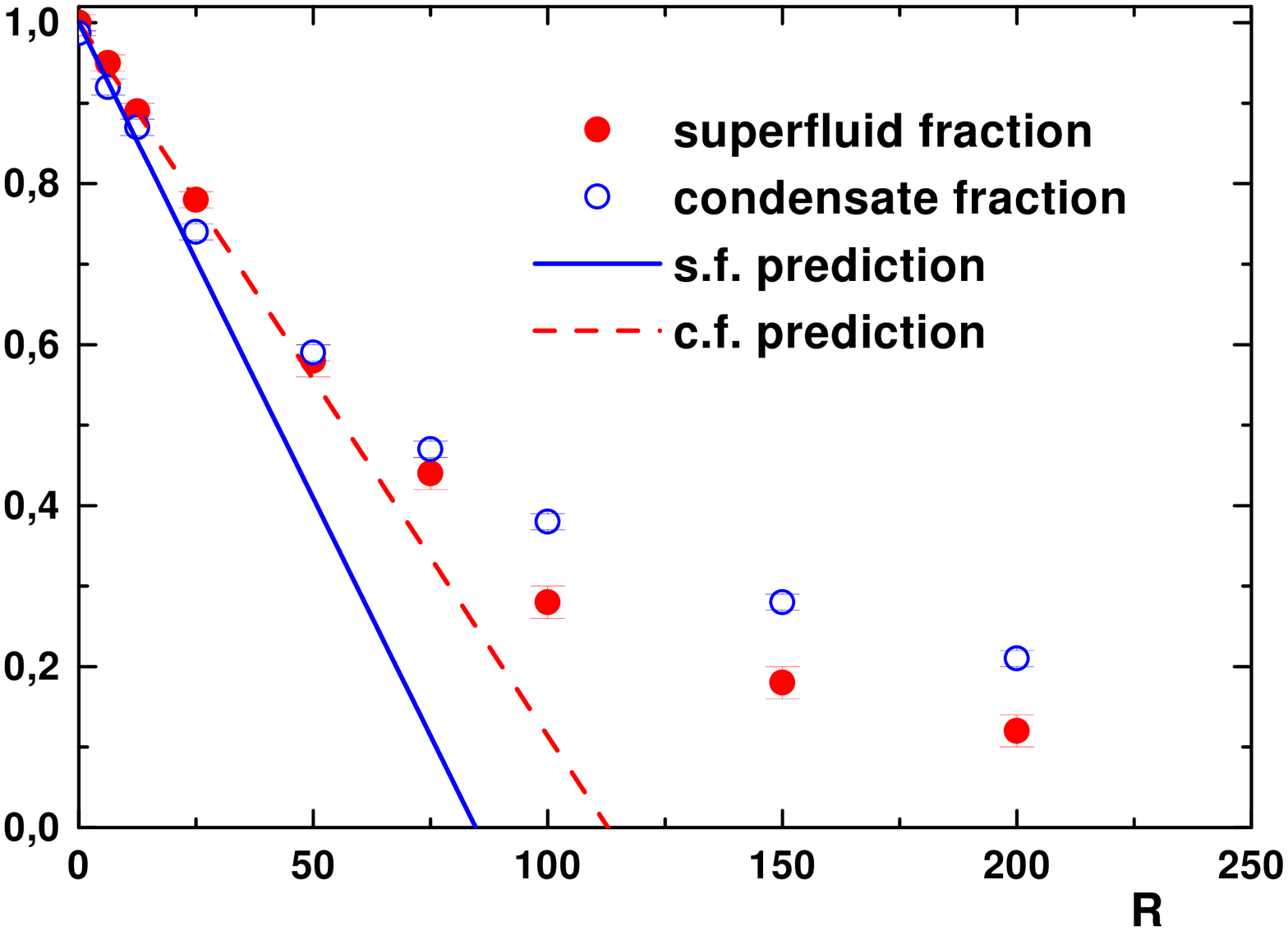}
\caption{Condensate fraction $N_0/N$ and superfluid fraction $\rho_s/\rho$
as a function of $R = \chi(b/a)^2$. Here $na^3 = 10^{-4}$ and $b/a = 5$.
The dashed and solid lines show Bogoliubov predictions for
$N_0/N$ and $\rho_s/\rho$ respectively.}
\label{Figure1}
\end{figure}

Let us now fix $na^3 = 10^{-4}$ and study the dependence of
$\rho_s/\rho$ and $N_0/N$ on the strength of disorder From
Fig.~\ref{Figure1} one sees that for very weak disorder (i.e. small
$R$) Bogoliubov results are valid. For larger disorder we find
deviations. Bogoliubov model predicts a linear dependence on $R$,
with a different slope for $\rho_s/\rho$ and $N_0/N$. We find
instead that the two decrease togrther up to the strong disorder
regime where $\rho_s/\rho < N_0/N$ as in Fig.~\ref{Figure5}.

\begin{figure}
\includegraphics[width=\textwidth]{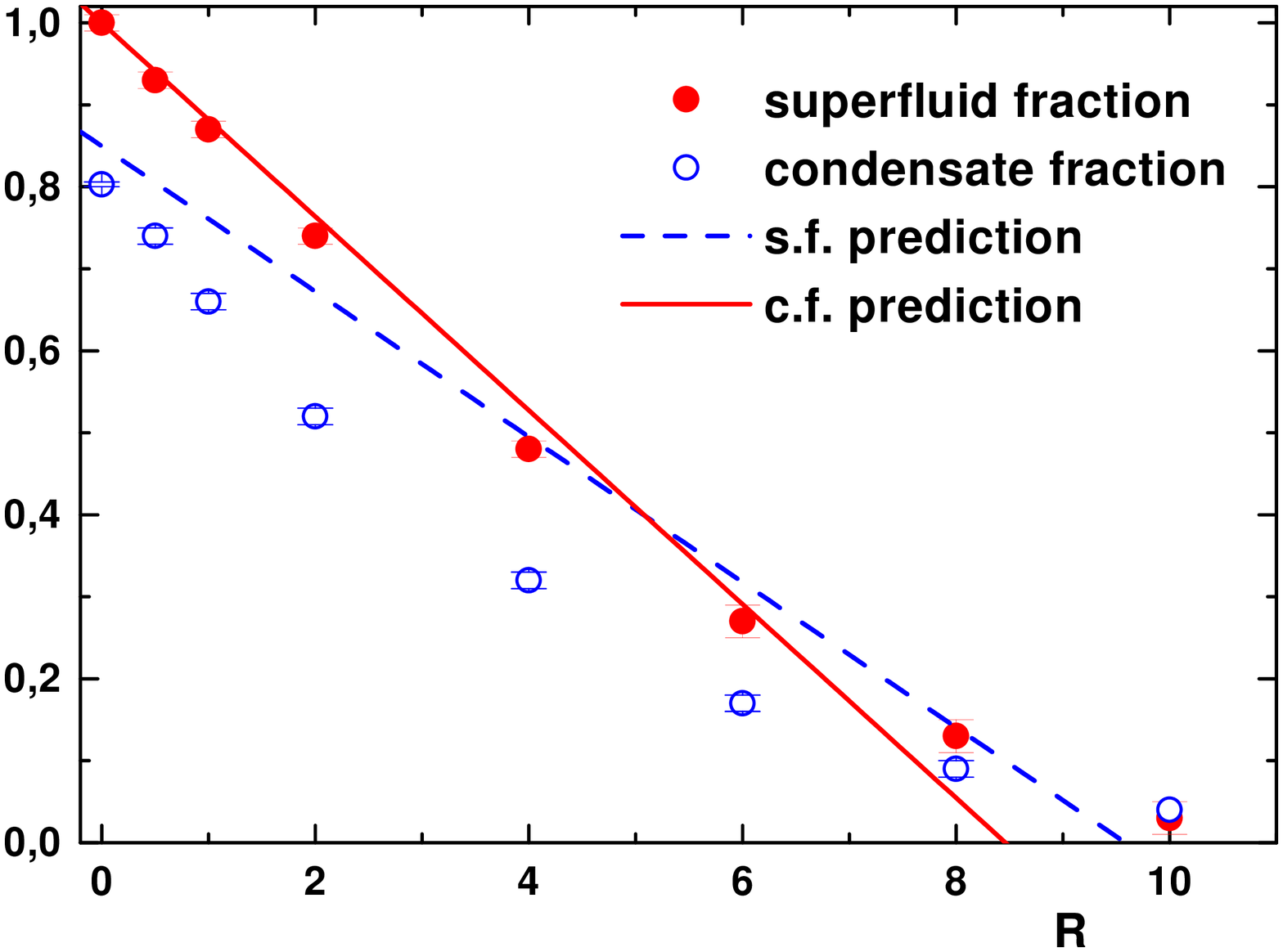}
\caption{Condensate fraction $N_0/N$ and superfluid fraction $\rho_s/\rho$
as a function of $R = \chi(b/a)^2$. Here $na^3 = 10^{-2}$ and $b/a = 2$.
The dashed and solid lines show Bogoliubov predictions for
$N_0/N$ and $\rho_s/\rho$ respectively.}
\label{Figure2}
\end{figure}

A different behavior exhibited at the larger density $na^3=10^{-2}$
as shown in Fig.\ref{Figure2}. Even in the pure case ($R=0$) the
condensate fraction does not agree with Bogoliubov prediction and
by increasing disorder deviations are more evident. On the
contrary, the superfluid density well agrees with Bogoliubov
prediction.

For small values of $R$, analogously to the case $na^3=10^{-4}$,
$\rho_s/\rho$ and $N_0/N$ decrease linearly with a similar slope.

Let us comment on the result $\rho_s/\rho < N_0/N$ which we find
for large values of $R$ (see Fig.~\ref{Figure5} and \ref{Figure1}).
This result is highly unusual since in general\footnote{ The only
exception\cite{StatPhys II} known to us takes place in the vicinity
of the $\lambda$-transition in liquid $^4$He, where
$N_0 \propto(T_\lambda -T)^{2\beta}$
while $\rho_s \propto(T_\lambda-T)^{(2-\alpha)/3}$,
and $2\beta =(2-\alpha)(1-\zeta)/3$ with $\zeta$ positive.
Still, for $^4$He both indices $\alpha$ and
$\zeta$ are very small, so with very good accuracy $\rho_s \sim N_0
\sim (T_\lambda -T)^{2/3}$ and the difference is so small that
there is no hope to measure it experimentally.}
$\rho_s/\rho <N_0/N$.
For example, in liquid $^4$He at
low temperatures only $10\%$ of the particles are
in the condensate, although the system is entirely superfluid.
An extreme example is provided by two-dimensional Bose systems at
$T\ne 0$ which do not exhibit Bose-Einstein condensation
(Hohenberg theorem), but do exhibit superfluidity below the
Kosterlitz-Thoules transition temperature.

It is interesting to understand how it is possible to realize a
system with $\rho_s/\rho<N_0/N$, or even to realize a normal system
(i.e. $\rho_s=0$) with nonzero condensate. A possible answer is by
realizing an adsorbing medium with isolated cavities of typical
size larger than the healing length. The gas gets trapped in the
cavities and a condensate can still exist in each of the cavities,
while the overall conductivity is absent. Let us estimate what is
the critical value for the excluded volume at which the gas the
critical parameters when the fluid can not flow from one side of
the box to another (i.e. the percolation threshold). The relative
excluded volume for the impurities can be estimated as

\begin{equation}
p = \frac{V^{excl}}{V} =
\frac{\frac{4}{3}\pi b^3N_{imp}}{V}
=\frac{4\pi}{3}\chi\left(\frac{b}{a}\right)^3 na^3
\end{equation}

The percolation threshold is given by $p_c=0.70$ \cite{Sok}.
In the case of the results of Fig.~\ref{Figure1}
(i.e. $na^3=10^{-4}$ and $b/a=5$) the percolation threshold
corresponds to $R_c = 350$.
This means that the system is approaching the percolation
limit and we can expect that $\rho_s/\rho < N_0/N$.

\subsection{Scaling behaviour}

The strength of disorder is described by two independent
parameters: the particle-impurity scattering length
$b$ and the concentration $\chi = N^{imp}/N$. One
of the important results of the Bogoliubov model is that a
single parameter $R = \chi (b/a)^2$ is sufficient to describe the
effect of disorder (see eqs. (\ref{E}),
(\ref{condensate depletion}), (\ref{SD})).

\begin{figure}
\includegraphics[width=\textwidth]{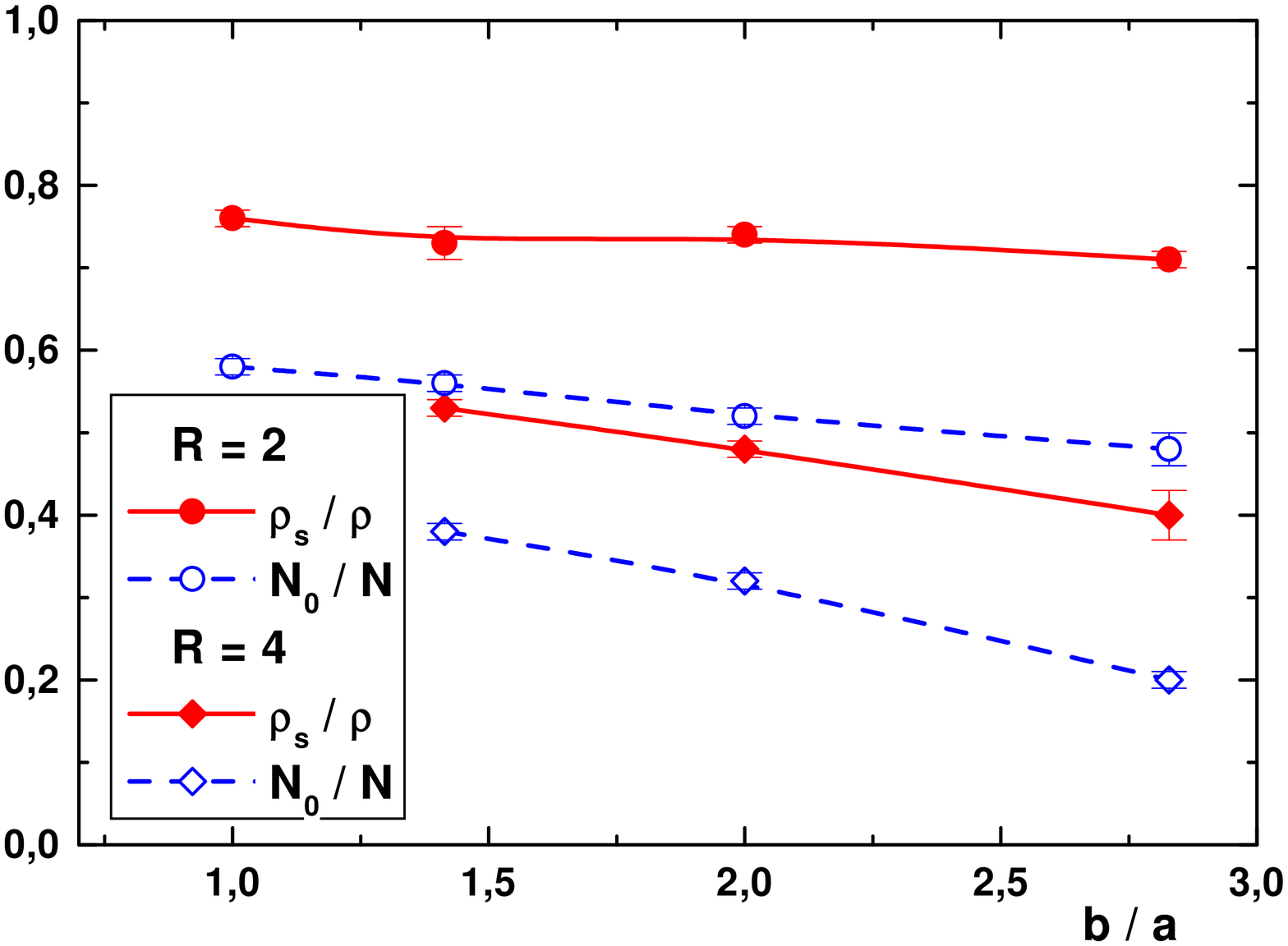}
\caption{Condensate fraction $N_0/N$ and superfluid fraction $\rho_0/\rho$
as functions of impurity size $b$ for two values of the scaling
parameter $R = 2$ and $R = 4$, and density $na^3 = 10^{-2}$}
\label{Figure3}
\end{figure}

\begin{figure}
\includegraphics[width=\textwidth]{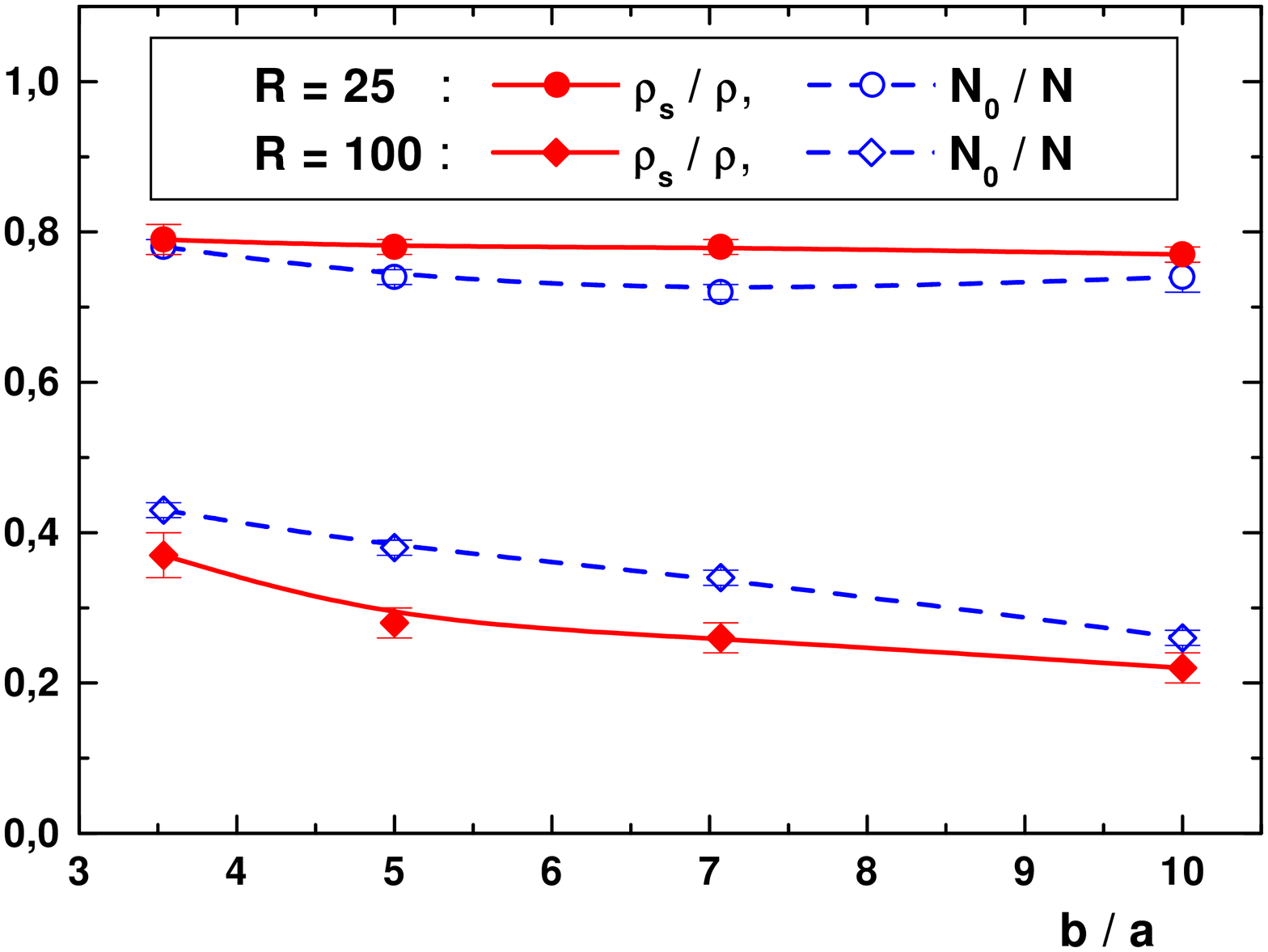}
\caption{Condensate fraction $N_0/N$ and superfluid fraction
as functions of impurity size $b$ for fixed value of the scaling
parameter $R = 25$ and $R=100$, and density $na^3 = 10^{-4}$}
\label{Figure4}
\end{figure}

We have calculated the condensate and superfluid fraction by
changing both $b$ and $\chi$ while keeping $R = \chi (b/a)^2$
constant. The results are shown in Fig.~\ref{Figure3} for
$na^3=10^{-2}$ and $R=2,~4$ and in Fig.~\ref{Figure4} for
$na^3=10^{-4}$ and $R=25,~100$.

Both at low and high density we see that the scaling behavior for
$N_0/N$ and $\rho_s\rho$ is well satisfied for the smallest values
of $R$ ($R =25$ for $na^3=10^{-4}$ and $R=2$ for $na^3=10^{-2}$).
It is worth noticing that these values of $R$ and $na^3$
correspond to regime where the results of first order perturbation
theory do not apply (see Figs. \ref{Figure1}, \ref{Figure2}).
This means that the scaling behavior in the parameter $R$ is valid
also beyond the region of applicability of the perturbation
expansion.

\subsection{Shape of the one-body density matrix}

The shape of the one-body density matrix (\ref{rho}) has been
calculated within the Bogoliubov model and is given by
(\ref{OBDM 1}) in the pure system and by (\ref{OBDM 2}) in the system
with disorder. The derivation of these results is valid
for dilute systems in the presence of weak
disorder, so one expects to find agreement with the DMC
results for small values of $na^3$ and $R$.
Since the Bogoliubov model neglects short-range correlation, we
also expect that result (\ref{OBDM 2}) is valid for large values
of $r$ ($r\gg a$, $r\gg b$).

The DMC simulation gives the mixed estimator for the OBDM and the
VMC gives the
variational estimate. By calculating these results using the
extrapolation technique (\ref{extrapolation}),
we can estimate the pure OBDM.

\begin{figure}
\includegraphics[width=\textwidth]{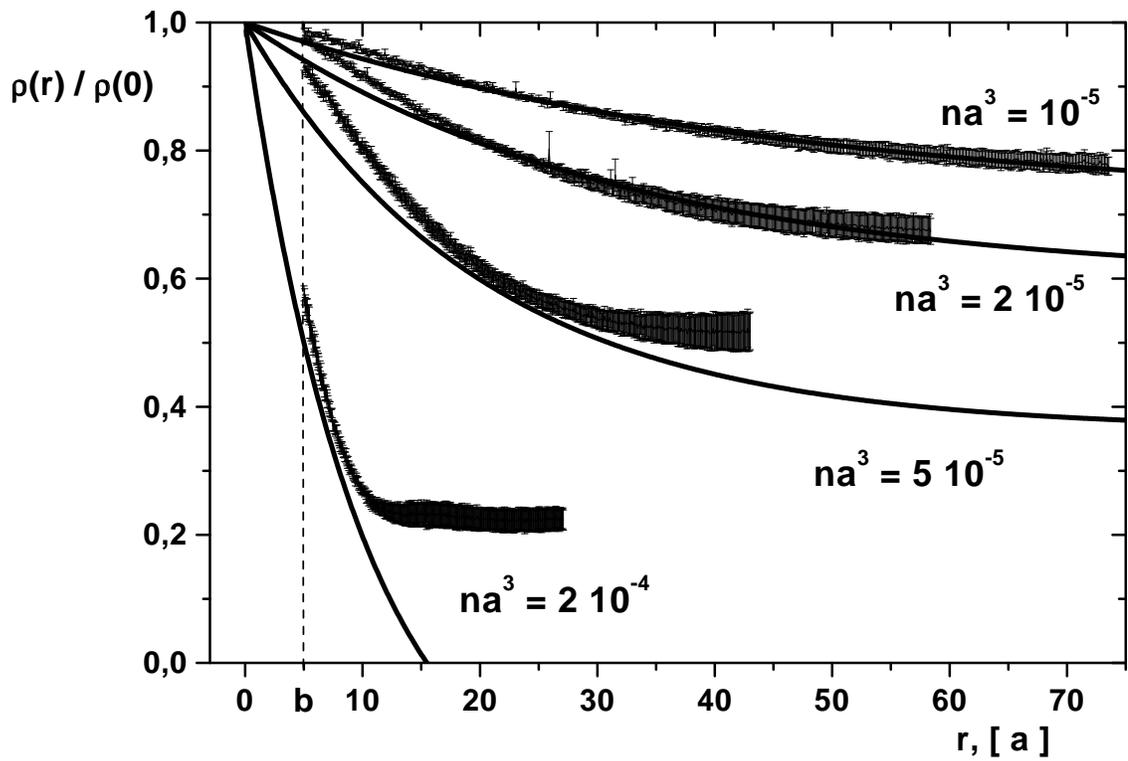}
\caption{One-body density matrix $\rho_s(r)/\rho(0)$ at different
densities $na^3 = 10^{-5}, 2\cdot 10^{-5}, 5\cdot 10^{-5}, 2\cdot 10^{-4}$,
and $R = \chi (b/a)^2 = 100$ with $b/a = 5$.
Solid lines correspond to result (\ref{OBDM 2}).}
\label{OBDMshape}
\end{figure}

We have calculated the OBDM at different densities $na^3$ and for a
fixed strength of the disorder $R=100$. The results are shown in
Fig.~\ref{OBDMshape}. Although the strength of the disorder is
large we find good agreement at small densities while by increasing
in the density we see significant differences.

\subsection{Quantum phase transition}

By adding disorder to the system the condensate and superfluid
density are depleted. One expects that at some critical amount of
disorder superfluidity vanishes and the system becomes normal. We
want to investigate the quantum phase transition of the hard-sphere
gas in the presence of hard-sphere impurities at $T=0$.

In the vicinity of the phase transition the correlation length becomes
large. This means that in the MC simulation one has finite size
errors (see \ref{Finite size errors}) and it is necessary to carry
out calculations with systems of different size and finally
extrapolate to the thermodynamic limit.

\begin{figure}
\includegraphics[width=\textwidth]{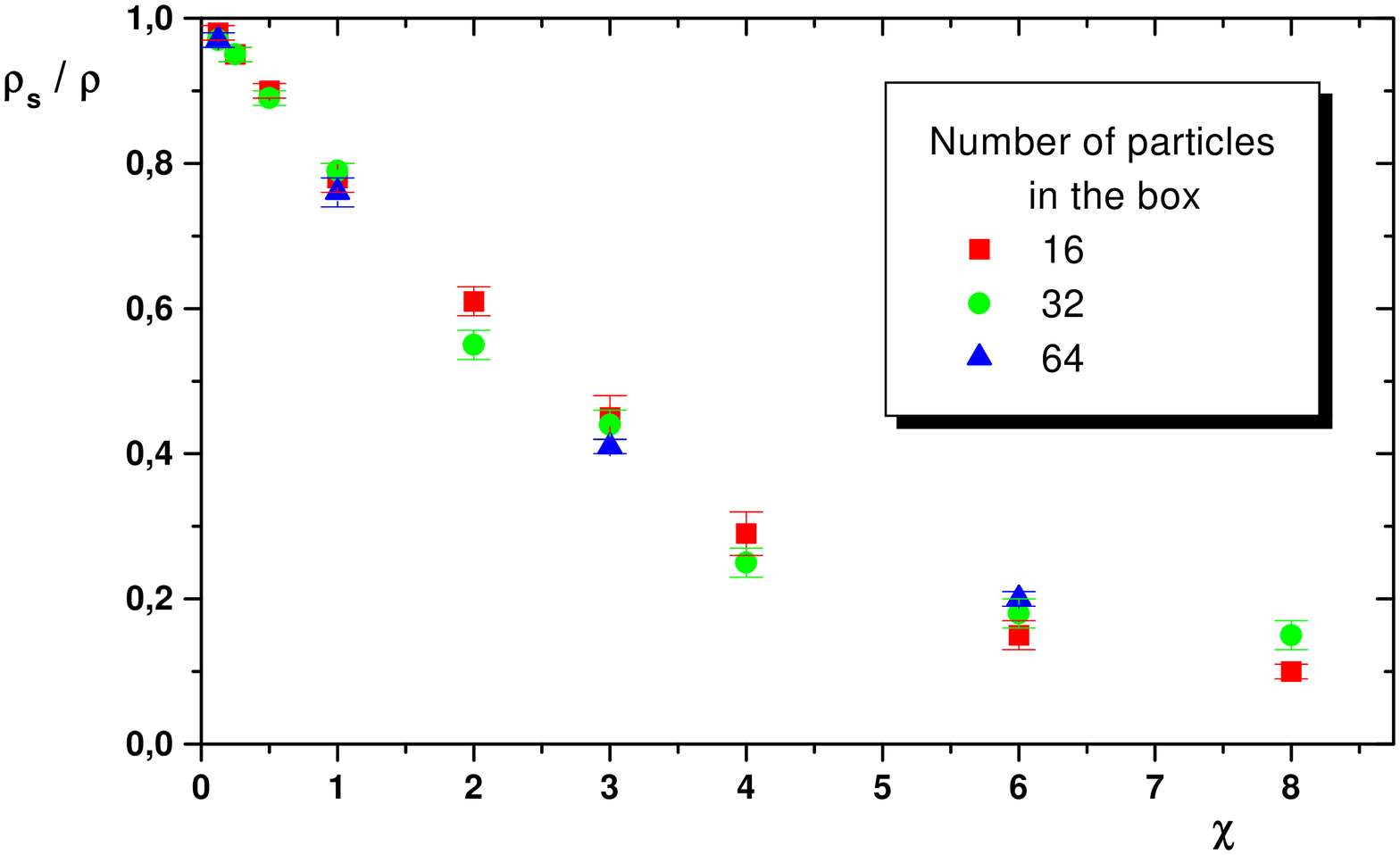}
\caption{Superfluid density measured as a function of $\chi$
at density $na^3 = 10^{-4}$ and $b/a = 5$ with 16, 32 and 64
particles in the simulation box}
\label{FiniteSizen1e_4}
\end{figure}

\begin{figure}
\includegraphics[width=\textwidth]{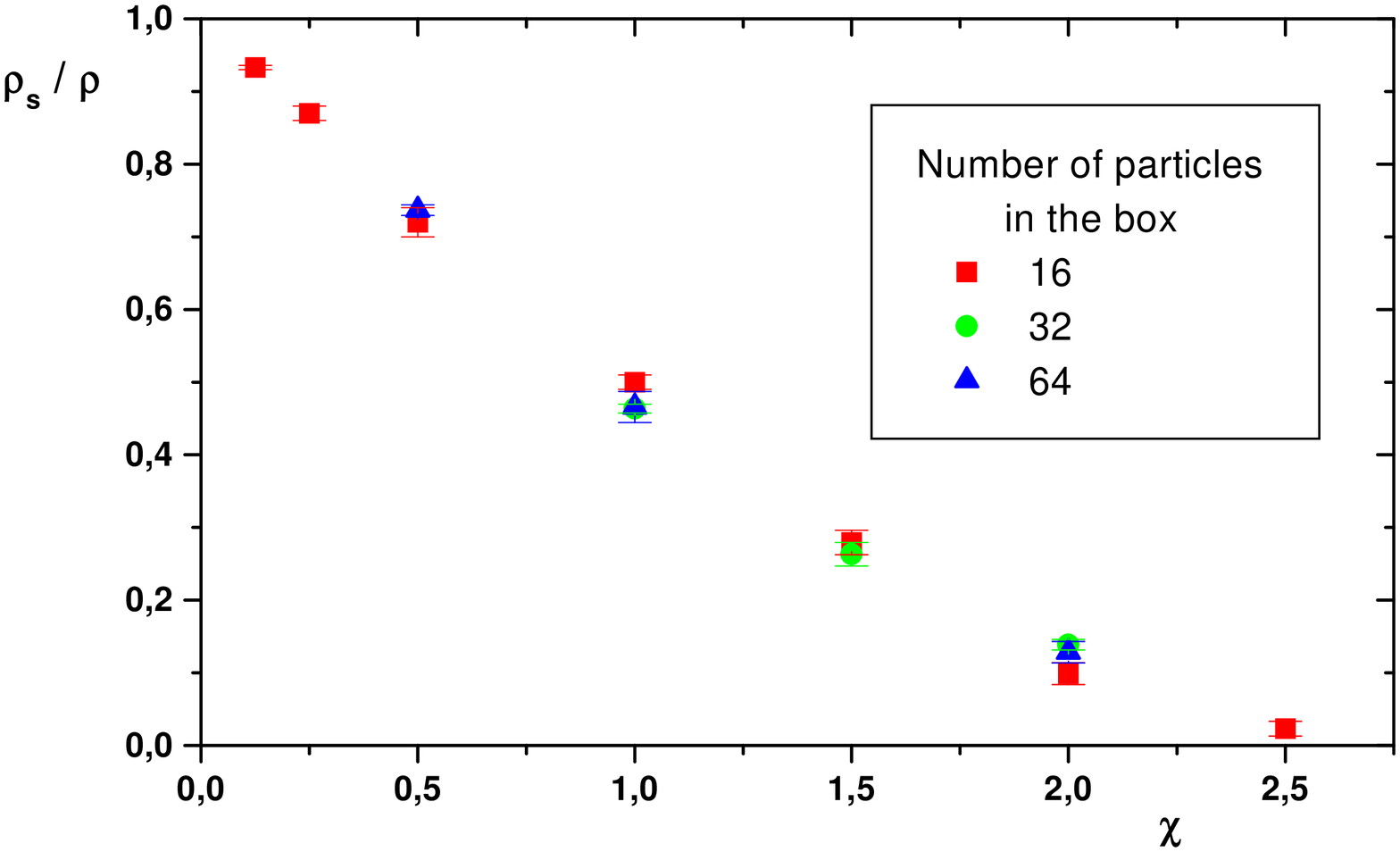}
\caption{Superfluid density measured as a function of $\chi$
 at density $na^3 = 10^{-2}$ and $b/a = 2$ with 16, 32 and 64
 particles in the simulation box}
\label{FiniteSizen1e_2}
\end{figure}

We calculate the superfluid density as a function of the
concentration of impurities $\chi$ while keeping the size of the
impurity constant. The simulation is carried out for systems of 16,
32 and 64 particles with periodic boundary conditions. The results
for the low density $na^3 = 10^{-4}$ is presented in
Fig.~\ref{FiniteSizen1e_4} and for the density $na^3 = 10^{-2}$ in
Fig.~\ref{FiniteSizen1e_2}.

The figures show that there is no significant finite size effect
present which means that we are still far from the critical region.

The further increase of the strength of
disorder in Fig.~\ref{FiniteSizen1e_4} and Fig.~\ref{FiniteSizen1e_4}
is impossible, because
of the constraint of non-overlapping impurities.
We conclude that within our model of non-overlapping impurities
the superfluid-insulator quantum transition is absent.

%% file: appendix.tex
\section{Notation for Fourier transforamtion}

\singlespacing

\begin{eqnarray}
f({\bf r}) = \int e^{i{\bf kr}} f_k \frac{d{\bf k}}{(2\pi)^3}
\end{eqnarray}

\begin{eqnarray}
f_k = \int e^{-i{\bf kr}} f({\bf r}) d{\bf r}
\end{eqnarray}

\begin{eqnarray}
\delta{\bf k} = \int e^{i{\bf kr}} d{\bf r}
\end{eqnarray}

\begin{eqnarray}
\int \delta {\bf k} \frac{d{\bf k}}{(2\pi)^3} = 1
\end{eqnarray}

\section{useful formulae}

\begin{eqnarray}
\begin{array}{lc}
E(p)&=\qquad\displaystyle\sqrt{\left(\frac{\hbar^2k^2}{2m}+(mc^2)^2\right)^2-(mc^2)^2}\quad= \\
&=\qquad\displaystyle\sqrt{\left(\frac{p^2}{2m}\right)^2+(pc)^2}\quad =\quad
mc^2 \left(\frac{p}{mc}\right)\sqrt{1+\frac{1}{4}\left(\frac{p}{mc}\right)^2}
\end{array}
\end{eqnarray}

\begin{eqnarray}
L_p = \frac{1}{mc^2} \left(E(p)-\frac{p^2}{2m} -mc^2\right)
\end{eqnarray}

\begin{eqnarray}
\begin{array}{lc}
L_p^2&\quad\displaystyle
=\frac{2E(p)}{(mc^2)^2} \left(E(p)-\frac{p^2}{2m}-mc^2\right) + 1\quad=\\
&\displaystyle
=\quad\frac{2E(p)}{(mc^2)^2} \left(E(p)-\sqrt{E(p)^2+(mc^2)^2}\right) + 1
\end{array}
\end{eqnarray}

\begin{eqnarray}
\begin{array}{lc}
u_k^2&\quad\displaystyle
= \frac{1}{1-L_p^2}
= \frac{\sqrt{E^2(k)+(mc^2)^2}}{2E(p)}+\frac{1}{2}\quad=\\
&\displaystyle=\quad
\frac{p^2/2m+mc^2}{2E(p)}+\frac{1}{2}
= \frac{(mc^2)^2}{2E(p)\left(\sqrt{E^2(k)+(mc^2)^2}-E(p)\right)}
\end{array}
\end{eqnarray}

\begin{eqnarray}
\begin{array}{lc}
v_k^2&\quad\displaystyle = \frac{L_p^2}{1-L_p^2}
= \frac{\sqrt{E^2(k)+(mc^2)^2}}{2E(p)}-\frac{1}{2}\quad=\\
&\displaystyle=\quad
\frac{p^2/2m+mc^2}{2E(p)}-\frac{1}{2}
= \frac{(mc^2)^2}{2E(p)\left(\sqrt{E^2(k)+(mc^2)^2}+E(p)\right)}
\end{array}\end{eqnarray}

\begin{eqnarray}
u_k v_k = \frac{L_p}{1-L_p^2}
=-\frac{mc^2}{2E(p)}
\end{eqnarray}

\begin{eqnarray}
\frac{1+L_p}{1-L_p} =
\frac{u_k+v_k}{u_k-v_k} =
\frac{\hbar^2k^2}{2mE(p)}
\end{eqnarray}

\doublespacing

\chapter{Series expansion of the OBDM \label{OBDM expansion}}
\newcommand{\K}{{\cal K}}

The one-body density matrix of the pure system at zero temperature
is given by (\ref{OBDM 1}). One can rewrite this integral as

\begin{eqnarray}
\rho^{(1)}(r) =
\frac{na}{\pi r} \int\limits_0^\infty F(\xi)
\sin\left(\frac{\xi r}{\sqrt{2}r_0}\right) d\xi
\label{OBDM integral}
\end{eqnarray}

with the function $F(\xi)$ defined as

\begin{eqnarray}
F(\xi) = 2\left(1+\frac{\xi^2}{4}\right)^{1/2}
-\left(1+\frac{\xi^2}{4}\right)^{-1/2}-\xi
\end{eqnarray}

and $r_0 = a / \sqrt{8\pi na^3}$ being the healing length.

Let us integrate (\ref{OBDM integral}) by parts $2\K$
times. All terms which have the form
$\sin(\xi r/\sqrt{2}r_0)F^{(k)}(\xi)\Bigl.\Bigr|^\infty_0$
 disappear, because
the sinus function is equal to zero in $\xi=0$ and $F^{(k)}(\infty)
= 0$, terms with cosine contribute
$\cos(\xi r/\sqrt{2}r_0)F^{(k)}(\xi)\Bigl.\Bigr|^\infty_0 = -F^{(k)}(0)$.
The result of the integration is the following

\begin{eqnarray}
\begin{array}{rcl}
\rho^{(1)}(r) &=&\displaystyle
\frac{na}{\pi r}
\sum\limits_{k=0}^{\cal K} (-1)^k\left(\frac{\sqrt{2}r_0}{r}\right)^{2k+1}
F^{(2k)}(0)\quad+\\
&&\displaystyle
+\quad(-1)^{\K+1}
\int\limits_0^\infty F^{(2\K+1)}(\xi)
\sin\left(\frac{\xi r}{\sqrt{2}r_0}\right)
\left(\frac{\sqrt{2}r_0}{r}\right)^{2\K+1}
d\xi
\end{array}
\label{integrated}
\end{eqnarray}

The $k$-th derivative of the function $F$ can be calculated from
the differentiation of the Taylor expansion of $F$ in zero

\begin{eqnarray}
F^{(2k)}(0) =
\frac{1+2k}{1-2k}
\frac{\sqrt{\pi}}{\Gamma\Bigl(\frac{1}{2}-k\Bigr)}
\frac{(2k)!}{2^{2k}\,k!}
\end{eqnarray}

As a result one has a representation of $\rho(r)$ in series of
powers of $1/r^2$.

By using the expansion (\ref{integrated}) one can calculate the
leading terms of $\rho^{(1)}(r)$ for $r\to\infty$

\begin{eqnarray}
\rho^{(1)}(r) = \frac{\sqrt{na}}{2\pi\sqrt{\pi}}\cdot\frac{1}{r^2}
-\frac{3\sqrt{na}\,r_0^2}{4\pi\sqrt{\pi}}\cdot\frac{1}{r^4} +
O\left(\frac{1}{r^6}\right)
\label{leading terms}
\end{eqnarray}

Which means that the asymptotic behavior $r\to\infty$ of $\rho(r)$
is one over distance squared\footnote{ The $1/r^2$ dependence can
be calculated directly from (\ref{rho}) by taking the limit $p\to
0$. Then $N_p \to mc/2p$ and integral (\ref{rho}) takes the form of
a Fourier transform of a Coulomb potential $\int\exp(i{\bf
kr})/r\,{\bf dr} = 4\pi/k^2$ and produces the first term of eq.
(\ref{leading terms}).}.
This behavior is correct at distances where the contribution from the
second term in (\ref{leading terms}) can be neglected, i.e.
$r\gg\sqrt{3/2}~r_0\approx r_0$ which means distances
much larger than the healing length.

%% file: acknow.tex

I would like to thank Prof. Sandro Stringari without whom this
thesis would not exist. He gave me the opportunity to join his
research group. Also I found invaluable his help in solving the
organizational problems.

I would like to gratefully acknowledge the enthusiastic supervision
of Dr. Stefano Giorgini and Prof. Lev P. Pitaevskii. They taught me
so many new things and were always ready to answer any of my
countless questions.

I would also like to thank Dr. Yuri E. Lozovik, my supervisor from
Moscow Institute of Physics and Technology,
who offered many insightful comments and suggestions
and always supported me.

I am grateful to many people whom I meet in the University of Trento
and who have assisted me so much in the course of this work.